\documentclass[manuscript]{BSLstyle} 

\nolinenumbers

\AuthAbbrv{b} 

\AuthorEmail[0000-0003-3214-0828]{Tim S. Lyon}[]{timothy\_stephen.lyon@tu-dresden.de}[]
\AuthorEmail[0000-0002-9384-1356]{Agata Ciabattoni}[]{agata@logic.at}[]
\AuthorEmail[0000-0002-9968-5323]{Didier Galmiche}[]{didier.galmiche@loria.fr}[]
\AuthorEmail[0000-0002-9384-1356]{Marianna Girlando}[]{m.girlando@uva.nl}[]
\AuthorEmail[0000-0001-9860-7203]{Dominique Larchey-Wendling}[]{dominique.larchey-wendling@loria.fr}[]
\AuthorEmail[0000-0002-1886-2106]{Daniel M\'ery}[]{daniel.mery@loria.fr}[]
\AuthorEmail[0000-0001-6254-3754]{Nicola Olivetti}[]{nicola.olivetti@univ-amu.fr}[]
\AuthorEmail[0000-0002-7940-9065]{Revantha Ramanayake}[]{d.r.s.ramanayake@rug.nl}[]

\Affiliation{Technische Universit\"at Dresden}{Germany}{}{}{1}
\Affiliation{Vienna University of Technology}{Austria}{}{}{2}
\Affiliation{Universit\'e de Lorraine, CNRS, LORIA}{France}{}{}{3,5,6}
\Affiliation{University of Amsterdam}{Netherlands}{}{}{4}
\Affiliation{Aix-Marseille University}{France}{}{}{7}
\Affiliation{University of Groningen}{Netherlands}{}{}{8}

\Title[Internal and External Calculi]{Internal and External Calculi: Ordering the Jungle without Being Lost in Translations}

\usepackage{hyperref}
\def\eg{{\em e.g.}}

\usepackage[english]{babel}
\usepackage{amssymb, bussproofs, prooftree, amsmath}
\usepackage{color}
\usepackage[all]{xy}
\usepackage{pgf}
\usepackage{changepage}
\usepackage{relsize, bm}
\usepackage{stmaryrd}
\usepackage{tikz}
\usepackage{comment}
\usetikzlibrary{trees}
\usetikzlibrary{fit}
\usepackage{tikz}
\usetikzlibrary{positioning,arrows,calc,decorations.markings}
\tikzset{
modal/.style={>=stealth',shorten >=1pt,shorten <=1pt,auto,node distance=1.5cm,semithick},
world/.style={circle,draw,minimum size=0.5cm,fill=gray!15},
point/.style={circle,draw,inner sep=0.5mm,fill=black},
reflexive above/.style={->,loop,looseness=7,in=120,out=60},
reflexive below/.style={->,loop,looseness=7,in=240,out=300},
reflexive left/.style={->,loop,looseness=7,in=150,out=210},
reflexive right/.style={->,loop,looseness=7,in=30,out=330}
}
\usepackage{dmebproof}
\newenvironment{ebp}[1][center=true,separation=1em,label separation=2pt,right label template=\footnotesize(\inserttext)]{%
	\begin{ebprooftree}[#1]%
	}{%
	\end{ebprooftree}%
}
\newenvironment{ebpd}[1][center=false,separation=1em,label separation=2pt,right label template=\footnotesize(\inserttext)]{%
	\begin{center}%
		\begin{ebp}[#1]%
		}{%
		\end{ebp}%
	\end{center}%
}

\usepackage{url}
\usepackage{proof}

\makeatletter
\AtBeginDocument{%
  \def\doi#1{\url{https://doi.org/#1}}}
\makeatother

\newcommand{\sect}{Section} 
\newcommand{\iffi}{\textit{iff} } 
\newcommand{\fig}{Figure} 
\newcommand{\dfn}{Definition}

\newcommand{\VLe}{{\mathsf V}}
\newcommand{\modele}{\models^{\exists}}
\newcommand{\modela}{\models^{\forall}}
\newcommand{\cless}{\preccurlyeq}

\newcommand{\conp}{\mathrm{CoNP}}

\definecolor{tim}{RGB}{0, 0, 250}

\newcommand{\rel}{\mathcal{R}} 

\newcommand{\lcalcsv}{\mathsf{L(\logicsv)}}

\newcommand{\ncalckt}{\mathsf{N(Kt)}}

\newcommand{\lcalcint}{\mathsf{L(IL)}}
\newcommand{\lcalcintii}{\mathsf{L'(IL)}}
\newcommand{\scalcint}{\mathsf{S(IL)}}
\newcommand{\scalc}{\mathsf{S(CP)}}
\newcommand{\hcalcsv}{\mathsf{H(S5)}}
\newcommand{\ncalcint}{\mathsf{N(IL)}}

\newcommand{\dcalckt}{\mathsf{D(Kt)}}
\newcommand{\lcalckt}{\mathsf{L(Kt)}}
\newcommand{\ns}{\Sigma}

\newcommand{\seq}{\Rightarrow}
\newcommand{\hh}{\mid}
\newcommand{\gseq}{S}

\newcommand{\langmod}{\mathcal{L}_{M}}
\newcommand{\langten}{\mathcal{L}_{T}}
\newcommand{\langint}{\mathcal{L}_{I}}

\newcommand{\logick}{\mathsf{K}} 
\newcommand{\logickt}{\mathsf{Kt}} 
\newcommand{\logicsiv}{\mathsf{S4}}
\newcommand{\logicsv}{\mathsf{S5}}

\newcommand{\logicint}{\mathsf{IL}}

\newcommand{\dia}{\Diamond}
\newcommand{\bldia}{\blacklozenge}
\newcommand{\blbox}{\blacksquare}
\newcommand{\imp}{\rightarrow}
\newcommand{\iimp}{\supset} 
\newcommand{\qdia}{\langle ? \rangle}
\newcommand{\qbox}{[?]}

\newcommand{\forma}{A} 
\newcommand{\formb}{B} 
\newcommand{\five}{\mathsf{5}}

\newcommand{\sar}{\Rightarrow} 
\newcommand{\hsar}{\Rightarrow} 
\newcommand{\lsar}{\Rightarrow} 
\newcommand{\dsar}{\Rightarrow} 
\newcommand{\nsar}{\Rightarrow} 
\newcommand{\seqcomp}{\odot}
\newcommand{\empseq}{\emptyset}
\newcommand{\rpath}{\leadsto_{\rel}}
\newcommand{\deriv}{\mathcal{D}}
\newcommand{\prf}{\mathcal{P}}
\newcommand{\wnest}[1]{{\circ}[#1]}
\newcommand{\bnest}[1]{{\bullet}[#1]}

\newcommand{\id}{(id)}
\newcommand{\botl}{(\bot_{l})}

\newcommand{\disr}{(\lor_{r})}
\newcommand{\negl}{(\neg_{l})}
\newcommand{\negr}{(\neg_{r})}
\newcommand{\disru}{(\lor)}
\newcommand{\conr}{(\land_{r})}
\newcommand{\conru}{(\land)}
\newcommand{\disl}{(\lor_{l})}
\newcommand{\conl}{(\land_{l})}
\newcommand{\impl}{(\rightarrow_{l})}
\newcommand{\impr}{(\rightarrow_{r})}

\newcommand{\diaru}{(\Diamond)}
\newcommand{\diarui}{(\Diamond_{1})}
\newcommand{\diaruii}{(\Diamond_{2})}
\newcommand{\bldiarui}{(\bldia_{1})}
\newcommand{\bldiaruii}{(\bldia_{2})}
\newcommand{\bldiaru}{(\bldia)}
\newcommand{\boxr}{(\Box_{r})}
\newcommand{\boxl}{(\Box_{l})}
\newcommand{\boxli}{(\Box_{l1})}
\newcommand{\boxlii}{(\Box_{l2})}

\newcommand{\boxru}{(\Box)}
\newcommand{\blboxru}{(\blbox)}
\newcommand{\cut}{(cut)}
\newcommand{\refl}{(ref)}
\newcommand{\trans}{(tra)}
\newcommand{\iimpl}{(\iimp_{l})}

\newcommand{\iimpr}{(\iimp_{r})}
\newcommand{\iimplpr}{(p_{\iimp_{l}})}
\newcommand{\idpr}{(r_{id})}
\newcommand{\wk}{(wk)}
\newcommand{\lsub}{(ls)}
\newcommand{\wkl}{(wk_{l})}
\newcommand{\wkr}{(wk_{r})}

\newcommand{\ctr}{(ctr)}
\newcommand{\lift}{(lift)}

\newcommand{\ew}{(ew)}

\newcommand{\rf}{(r\!f)}
\newcommand{\rp}{(rp)}

\newcommand{\lnint}{n} 
\newcommand{\ldkt}{d} 

\newcommand{\R}{\mathcal{R}}

\newcommand{\Ra}{\Rightarrow}

\renewcommand{\P}{\mathcal{P}}

\newcommand{\dual}[1]{\overline{#1}}
\newcommand{\labset}{\mathsf{Lab}}
\newcommand{\set}[1]{\{#1\}}
\newcommand{\lcalcv}{\mathsf{G3V}}

\newcommand{\ie}{\emph{i.e.}}

\newcommand{\wrt}{w.r.t.}

\newcommand{\NAT}{\mathbb{N}} 

\newcommand{\mathacro}[1]{\ensuremath{\mathsf{#1}}}

\newcommand{\BI}{\mathacro{BI}}

\newcommand{\LBI}{\mathacro{LBI}}
\newcommand{\GBI}{\mathacro{GBI}}

\newcommand{\VPA}{A} 
\newcommand{\VPB}{B} 
\newcommand{\VPC}{C} 
\newcommand{\VPP}{p} 
\newcommand{\VPQ}{q} 
\newcommand{\VPR}{r} 
\newcommand{\Prop}{\ensuremath{\mathsf{Prop}}}
\newcommand{\Fm}{\ensuremath{\mathsf{Fm}}}

\newcommand{\LLD}{\delta}
\newcommand{\aneuL}{\mathrm{a}}
\newcommand{\mneuL}{\mathrm{m}}
\newcommand{\rneuL}{\mathrm{r}}
\newcommand{\abotL}{\varpi}
\newcommand{\lab}[1]{\ensuremath{#1}}

\newcommand{\labl}{\lab{\ell}}
\newcommand{\addL}{\mathfrak{a}}
\newcommand{\ADDL}[2]{\ensuremath{\addL(\lab{#1},\lab{#2})}}
\newcommand{\mulL}{\mathfrak{m}}
\newcommand{\MULL}[2]{\ensuremath{\mulL(\lab{#1},\lab{#2})}}
\newcommand{\ronL}{\mathfrak{r}}
\newcommand{\RONL}[2]{\ensuremath{\ronL(\lab{#1},\lab{#2})}}
\newcommand{\leqL}{\leqslant}
\newcommand{\LEQL}[2]{{\lab{#1}\mathrel{\leqL}\lab{#2}}}
\newcommand{\lf}[2]{\ensuremath{#2\mathbin{:}\lab{#1}}}
\newcommand{\Labs}{\mathcal{L}}
\newcommand{\LabsU}{U}
\newcommand{\LabsL}{L}

\newcommand{\BImtop}{\makebox[2.5ex]{\ensuremath{\top\hspace{-0.75ex}{\scriptscriptstyle\mulL}}}} 
\newcommand{\BIatop}{\makebox[2.5ex]{\ensuremath{\top\hspace{-0.5ex}{\scriptscriptstyle\addL}}}} 
\newcommand{\BIabot}{\bot} 
\newcommand{\BImand}{\mathbin{\makebox[1.50ex]{\ensuremath{\ast}}}}
\newcommand{\BImimp}{\mathbin{\makebox[2.25ex]{\ensuremath{-\hspace{-2.75pt}\ast}}}}
\newcommand{\BIaand}{\mathbin{\makebox[1.50ex]{\ensuremath{\wedge}}}}
\newcommand{\BIaimp}{\mathbin{\makebox[2.25ex]{\ensuremath{\supset}}}}
\newcommand{\BIaor}{\mathbin{\makebox[1.50ex]{\ensuremath{\vee}}}}

\newcommand{\BIanul}{\mathrm{\varnothing_{\hspace{-0.3ex}\scriptscriptstyle \addL}}}
\newcommand{\BImnul}{\mathrm{\varnothing_{\hspace{-0.3ex}\scriptscriptstyle \mulL}}}
\newcommand{\BIasep}{\ensuremath{\mathop{;}}}
\newcommand{\BImsep}{\ensuremath{\mathop{,}}}
\newcommand{\BIequi}{\mathbin{\equiv}}

\newcommand{\SEQ}[3][]{\ensuremath{{#2\Rightarrow#3}}}

\newcommand{\SEQZ}[4][]{\ensuremath{{%
{\begin{array}[b]{l}#2\end{array}}%
{\begin{array}[b]{l}#3\end{array}}%
\Rightarrow{#4}}}}

\newcommand{\SG}{\Gamma}
\newcommand{\SD}{\Delta}
\newcommand{\ST}{\Theta}
\newcommand{\PRN}[1]{\ensuremath{\mathrm{#1}}}
\newcommand{\PRNL}[1]{\ensuremath{\mathrm{#1_\mathit{l}}}}
\newcommand{\PRNR}[1]{\ensuremath{\mathrm{#1_\mathit{r}}}}

\newcommand{\gell}{\ldots}

\newcommand{\ded}[1][]{\mathrel{\vdash}}

\newcommand{\frc}[1][]{\models_{#1}} 
\newcommand{\CSM}[3][]{\ensuremath{{#2\frc[#1]#3}}}

\newcommand{\ens}[1]{\{\,#1\/\,\}} 
\newcommand{\mpipe}{\mid}
\newcommand{\ensc}[2]{\ens{#1\mpipe #2}} 
\newcommand{\quo}[1]{``#1''} 

\newcommand{\pki}[2][]{\mathop{[}#2\mathop{]}_{#1}}
\newcommand{\kri}{\pki}
\newcommand{\krm}{M}
\newcommand{\km}{\mathcal{K}}

\newcommand{\Wld}[1][M]{#1} 

\newcommand{\wldm}{w}
\newcommand{\wldn}{u}

\newcommand{\neuS}{1} 
\newcommand{\absS}{\infty} 
\newcommand{\topS}{0} 
\newcommand{\mulS}{\otimes} 
\newcommand{\MULS}[2]{#1\mathbin{\mulS}#2}
\newcommand{\addS}{\oplus}
\newcommand{\ADDS}[2]{#1\mathbin{\addS}#2}
\newcommand{\leqS}{\sqsubseteq}
\newcommand{\LEQS}[2]{#1\mathrel{\leqS}#2}

\newcommand{\BT}[2]{\ensuremath{#2:#1}}
\newcommand{\btol}{\mathfrak{L}}
\newcommand{\BTOL}[2]{\btol(#1,#2)}

\newcommand{\ltob}{\mathfrak{B}}

\newcommand{\Fig}{Figure}
\newcommand{\myskip}{\vspace{8pt}}

\newcommand{\dmKRS}{\mathacro{dmKRS}}
\newcommand{\smKRS}{\mathacro{smKRS}}

\begin{document}

\begin{abstract}
\noindent This paper gives a broad account of the various sequent-based proof formalisms in the proof-theoretic literature. We consider formalisms for various modal and tense logics, intuitionistic logic, conditional logics, and bunched logics. After providing an overview of the logics and proof formalisms under consideration, we show how these sequent-based formalisms can be placed in a hierarchy in terms of the underlying data structure of the sequents. We then discuss how this hierarchy can be traversed using translations. Translating proofs up this hierarchy is found to be relatively straightforward while translating proofs down the hierarchy is substantially more difficult. Finally, we inspect the prevalent distinction in structural proof theory between `internal calculi' and `external calculi.' We discuss the ambiguities involved in the informal definitions of these categories, and we critically assess the properties that (calculi from) these classes are purported to possess.
\end{abstract}


\section*{Introduction}\label{sec:intro}

The widespread application of logical methods in computer science, epistemology, and artificial intelligence has resulted in an explosion of new logics. These logics are more expressive than classical logic, allowing for finer distinctions and a direct representation of notions that cannot be well-stated in classical logic. For instance, they are used to express different modes of truth (e.g., modal logics~\cite{BlaRijVen01}) and to study different types of reasoning, e.g., hypothetical or plausible reasoning (e.g., conditional logics~\cite{Lew73}) or reasoning about the separation and sharing of resources (e.g., bunched implication logics~\cite{Pym02a}). In addition to formalizing reasoning, these logics are also used to model systems and prove properties about them, leading, for example, to applications in software verification (e.g.,~\cite{McM18}).

These applications require the existence of analytic calculi. Analytic calculi consist solely of rules that compose (decompose, in the case of tableau calculi) the formulae to be proved in a stepwise manner, and in particular, the key rule of cut---used to simulate \textit{modus ponens}---is not needed. As a result, the proofs from an analytic calculus possess the subformula property: every formula that appears (anywhere) in the proof is a subformula of the formulae proved. This is a powerful restriction on the form of proofs, which can be exploited to develop automated reasoning methods~\cite{Sla97} and establish important properties of logics such as consistency~\cite{Gen35a,Gen35b}, decidability~\cite{Dyc92}, and interpolation~\cite{Mae60}.

Since its introduction by Gentzen in the 1930s and his seminal proof of the \textit{Hauptsatz} (see~\cite{Gen35a,Gen35b}), the sequent calculus formalism has become one of the preferred frameworks for constructing analytic calculi. This is because such systems are relatively simple and do not require much technical machinery (that is, `bureaucracy') to enable analyticity. The downside of this simplicity is that Gentzen sequent calculi are often not expressive enough to capture many logics of interest in an analytic manner. In response, many proof-theoretic formalisms extending Gentzen's formalism have been proposed over the last 50 years to recapture analyticity for more expressive logics. Such formalisms include hypersequent calculi~\cite{Avr96,Min68}, nested sequent calculi~\cite{Bul92,Kas94}, bunched sequent calculi~\cite{Pym02a}, display calculi~\cite{Bel82,Wan94}, and labeled sequent calculi~\cite{Sim94,Vig00}. Each of these proof-theoretic formalisms (or, \emph{formalisms} for short) is characterized in terms of the standard notation it uses, the data structures employed in sequents, the types of inference rules that normally appear, and the types of properties ordinarily shared by the proof calculi thereof (which serve as \emph{instances} of a formalism). As the notion is central to this paper, we further remark that a proof-theoretic formalism is a \emph{paradigm} in which calculi are built or defined, i.e., a formalism constitutes the \emph{way} in which calculi are constructed, giving rise to a family resemblance shared by systems within the same formalism.

In the literature, proof-theoretic formalisms and calculi have often been classified into \emph{internal} or \emph{external} (e.g.,~\cite{CiaLyoRamTiu21,DalLelOliPim20,Rea15}). There is no formal definition of these properties, and the proof-theoretic community lacks consensus on how each term should be precisely defined. Nevertheless, the literature abounds with informal definitions of internal and external calculi. Typically, internal calculi are described along multiple (sometimes intersecting, sometimes conflicting) lines; e.g., internal calculi have been qualified as proof systems omitting semantic elements from their syntax, or proof systems where every sequent is translatable into an equivalent logical formula. Often times, external calculi are defined as the opposite of internal calculi and, therefore, have also been qualified in multiple ways. 
It has been claimed that internal calculi are better suited than external calculi for establishing properties such as termination, interpolation, and optimal complexity, while it is purported that external calculi are better suited for counter-model generation and permit easier proofs of cut-admissibility and completeness. We will challenge this divide in this paper. 



Due to the diverse number of proof-theoretic formalisms,  
a large body of work has been dedicated to investigating the relationships between calculi within distinct formalisms by means of \emph{translations}. A translation is a function from one proof formalism into another which extends in a natural way to yield structure-preserving maps between derivations in the concrete calculi instantiating the formalisms. Translations represent a useful tool to formally compare and classify different kinds of proof systems.

The goal of this paper is threefold: (1) we discuss the various sequent-style formalisms that have come to prominence in structural proof theory, (2) we map out the relationships between various proof-theoretic formalisms by means of translations, and (3) we investigate the internal and external distinction in light of these relationships.\footnote{We restrict our study to sequent-based formalisms in this paper. Nevertheless, our study retains generality as other types of proof systems, e.g., tableau systems and natural deduction systems, can be transformed into sequent calculi.} What we find is that proof-theoretic formalisms sit within a hierarchy that increases in complexity from Gentzen sequents up to labeled sequents, and is based upon the underlying data structure of the sequents used in the system. We will argue that it is `easier' to translate proofs up this hierarchy than down this hierarchy. Furthermore, we will explain the ambiguities involved in the terms `internal' and `external,' and dispel myths about the properties such calculi are purported to possess. To provide a broad account of sequent-based systems 
we consider a large number of formalisms and systems for a wide array of logics, including modal and tense logics, intuitionistic logic, conditional logics, and bunched implication logic.

This paper is organized as follows: In \sect~\ref{sec:prelims}, we introduce the various families of logics we consider and their semantics, including modal and tense logics, intuitionistic logic, conditional logics, and bunched logics. In \sect~\ref{sec:formalisms}, we explain the various sequent-based formalisms and specific systems that have been introduced for these logics, giving a broad account of the types of sequent systems that appear in the literature. In the subsequent section (\sect~\ref{sec:organizing-jungle}), we organize these proof-theoretic formalisms and systems into a hierarchy and explain how to traverse this hierarchy by means of translations. Lastly, in \sect~\ref{sec:int-ext}, we discuss the internal and external distinction, and clarify what properties `internal' and `external' calculi can be expected to satisfy.

\section{Logical Preliminaries}\label{sec:prelims}

To keep the paper self contained and make for a more general approach, we introduce a variety of logics: modal, tense, intuitionistic, conditional, and bunched logics. We will discuss various sequent-style systems for these logics in the sequel. 

All logics we consider as propositional, and thus, rely on a set $\Prop := \{p, q, r, \ldots\}$ of \emph{propositional atoms} (which are occasionally annotated). For convenience, we will make use of the following two (equivalent) languages:
\begin{eqnarray}
\label{eq:lang:imp}
\forma &::=& p \ | \ \bot \ | \ \forma \lor \forma \ | \ \forma \land \forma \ | \ \forma \imp \forma\\
\label{eq:lang:and-or}
\forma &::=& p \ | \ \overline{p} \ | \ \forma \lor \forma \ | \ \forma \land \forma \     
\end{eqnarray}
When adopting \eqref{eq:lang:imp}, we define $\neg \forma$ as $\forma \imp \bot$ and $\top$ as $ \neg \bot$. In \eqref{eq:lang:and-or}, implication is not a primitive operator and the dual  $\dual{\cdot}$ is allowed to occur only on propositional atoms. However, by taking $\dual{\forma \lor \formb} := \dual{\forma} \land \dual{\formb}$ and $\dual{\forma \land \formb} := \dual{\forma} \lor \dual{\formb}$, we can define $\forma \imp \formb$ as $\dual \forma \lor \formb$. 
The propositional language \eqref{eq:lang:imp} is traditionally used to define \emph{two-sided} sequents, while \eqref{eq:lang:and-or} is convenient when working with \emph{one-sided} sequents. This distinction will become clear in Section~\ref{sec:formalisms}; for the moment, observe that the two formulations are equivalent, as the connectives are interdefinable. For the logics based on classical propositional language, we shall sometimes use \eqref{eq:lang:imp} and sometimes \eqref{eq:lang:and-or}, depending on the corresponding proof system we consider. When introducing intuitionistic logic, we instead need to use \eqref{eq:lang:imp}. To avoid any confusion, we shall use $\iimp$ to denote intuitionistic implication. 


\subsection{Modal Logics}

Even though a study of modalities dates back to Aristotle, modal logic as we know it originates within the work of C.I. Lewis~\cite{Lew18}, who formulated a notion of strict implication in an attempt to resolve certain paradoxes of material implication. Since then, various modal logics have been defined by expanding a base logic (e.g., classical or intuitionistic logic) with \emph{modalities}, that is, logical operators that qualify the truth of a proposition. \emph{Normal modal logics} extend classical propositional logic through the incorporation of \emph{alethic} modalities, namely, ``it is possible that'' (denoted by $\Diamond$) and ``it is necessary that'' (denoted by $\Box$). For an in-depth treatment and presentation of such logics, see Blackburn et al.~\cite{BlaRijVen01}.

For $p$ ranging over $\Prop$, we define the language $\langmod$ of modal logics by adding the $\Box$ modality to  \eqref{eq:lang:imp} above:\footnote{This choice is functional to the choice of the proof systems we will introduce in Section~\ref{sec:formalisms}; the language could have been defined by adding $\Box$ and $\Diamond$ to \eqref{eq:lang:and-or} instead.}
$$
\forma ::= p \ | \ \bot \ | \ \forma \lor \forma \ | \ \forma \lor \forma  \ | \ \forma \imp \forma \ | \ \Box \forma  
$$
We then set $\Diamond A := \lnot \Box \lnot A$.  
Modal formulae are interpreted over \emph{modal Kripke models}. We define such models below, and afterward, define how formulae are interpreted over them.


\begin{definition}[Modal Kripke Model]\label{def:modal-kripke-model}  A \emph{Kripke frame} is defined to be an ordered pair $F := (W,R)$ such that $W$ is a non-empty set of points, called \emph{worlds}, and $R \subseteq W \times W$ is the \emph{accessibility relation}. A \emph{modal Kripke model} is defined to be a tuple $M = (F,V)$ such that $F$ is a Kripke frame and $V : \Prop \to 2^{W}$ is a \emph{valuation function} mapping propositions to sets of worlds.
\end{definition}

\begin{definition}[Semantic Clauses]
\label{def:modal-logic-semantics} 
Let $M = (W,R,V)$ be a modal Kripke model. We define a forcing relation $\models$ such that
\begin{description}

\item[$\bullet$] $M, w \models p$ \iffi $w \in V(p)$;


\item[$\bullet$] $M, w \not\models \bot$;

\item[$\bullet$] $M, w \models \forma \lor \formb$ \iffi $M, w \models \forma$ or $M, w \models \formb$;

\item[$\bullet$] $M, w \models \forma \land \formb$ \iffi $M, w \models \forma$ and $M, w \models \formb$;

\item[$\bullet$] $M, w \models \forma \imp \formb$ \iffi if $M, w \models \forma$, then $M, w \models \formb$;



\item[$\bullet$] $M, w \models \Box \forma$ \iffi for every $u \in W$, if $wRu$, then $M, u \models \forma$;


\item[$\bullet$] $M \models \forma$ \iffi for every $w \in W$, $M, w \models \forma$.

\end{description}
We define a formula $\forma \in \langmod$ to be \emph{$\langmod$-valid} \iffi for all modal Kripke models $M$, $M \models \forma$. We define the minimal normal modal logic $\logick$ to be the set of $\langmod$-valid formulae.
\end{definition}

\begin{figure}[t]
\begin{center}
\bgroup
\setlength{\tabcolsep}{5pt}
\def\arraystretch{1.2}
\begin{tabular}{| c | c | c |}
\hline
Name & Frame Property & Modal Axiom\\
\hline
Reflexivity & $\forall w wRw$ & $\Box \forma \rightarrow \forma$ \\
Symmetry & $\forall w, u (wRu \imp uRw)$ & $\forma \rightarrow \Box \neg \Box \neg \forma$ \\
Transitivity & $\forall w, v, u (wRv \land vRu \imp wRu)$ & $\Box \forma \rightarrow \Box \Box \forma$ \\
Euclideanity & $\forall w, v, u (wRv \land wRu \imp vRu)$ & $\neg \Box \forma \rightarrow \Box \neg \Box \forma$ \\
\hline
\end{tabular}
\egroup
\end{center}
\caption{Frame properties and corresponding axioms.}
\label{fig:correspondences}
\end{figure}

The truth condition for  $\Diamond$ formulae, which is not included in the definition above, is the following: $M, w \models \dia A$ \iffi there exists $u \in W$ such that $M, u \models \forma$. As is well-known in the domain of modal logics, certain formulae are valid on a class of modal Kripke frames if and only if the accessibility relation of those frames satisfies a certain property. This discovery led to the formulation of \emph{correspondence theory}~\cite{BlaRijVen01}, which investigates relationships between modal axioms and the properties possessed by modal Kripke frames. In \fig~\ref{fig:correspondences}, we display some popular and well-studied correspondences, and define the two prominent modal logics $\logicsiv$ and $\logicsv$ accordingly:
 
\begin{definition}[$\logicsiv$ and $\logicsv$] The modal logic $\logicsiv$ is defined to be the set of $\langmod$-valid formula over modal Kripke frames whose relation is reflexive and transitive. The modal logic $\logicsv$ is defined to be the set of $\langmod$-valid formula over modal Kripke frames whose relation is reflexive and Euclidean.
\end{definition}

 






\subsection{Tense Logics}

Tense logics were invented by Prior in the 1950s~\cite{Pri57}, and are types of normal modal logics that not only include the $\dia$ and $\Box$ modalities, but the \emph{converse modalities} $\bldia$ and $\blbox$. These modalities are interpreted in a temporal manner, that is, $\Diamond$ is read as ``in some future moment,'' $\Box$ is read as ``in every future moment,'' $\bldia$ is read as ``in some past moment,'' and $\blbox$ is read as ``in every past moment.'' In this paper, we consider the \emph{minimal tense logic $\logickt$}~\cite{BlaRijVen01}, whose language $\langten$ is defined by adding tense modalities to \eqref{eq:lang:and-or}:
$$
\forma ::= p \ | \ \dual p \ | \ \forma \lor \forma \ | \ \forma \land \forma \ | \ \qdia \forma \ | \ \qbox \forma
$$
where $p$ ranges over $\Prop$, $\qdia \in \{\dia,\bldia\}$, and $\qbox \in \{\Box,\blbox\}$. Formulae from $\langten$ are in negation normal form as this will simplify the sequent systems we consider later on. Note that negation $\neg$ and implication $\imp$ can be defined as usual; see e.g.~\cite{CiaLyoRamTiu21}. Like formulae in $\langmod$, we interpret formulae from $\langten$ over modal Kripke models (\dfn~\ref{def:modal-kripke-model}).

\begin{definition}[Semantic Clauses]\label{def:tense-logic-semantics} Let $M = (W,R,V)$ be a modal Kripke model. We define the forcing relation $\models$ as follows, where the clauses for $\land$ and $\lor$ are as in \dfn~\ref{def:modal-logic-semantics}:
\begin{description}

\item[$\bullet$] $M, w \models p$ \iffi $w \in V(p)$;

\item[$\bullet$] $M, w \models \dual p$ \iffi $w \not\in V(p)$;



\item[$\bullet$] $M, w \models \dia \forma$ \iffi there exists a $u \in W$ such that $wRu$ and $M, u \models \forma$;

\item[$\bullet$] $M, w \models \bldia \forma$ \iffi there exists a $u \in W$ such that $uRw$ and $M, u \models \forma$;

\item[$\bullet$] $M, w \models \Box \forma$ \iffi for every $u \in W$, if $wRu$, then $M, u \models \forma$;

\item[$\bullet$] $M, w \models \blbox \forma$ \iffi for every $u \in W$, if $uRw$, then $M, u \models \forma$;

\item[$\bullet$] $M \models \forma$ \iffi for every $w \in W$, $M, w \models \forma$.

\end{description}
We define a formula $\forma \in \langten$ to be \emph{$\langten$-valid} \iffi for all modal Kripke models $M$, $M \models \forma$. We define the minimal tense logic $\logickt$ to be the set of $\langten$-valid formulae.
\end{definition}

\subsection{Conditional Logics}\label{condlog}

Conditional logics formalize hypothetical statements that cannot be faithfully represented using material implication and/or the modal operator $\Box$. Examples of such sentences are \emph{counterfactual conditionals}, e.g., ``if $A$ were the case, then $B$ would be the case,'' and non-monotonic statements, such as ``Normally, if $A$ then $B$.'' To represent counterfactuals and non-monotonic sentences, conditional logics introduce in a classical propositional language a binary modal operator, the \emph{conditional}, which we denote by $A > B$. Although the family of conditional logics contains over 50 systems, we concentrate on the conditional logic $\VLe$, which is the basic logic of counterfactual reasoning as introduced by D.~Lewis~\cite{Lew73}. 
We choose to focus our attention on this conditional logic as its proof-theoretical treatment, while being simpler than for other systems, illustrates the methods needed to capture conditionals.



In Lewis's account, the conditional operator is defined in terms of another operator, referred to as \emph{comparative plausibility} and denoted $\cless$. The formula $A\cless B$ states that ``$A$ is at least as plausible as $B$.'' The conditional operator $A > B$ may then be defined as $(\bot \cless A) \lor \neg ((A \land \neg B) \cless (A \land B))$, meaning that ``either $A$ is impossible or $A \land \neg B$ is less plausible than $A \land B$.'' This definition can be simplified by replacing $A \land B$ by $A$ in the second disjunct, yielding: $A > B := (\bot \cless A) \lor \neg ((A \land \neg B) \cless A).$
 
Conversely, the comparative plausibility $\cless$ can be defined in terms of the conditional operator $>$. Our full language is defined by adding 
$\cless$ to the language \eqref{eq:lang:imp} above, for $p$ ranging over $\Prop$:
$$
 \forma ::= p \ | \ \bot \ | \ \forma \lor \forma \ | \ \forma \land \forma \ | \ \forma \imp \forma \ | \ \forma \cless \forma
$$


From a semantic point of view, logic $\VLe$ is characterized by special kinds of neighborhood models, introduced by Lewis and called \emph{sphere models}. In this semantics, each world is assigned a \emph{system of spheres}, i.e., a set of nested neighborhoods. The intuition is that spheres represent degrees of plausibility, so that worlds in smaller/innermost spheres are considered more plausible than worlds contained solely in larger/outermost spheres. 

\begin{definition}[Sphere Model] A sphere model $M = (W, S, V)$ is a triple such that $W$ is a non-empty set of worlds, $S: W \to 2^{2^W}$, $V: \Prop \to 2^W$ is a valuation function, and the following conditions are satisfied, for every $w\in W$:
\begin{description}

\item[$\bullet$] \emph{Non-emptiness}: For every $\alpha \in S(w)$, $\alpha\not= \emptyset$;

\item[$\bullet$] \emph{Nesting}: For every $\alpha,\beta\in S(w)$, either $\alpha\subseteq\beta$ or $\beta \subseteq\alpha$.

\end{description}
The elements of a system of spheres $S(w)$ are called \emph{spheres} and we use $\alpha, \beta, \ldots$ to denote them.
\end{definition}

\begin{definition}[Semantic Clauses] Given a sphere model $M = (W, S, V)$, the forcing relation $\models$ is defined by adding to the forcing relation defined in \dfn~\ref{def:modal-logic-semantics} the following clause for $\cless$:
\begin{description}





\item[$\bullet$] $M, w \models A\cless B$ \iffi for all $\alpha \in S(w)$, if there exists $u \in \alpha$ such that $M, u \models B$, then there exists $v \in \alpha$ such that $M, v \models A$.

\end{description}
We define a formula $\forma$ to be \emph{valid} \iffi for any sphere model $M$, $M \models \forma$, and we define the conditional logic $\VLe$ to be the set of all valid formulae over the class of sphere models.
\end{definition}

 Given a sphere model  $M = (W, S, V)$, we find it useful to introduce the following notation for a sphere $\alpha\in S(w)$:
\begin{description}

\item[$\bullet$] $\alpha\modele A$ \iffi there is $ w\in \alpha$ such that $ M, w\models A$;
	
\item[$\bullet$] $\alpha\modela A$ \iffi for all $ w\in \alpha$, it holds that $ M, w\models A$.
\end{description}
With this notation, the semantic clause for the comparative plausibility operator becomes the following: 
\begin{description}

\item[$\bullet$] $M, w \models A\cless B$ \iffi for all $\alpha \in S(w)$, if $\alpha\modele B$, then $\alpha\modele A$.
\end{description}

For completeness, we report the truth condition of the conditional operator, which is defined in our language: 
\begin{description}
    \item[$\bullet$] $M, w \models A > B$ \iffi either for all $\alpha \in S(w)$, $\alpha\not\modele A$, or there is an $\alpha\in S(w)$ such that $\alpha\modele A$ and $\alpha\modela A \to B$.
\end{description}

By imposing additional properties on sphere models, we obtain Lewis's family of conditional logics. Later on, we will define calculi for conditional logics that incorporate inference rules for the comparative plausibility operator.

\subsection{Intuitionistic Logic}

Intuitionistic logic aims to capture the notion of constructive proof, something which classical logic fails to do~\cite{Hey30}. For this reason intuitionistic logic does not contain familiar classical axioms such as the law of the excluded middle ($p\lor \lnot p$) and double negation elimination ($\lnot\lnot p \iimp p$). The language of intuitionistic logic $\langint$ is just the language of classical logic \eqref{eq:lang:imp}, where classical implication is replaced by intuitionistic implication $\iimp$:
$$
\forma ::= p \ | \ \bot \ | \ \forma \lor \forma \ | \ \forma \land \forma \ | \ \forma\iimp\forma
$$
 where $p$ ranges over $\Prop$. As usual, we define $\lnot A := A \iimp \bot$. Contrary to the classical case, the connectives $\land$ and $\lor$ are not inter-definable. Intuitionistic formulae are interpreted over \emph{intuitionistic Kripke models}.
 
\begin{definition}[Intuitionistic Kripke Model]  An \emph{intuitionistic Kripke frame} is defined to be an ordered pair $F := (W,\leq)$ such that $W$ is a non-empty set of points, called \emph{worlds}, and the \emph{accessibility relation} $\leq \ \subseteq W \times W$ is reflexive and transitive. A \emph{intuitionistic Kripke model} is defined to be a tuple $M = (F,V)$ such that $F$ is an intuitionistic Kripke frame and $V : \Prop \to 2^{W}$ is a \emph{valuation function} satisfying the \emph{persistence} condition, that is, if $w \in V(p)$ and $w \leq u$, then $u \in V(p)$.
\end{definition}

\begin{definition}[Semantic Clauses] Given an intuitionistic Kripke model $M = (W,\leq,V)$, we define a forcing relation $\models$ for propositional atoms, $\bot$, $\lor$, and $\land$ as in \dfn~\ref{def:modal-logic-semantics}, but replace the clause for $\imp$ with the following: 
\begin{description}




\item[$\bullet$] $M, w \models \forma \iimp \formb$ \iffi for all $u\in W$, if $w \leq u$ and $M,u\models\forma$, then $M,u\models\formb$;


\end{description}
We define a formula $\forma \in \langint$ to be \emph{intuitionistically-valid} \iffi for all intuitionistic Kripke models $M$, $M \models \forma$. We define \emph{intuitionistic logic} $\logicint$ to be the set of intuitionistically-valid formulae.
\end{definition}

A basic fact of intuitionistic logic is that it forms a proper subset of classical logic. Conversely, via the double negation translation, classical logic can be embedded into $\logicint$~\cite{Bus98}. Moreover, there is also a natural embedding of (axiomatic extensions of) intuitionistic logic (called \emph{intermediate logics}) into (axiomatic extensions of) the modal logic $\logicsiv$~\cite{Kri65}.

\subsection{Bunched Logics}

\label{sec:bunched-logics}

Bunched logics are substructural logics\footnote{Logics that include connectives for which at least one of the usual structural rules (weakening, contraction, exchange, associativity) does not hold.} arising from mixing different kinds of connectives associated with a resource aware interpretation. In this paper we focus on the logic of bunched implications ({\BI})~\cite{OHearn99,Pym02a}, which combines propositional intuitionistic logic with intuitionistic multiplicative linear logic. More formally, the set of formulae of {\BI}, denoted $\Fm$, is given by the following grammar in BNF:
 \[
   \VPA ::= \VPP \mpipe 
      \underbrace{%
      \BImtop \mpipe 
      \VPA \BImand \VPA  \mpipe 
      \VPA \BImimp \VPA
  	  }_{\text{multiplicatives}}  
      \mpipe
      \underbrace{%
      \BIatop \mpipe 
      \BIabot \mpipe
      \VPA \BIaand \VPA \mpipe
      \VPA \BIaor \VPA \mpipe
      \VPA \BIaimp \VPA
    }_{\text{additives}}
 \]
 where $p$ ranges over $\Prop$.
 
{\BI} admits various forcing semantics~\cite{Pym02a}, called \emph{resource semantics}, which use more elaborate models than those used for intuitionistic logic or  modal logics. The most intuitive and widespread resource semantics for~{\BI} is the monoid based Kripke semantics that arises from the definition of a (multiplicative) resource composition $\mulS$ on worlds, viewed as resources.
The monoid based Kripke semantics can be generalized to a relational semantics~\cite{Gal05a} replacing both the accessibility $\leqS$ and the monoidal composition $\mulS$ with a ternary relation~$R$ on worlds à la Routley-Meyer (thus reading $\LEQS{\MULS{\wldm}{\wldm'}}{\wldn}$ as a particular case of $R\wldm\wldm'\wldn$).

The standard monoid based Kripke semantics~\cite{Gal05a} requires only one resource composition reflecting the properties of the multiplicative connectives. The specifics of the additive connectives are implicitly reflected in their forcing clauses using the properties of the accessibility relation.
However, in this paper, we follow~\cite{Gal19a} and use a monoid based Kripke semantics in which we add a second (additive) resource composition~$\addS$ that explicitly reflects the syntactic behaviour of $\BIaand$ into the semantics.

\begin{definition}[Resource Monoid]
\label{def:krm}
A \emph{resource monoid (RM)} is a structure 
$\krm=(\Wld,\mulS,\neuS,\addS,\topS,\absS,\leqS)$ where 
$(\Wld, \mulS, \neuS)$, $(\Wld, \addS, \topS)$ are commutative monoids
and $\leqS$ is a preordering relation on $\Wld$ such that: 
\begin{description}
\item[$\bullet$] for all $\wldm \in \Wld$, $\LEQS{\wldm}{\absS}$ and $\LEQS{\absS}{\MULS{\absS}{\wldm}}$;
\item[$\bullet$] for all $\wldm, \wldn \in \Wld$, $\LEQS{\wldm}{\ADDS{\wldm}{\wldn}}$ and 
  $\LEQS{\ADDS{\wldm}{\wldm}}{\wldm}$;
\item[$\bullet$]
$
 \text{if } \LEQS{\wldm}{\wldn} \text{ and } \LEQS{\wldm'}{\wldn'}
 \text{, then }  
  \LEQS{\MULS{\wldm}{\wldm'}}{\MULS{\wldn}{\wldn'}} \text{ and }
  \LEQS{\ADDS{\wldm}{\wldm'}}{\ADDS{\wldn}{\wldn'}}
$.
\end{description}
\end{definition}

Let us remark that the conditions of Definition~\ref{def:krm} imply that
$\absS$ and $\topS$ respectively are greatest and least elements and 
that $\addS$ is idempotent.

\begin{definition}[Resource Interpretation]
\label{def:kri}
Given a resource monoid $\krm$, a \emph{resource interpretation (RI) for $\krm$}, 
is a function $\kri{-}: \Fm \longrightarrow 2^{\Wld}$ satisfying
$\forall\,\VPP \in \Prop$, $\absS \in \kri{\VPP}$
and $\forall\,\wldm, \wldn \in \Wld$, if $\wldm \in \pki{\VPP}$ and $\LEQS{\wldm}{\wldn}$, then $\wldn \in \kri{\VPP}$.
\end{definition}

\begin{definition}[Kripke Resource Model]
\label{def:krmodel}
A \emph{Kripke resource model (KRM)} is a structure 
$\km = (\krm, \frc, \kri{-})$ 
where $\krm$ is a resource monoid, 
$\kri{-}$ is a resource interpretation
and $\frc$ is a forcing relation such that:
 \begin{description}
\item[$\bullet$] $\CSM{\krm,\wldm}{\VPP}$ \iffi $\wldm \in \kri{\VPP}$;

        \item[$\bullet$] $\CSM{\krm,\wldm}{\BIabot}$ \iffi $\LEQS{\absS}{\wldm}$;
		      $\CSM{\krm,\wldm}{\BIatop}$ \iffi $\LEQS{\topS}{\wldm}$;
		      $\CSM{\krm,\wldm}{\BImtop}$ \iffi $\LEQS{\neuS}{\wldm}$;

        \item[$\bullet$] $\CSM{\krm,\wldm}{\VPA \BImand \VPB}$ \iffi for some
                  $\wldn, \wldn'$ in $\Wld$, 
                  $\LEQS{\MULS{\wldn}{\wldn'}}{\wldm}$,
                  $\CSM{\krm,\wldn}{\VPA}$ and $\CSM{\krm,\wldn'}{\VPB}$;

        \item[$\bullet$] $\CSM{\krm,\wldm}{\VPA \BIaand \VPB}$ \iffi for some
                  $\wldn, \wldn'$ in $\Wld$, 
                  $\LEQS{\ADDS{\wldn}{\wldn'}}{\wldm}$,
                  $\CSM{\krm,\wldn}{\VPA}$ and $\CSM{\krm,\wldn'}{\VPB}$;
		\item[$\bullet$] $\CSM{\krm,\wldm}{\VPA \BImimp \VPB}$ \iffi for all
                  $\wldn, \wldn'$ in $\Wld$ such that $\CSM{\krm,\wldn}{\VPA}$
                  and $\LEQS{\MULS{\wldm}{\wldn}}{\wldn'}$, 
                  $\CSM{\krm,\wldn'}{\VPB}$;
		\item[$\bullet$] $\CSM{\krm,\wldm}{\VPA \BIaimp \VPB}$ \iffi for all
                  $\wldn, \wldn'$ in $\Wld$ such that $\CSM{\krm,\wldn}{\VPA}$
                  and $\LEQS{\ADDS{\wldm}{\wldn}}{\wldn'}$, 
                  $\CSM{\krm,\wldn'}{\VPB}$; 
		\item[$\bullet$] $\CSM{\krm,\wldm}{\VPA \BIaor \VPB}$ \iffi
                  $\CSM{\krm,\wldm}{\VPA}$ or $\CSM{\krm,\wldm}{\VPB}$. 
 \end{description}

A formula $\VPA$ is valid in the Kripke resource semantics \iffi
$\CSM{\krm,\neuS}{\VPA}$ in all Kripke resource models.
\end{definition}

Let us call {\dmKRS} the Kripke resource semantics based on double monoids as defined in this section (and first introduced in~\cite{Gal19a}) and call {\smKRS} the standard one based on single monoids (as defined in~\cite{Gal05a}).
The {\smKRS} is recovered from the {\dmKRS} by erasing all references to $\addS$ in Definition~\ref{def:krm} and replacing the forcing clauses for the additive connectives in
Definition~\ref{def:krmodel} with the following ones:
\begin{description}

    \item[$\bullet$] $\CSM{\krm,\wldm}{\BIatop}$ \iffi always;

    \item[$\bullet$] $\CSM{\krm,\wldm}{\VPA \BIaand \VPB}$ \iffi 
          $\CSM{\krm,\wldm}{\VPA}$ and $\CSM{\krm,\wldm}{\VPB}$;

    \item[$\bullet$] $\CSM{\krm,\wldm}{\VPA \BIaimp \VPB}$ \iffi
          for all $\wldn$ in $\Wld$ such that $\LEQS{\wldm}{\wldn}$, 
          if $\CSM{\krm,\wldn}{\VPA}$ then $\CSM{\krm,\wldn}{\VPB}$. 
\end{description}

The forcing clauses for the additive connectives might seem more natural in the {\smKRS} as they convey their intuitive interpretation in terms of resource sharing (see ~\cite{Pym02a} for details).
Indeed, it is immediately seen that $\VPA \BIaand \VPB$ is about $\VPA$ and $\VPB$ sharing the same resource (namely, the resource $\wldm$ in $\CSM{\krm,\wldm}{\VPA \BIaand \VPB})$.
Let us remark that the interpretation of the multiplicative connectives in terms of resource separation 
($\VPA \BImand \VPB$ holds for resource $\wldm$ if it can be split into two resources $\wldn$ and $\wldn'$, one satisfying $\VPA$ and the other satisfying $\VPB$) remains the same in both semantics.
Let us also mention that the interpretation of {\BI} formulae in terms of resource sharing and separation is one of the key differences with Linear Logic \cite{Gir87} and its interpretation of formulae in terms of resource accounting (consumption and production).

Although (arguably) less intuitive, the {\dmKRS} clearly makes the presentation of the semantics more uniform as the differences between the additive and multiplicative connectives are captured at the level of the algebraic properties that the corresponding monoidal operators should satisfy (e.g., idempotence for $\addS$ but not for $\mulS$) and not at the level of the forcing clauses which can therefore be formulated in a similar way.
As we shall see later in Section~\ref{sec:bunched-sequents} it also makes the {\dmKRS} more in tune with the bunched sequent calculus of {\BI} in which the differences between the additive and multiplicative connectives are handled at the level of the structural rules that the connectives should satisfy and not at the level of the logical rules (which share a similar form). 

\newcommand{\hide}[1]{}
\hide{%
Definition~\ref{def:krmodel} clearly shows that adding an explicit additive composition allows us to improve the symmetry and uniformity of the presentation as the differences between the additive and multiplicative connectives are captured at the level of the algebraic properties that the corresponding monoidal operators satisfy (e.g., idempotence for $\addS$ but not for $\mulS$) and not at the level of the forcing clauses themselves (they share a similar form), just like the bunched sequent calculus introduced later in Section~\ref{sec:bunched-sequents} handles the differences at the level of the structural rules that the connectives satisfy, the shape of the logical rules themselves being similar. 
Therefore, the resource semantics presented in this section better reflects the behaviour of the calculi that we shall discuss in the forthcoming sections than the traditional monoid based resource semantics, which on the other hand better reflects the intuitive interpretation of the additives connectives in terms of resource sharing.
Indeed, consider the traditional clauses for the additive connectives given below:
 \begin{itemize}
		\item $\CSM{\wldm}{\VPA \BIaand \VPB}$ iff 
                  $\CSM{\wldm}{\VPA}$ and $\CSM{\wldm}{\VPB}$;
		\item $\CSM{\wldm}{\VPA \BIaimp \VPB}$ iff for all
                  $\wldn$ in $\Wld$ such that 
                  $\LEQS{\wldm}{\wldn}$, 
                  if $\CSM{\wldn}{\VPA}$ then $\CSM{\wldn}{\VPB}$; 
 \end{itemize}
It is immediately seen that the truth of $\VPA \BIaand \VPB$ at a resource $\wldm$ expresses the sharing of the resource $\wldm$ by $\VPA$ and $\VPB$ and that $\VPA \BIaimp \VPB$ is about preserving the truth of $\VPB$ whenever $\VPA$ is true in all resources $\wldn$ ``bigger'' than $\wldm$.
}

\section{An Overview of the Proof-Theoretic Jungle}\label{sec:formalisms}

\begin{figure}[t]
\begin{center}
\bgroup
\setlength{\tabcolsep}{5pt}
\def\arraystretch{1.1}
\begin{tabular}{|c|c|c|}

\hline
\textbf{System Type} & \textbf{Data Structure of Sequent} \\
\hline
\hline
Labeled Sequents & Graphs of Gentzen Sequents  \\
\hline
 Display Sequents & (Pairs of)(Poly-)Tree(s) of\\
 & Gentzen Sequents  \\
\hline
 Nested, Tree-hypersequents, & Trees of Gentzen Sequents\\
 \& Bunched Sequents &  \\
\hline
2-Sequents & Lines of Gentzen Sequents \\
 \& Linear Nested Sequents &  \\
\hline
Hypersequents & (Multi-)Set of Gentzen Sequents \\
\hline
Gentzen Sequents & (Pairs of)(Multi-)Set(s) \\
\hline
\end{tabular}
\egroup
\end{center}
\caption{Common sequent formalisms and their data structure.
\label{fig:hierarchy}}
\end{figure}

In this section, we give a broad overview of the various sequent-based formalisms that have come to prominence as generalizations of Gentzen's sequent formalism~\cite{Gen35a,Gen35b}. Each formalism enriches the data structure employed in Gentzen sequents. Figure~\ref{fig:hierarchy} summarizes the formalisms we will consider, and the data structure used in the sequents of the formalism. These formalisms form a hierarchy, starting from Gentzen sequents at the bottom and increasing in complexity up to labeled sequents at the top.

\subsection{Gentzen System: Classical Logic}
\label{subsec-gentzen-sequent}

Gentzen~\cite{Gen35a,Gen35b} introduced the sequent formalism to proof theory by defining sequent calculi for classical and intuitionistic logic. We begin by recalling the sequent calculus for classical logic. A sequent is an object of the form $\Gamma \sar \Delta$ where $\Gamma$ and $\Delta$ are (possibly empty) \emph{multisets} of formulae from language (\ref{eq:lang:imp}). 
We call $\Gamma$ the \emph{antecedent} and $\Delta$ the \emph{consequent} of the sequent. Each sequent $\Gamma \sar \Delta$ with  $\Gamma = A_1,\ldots, A_m$ and $\Delta = B_1,\ldots, B_n$ can be interpreted as a formula of the following form:
$$
\tau(\Gamma \sar \Delta) := (A_1 \land \cdots \land A_m) \rightarrow (B_1 \lor \cdots \lor B_n)
$$
When $m=0$, the empty conjunction is interpreted as $\top$, and when $n =0$, the empty disjunction is interpreted as $\bot$. 

A  sequent calculus contains axioms, also called \emph{initial sequents}, and rules that let one derive sequents from sequents. The latter are divided into \emph{logical} rules that introduce complex formulae in either the antecedent or consequent of a sequent, and \emph{structural} rules which modify the structure of the antecedent/consequent, without changing the formulae themselves. \fig~\ref{fig:sequent-calculus-CL-log} contains the axioms $\id$ 
and $\botl$, and the logical rules for the sequent calculus $\scalc$ for classical propositional logic. 

\begin{figure}[t]

\begin{center}
\begin{tabular}{c c }

\AxiomC{$\phantom{\Gamma}$}
\RightLabel{$\id$}
\UnaryInfC{$\Gamma, p \sar p, \Delta$}
\DisplayProof

&

\AxiomC{$\phantom{\Gamma}$}
\RightLabel{$\botl$}
\UnaryInfC{$\bot, \Gamma \sar \Delta$}
\DisplayProof






\\[0.5cm]

	\AxiomC{$\Gamma, A, B \sar \Delta$}
		\RightLabel{$\conl$}
		\UnaryInfC{$\Gamma, A\land B \sar \Delta$}
		\DisplayProof
		&
		\AxiomC{$\Gamma \sar A, \Delta$}
		\AxiomC{$\Gamma \sar B, \Delta$}
		\RightLabel{$\conr$}
		\BinaryInfC{$\Gamma  \sar \forma \land \formb, \Delta$}
		\DisplayProof

\\[0.5cm]

	 \AxiomC{$\Gamma, \forma \sar \Delta$}
		\AxiomC{$\Gamma, \formb \sar \Delta$}
		\RightLabel{$\disl$}
		\BinaryInfC{$\Gamma, \forma \lor \formb \sar \Delta$}
		\DisplayProof 
& 
\AxiomC{$\Gamma  \sar \forma, \formb, \Delta$}
\RightLabel{$\disr$}
\UnaryInfC{$\Gamma \sar \forma \lor \formb, \Delta$}
\DisplayProof

\\[0.5cm]  
		\AxiomC{$\Gamma\sar  \forma,  \Delta$}
		\AxiomC{$\Gamma, \formb \sar \Delta$}
		\RightLabel{($\rightarrow_l$)}
		\BinaryInfC{$\Gamma, \forma \rightarrow \formb \sar \Delta$}
		\DisplayProof

		&
		
		\AxiomC{$\Gamma, \forma\sar \formb, \Delta$}
		\RightLabel{($\rightarrow_r$)}
		\UnaryInfC{$\Gamma \sar \forma \rightarrow \formb, \Delta$}
		\DisplayProof

	\end{tabular}
\end{center}

	\caption{Initial sequents and logical rules of $\scalc$.\label{fig:sequent-calculus-CL-log}}
\end{figure}

We define a \emph{derivation} $\deriv$ of a sequent $\gseq$ to be a (potentially infinite) tree whose nodes are sequents satisfying the following conditions: (1) the root of $\deriv$ is the sequent $\gseq$, and (2) every parent node is the instance of the conclusion of a rule with its children the corresponding premises. We say that a derivation $\deriv$ of $\gseq$ is a \emph{proof} of $\gseq$ if all the leaves of $\deriv$ are axioms. We say that a sequent $\gseq$ is \emph{provable} \iffi it has a proof. The \emph{height} of a derivation is equal to the number of sequents along a maximal path from the root to a leaf. In a rule, we define the \emph{principal formulae} to be those explicitly introduced in the conclusion, and the \emph{auxiliary formulae} to be those explicitly used in the premise(s) to derive the conclusion. For example, $A \rightarrow B$ is the principal formula in $\impl$ and $A$ and $B$ are the auxiliary formulae. By \emph{proof-search} we mean an algorithm that builds a derivation by applying inference rules bottom-up.

The structural rules for $\scalc$ are displayed in \fig~\ref{struct}. 
The \emph{weakening} rules ($wk_l$) and ($wk_r$) introduce formulae into the antecedent and consequent of a sequent, while the \emph{contraction} rules ($cr_l$) and ($cr_r$) remove additional copies of formulae. The $\cut$ rule can be seen as a generalization of modus ponens and has a special status, namely, it encodes the transitivity of deduction. Observe that the $\cut$ rule is not analytic, as the premises contain an arbitrary formula that disappears in the conclusion. It is important to notice that all structural rules, and in particular the $\cut$ rule, are \emph{admissible} in the calculus $\scalc$, meaning that if instances of the premises are provable, then so is the corresponding conclusion. By this fact, structural rules are recognized to be unnecessary for completeness.

The logical rules for negation, which we have chosen not to include as primitive rules, are also admissible in $\scalc$, and we will sometimes use them in derivations:
\begin{center}

\AxiomC{$\Gamma \sar A, \Delta$}
\RightLabel{$\negl$}
\UnaryInfC{$\Gamma, \neg A \sar \Delta$}
\DisplayProof
\qquad 
\AxiomC{$\Gamma, A \sar \Delta$}
\RightLabel{$\negr$}
\UnaryInfC{$\Gamma \sar \neg A, \Delta$}
\DisplayProof
\end{center}

\begin{figure}[t]

\begin{center}
\begin{tabular}{c c c }




\AxiomC{$\Gamma \sar \Delta$}
\RightLabel{($wk_l$)}
\UnaryInfC{$\Gamma, A \sar \Delta$}
\DisplayProof

&

\AxiomC{$\Gamma \sar \Delta$}
\RightLabel{($wk_r$)}
\UnaryInfC{$\Gamma\sar \Delta, A$}
\DisplayProof

&

\AxiomC{$\Gamma, A, A \sar  \Delta$}
\RightLabel{($cr_l$)}
\UnaryInfC{$\Gamma, \forma \sar \Delta$}
\DisplayProof

\\[0.5cm]

\multicolumn{3}{c}{
\AxiomC{$\Gamma\sar A, A,\Delta$}
	\RightLabel{($cr_r$)}
	\UnaryInfC{$\Gamma \sar A, \Delta$}
	\DisplayProof

\qquad 

\AxiomC{$\Gamma\sar  \forma,  \Delta$}
\AxiomC{$\Gamma, \forma \sar \Delta$}
\RightLabel{$\cut$}
\BinaryInfC{$\Gamma  \sar \Delta$}
\DisplayProof
}
\end{tabular}
\end{center}

\caption{Structural rules for the sequent calculus $\mathsf{S(CP)}$.\label{struct}}
\end{figure}

The Gentzen calculus $\scalc$ is a `two-sided' proof system, meaning that sequents are composed of an antecedent and consequent, and consequently the proof system is constituted by left and right logical rules. By taking the language of classical propositional logic to be \eqref{eq:lang:and-or}, it is possible to define a more compact `one-sided' version of $\scalc$. In this case, a sequent is just a multiset of formulae $\Delta = B_1 \lor \cdots \lor B_n$, and it is interpreted as the formula $\tau(\Delta) :=  B_1 \lor \cdots \lor B_n$. The rules of the one-sided calculus are displayed in Figure~\ref{fig:cl:one-sided}. It is easy to see that the one-sided and the two-sided versions of $\scalc$ are equivalent.  

The Gentzen calculus $\scalc$ has some important properties, discussed below, that set  it as an `ideal' proof system. The following terminology will also be applied to the other kinds of sequent-style systems we consider later on.
\begin{enumerate}[wide, labelwidth=!, labelindent=0pt]

\item[$\bullet$]  \emph{Analyticity}: the premises of each rule only contain subformulae of the conclusion. Thus, if we do not consider multiple occurrences of the same formulae (i.e., we consider a sequent as a pair of sets), given a proof of a sequent $\gseq$, there are only finitely many different sequents that can occur in $\prf$. This follow from the admissibility of the cut rule (i.e., \emph{cut-elimination}), which means that every proof containing applications of $\cut$ can be transformed into a cut-free proof of the same conclusion~\cite{Gen35a,Gen35b}.

\item[$\bullet$] \emph{Termination}: the premises of each rule are less complex than the conclusion. This property holds in $\scalc$ (without structural rules) since the auxiliary formulae are always less complex than the principal formulae. This property together with analyticity ensures that the process of building a derivation (bottom-up) always terminates, that is, every branch of a derivation $\deriv$ terminates at an axiom or an unprovable sequent (usually containing only atoms).

\item[$\bullet$] \emph{Invertibility}: if any instance of the conclusion of a rule is provable, then its corresponding premises are provable. By this property, the order of bottom-up applications of rules during proof-search does not matter: either we obtain a proof of the root sequent, or we get a (finite) derivation containing an unprovable sequent as a leaf. When this property is present, backtracking (i.e., searching for alternative proofs) is unnecessary during proof-search.

\item[$\bullet$] \emph{Counter-model generation}: if proof-search yields a derivation $\deriv$ that is not a proof of the conclusion, then there exists an unprovable sequent as a leaf which can be used to define a counter-model of the conclusion. In the case of $\scalc$, if $\Gamma \sar \Delta$ is such a leaf in a derivation $\deriv$, then $\Gamma \cap \Delta = \emptyset$, and one can define a propositional evaluation `$V(p) = \textbf{t}$ \iffi $p\in \Gamma$' that falsifies the conclusion of $\deriv$.

\item[$\bullet$] \emph{Complexity-optimal}: the proof system admits a (relatively straightforward) proof-search algorithm that decides the (in)validity of formulae in the complexity of the logic.

\end{enumerate}

\begin{figure}[t]
\begin{center}
\begin{tabular}{c c c}

\AxiomC{$\phantom{\Gamma}$}
\RightLabel{$\id$}
\UnaryInfC{$\Delta, p, \dual p$}
\DisplayProof
 & 
		\AxiomC{$ A, \Delta$}
		\AxiomC{$ B, \Delta$}
		\RightLabel{$\conr$}
		\BinaryInfC{$\forma \land \formb, \Delta$}
		\DisplayProof

& 
\AxiomC{$ \forma, \formb, \Delta$}
\RightLabel{$\disr$}
\UnaryInfC{$ \forma \lor \formb, \Delta$}
\DisplayProof

	\end{tabular}
\end{center}

    \caption{One-sided rules of $\scalc$. }
    \label{fig:cl:one-sided}
\end{figure}

The calculus $\mathsf{S(CP)}$ satisfies the above four properties; as a consequence, the calculus provides a decision procedure for classical propositional logic. Proof-search is carried out by building \emph{just one} derivation that will either be a proof, or from which a counter-model of the conclusion can be extracted. Furthermore, the decision procedure based on the calculus has an optimal complexity ($\conp$).

\subsection{Gentzen System: Intuitionistic Logic}

Gentzen's sequent systems are flexible enough to capture other logics. For example, intuitionistic logic can be provided a sequent calculus by making simple modifications to $\scalc$. The calculus $\scalcint$ for intuitionistic logic is obtained by replacing the $\impl$ and $\impr$ rules in $\scalc$ with the $\iimpl$ and $\iimpr$ rules shown in \fig~\ref{IntNUOVO}. Originally, Gentzen obtained a sequent calculus for intuitionistic logic by imposing a restriction on the sequent calculus $\mathsf{S(CP)}$ for classical logic, namely, only sequents with at most one formula in the consequent (i.e., sequents $\Gamma \sar \Delta$ such that $|\Delta| \leq 1$) could be used in derivations. However, Gentzen's restriction invalidates certain admissibility and invertibility properties, which can be regained allowing multiple formulae to occur in the conclusion. 
The calculus $\scalcint$, due to Maehara~\cite{Mae54}, is a variant of Gentzen's sequent calculus for intuitionistic logic that has all structural rules, including $\cut$, admissible.

\begin{figure}[t]

\begin{center}
\begin{tabular}{c c}

\AxiomC{$\Gamma, \forma \iimp \formb, \formb \sar \Delta$}
\AxiomC{$\Gamma, \forma \iimp \formb \sar \forma, \Delta$}
\RightLabel{$\iimpl$}
\BinaryInfC{$\Gamma, \forma \iimp \formb \nsar \Delta$}
\DisplayProof

&

\AxiomC{$\Gamma, \forma \sar \formb$}
\RightLabel{$\iimpr$}
\UnaryInfC{$\Gamma \sar \forma \iimp \formb, \Delta$}
\DisplayProof

\end{tabular}
\end{center}

\caption{Intuitionistic implication rules for $\scalcint$.\label{IntNUOVO}
}
\end{figure}

The calculus $\scalcint$ is analytic, though not terminating as the premises of a rule may be as complex as the conclusion as witnessed by the premises of ($\iimp_l$). Additionally, the ($\iimp_r$) rule is not invertible, which is an impediment to proof-search. In particular, if proof-search constructs a derivation that is not a proof, then a proof may still exist and the constructed derivation may not provide a counter-model of the conclusion. Due to analyticity a decision procedure can still be obtained; 
however, the procedure will also require loop checking which diminishes its efficiency. For an overview of various proof systems for intuitionistic logic and associated decision procedures, see~\cite{Dyc16}.

Although Gentzen systems have been provided for many logics, the formalism is still not general enough to yield cut-free systems for many logics of interest (e.g., $\logicsv$ and bi-intuitionistic logics~\cite{LelPat13,BuiGor07}). This motivates the search for more expressive formalisms that enrich Gentzen sequents to recapture analyticity and other properties.

\subsection{Beyond Gentzen's Formalism}

In the previous section we highlighted some desirable properties of proof systems, which Gentzen sequent systems often times satisfy. However, we are here interested in the definition of formalisms satisfying desirable properties for large families of logics. Thus, we identify five desiderata for proof-theoretic formalisms:\footnote{For discussions of other desiderata for proof systems and formalisms, see \cite{Wan94,Avr96}.} 
\begin{enumerate}
    \item[(1)] \emph{Generality}: the formalism covers a sizable class of logics with proof systems sharing desirable properties; 
    \item[(2)] \emph{Uniformity}: the formalism need not be enriched to obtain a system for a logic within a given class; 
    \item[(3)] \emph{Modularity}: a system for one logic within the considered class can be transformed into a system for another, with properties preserved, by adding/deleting rules or modifying the functionality of rules;
    \item[(4)] \emph{Constructibility}: a method is known for constructing a calculus for a given logic in the considered class;
    \item[(5)] \emph{Syntactic Parsimony}: the data structures employed are as simple as required by the logic or purpose of the proof systems.
\end{enumerate}

When the desiderata (1)-(4) are satisfied, a proof formalism is expected to generate large classes of proof calculi for logics 
without requiring substantial work on the side of the logician. According to requirement~(5), a formalism should employ sequents that are as simple as possible, in order to maintain their interpretation as formulae of the language and simplify derivations. 


It is not to be taken for granted that a single proof formalism can fulfill all of the above requirements, which justifies the study of alternative proof systems and formalisms with different properties and applications. For instance, although Gentzen's sequent formalism satisfies syntactic parsimony to a high degree, the formalism lacks uniformity and modularity, since simple modifications to a calculus can nullify key properties such as analyticity. Similarly, although nested sequents employ trees of Gentzen sequents, they are better suited for counter-model extraction than Gentzen sequent calculi, and so, if we aim to use our systems to extract counter-models of formulae, then it is sensible to trade the simple structure of Gentzen sequents for nested sequents.

In the next subsections we will present a number of formalisms that are less parsimonious that Gentzen sequents, but are more satisfactory than Gentzen sequents regarding requirements (1)-(4). 



\subsection{Hypersequents}
\label{Subs:hypersequents}

Introduced independently by Mints~\cite{Min68}, Pottinger~\cite{Pot83}, and Avron~\cite{Avr87}, the hypersequent formalism is a simple generalization of Gentzen's sequent formalism. A \emph{hypersequent} is  an expression of the form $\Gamma_1 \hsar \Delta_1 \mid \cdots \mid \Gamma_n \hsar \Delta_n$ such that each \emph{component} $\Gamma_i \hsar \Delta_i$ is a Gentzen sequent. That is, a hypersequent is a (multi)set of Gentzen sequents, where each element of the (multi)set is separated by the `$\mid$' operator. Usually, we interpret the `$\mid$' operator disjunctively, meaning, hypersequents are interpreted as disjunctions of Gentzen sequents. We use $G$, $H$, $\ldots$ to denote hypersequents.

\begin{figure}[t]

\begin{center}
\begin{tabular}{c c c}
\AxiomC{$\phantom{\Gamma}$}
\RightLabel{$\id$}
\UnaryInfC{$G \mid \Gamma, p \hsar p, \Delta$}
\DisplayProof

&

\AxiomC{$\phantom{\Gamma}$}
\RightLabel{$\botl$}
\UnaryInfC{$G \mid \Gamma, \bot \hsar \Delta$}
\DisplayProof

&

\AxiomC{$G \mid \Gamma \hsar \Delta \mid \ \hsar \forma$}
\RightLabel{$\boxr$}
\UnaryInfC{$G \mid \Gamma \hsar \Box \forma, \Delta$}
\DisplayProof
\end{tabular}
\end{center}

\begin{center}
\begin{tabular}{c c}
\AxiomC{$G \mid \Gamma, \Box \forma, \forma \hsar \Delta$}
\RightLabel{$\boxli$}
\UnaryInfC{$G \mid \Gamma, \Box \forma \hsar \Delta$}
\DisplayProof

&

\AxiomC{$G \mid \Gamma, \Box \forma \hsar \Delta \mid \Sigma, \forma \sar \Pi$}
\RightLabel{$\boxlii$}
\UnaryInfC{$G \mid \Gamma, \Box \forma \hsar \Delta \mid \Sigma \sar \Pi$}
\DisplayProof
\end{tabular}
\end{center}

\caption{Rules for a hypersequent calculus $\hcalcsv$ for $\logicsv$.\label{fig:hyperseq-sfive}
}
\end{figure}

To demonstrate the hypersequent formalism, we provide an example of a  hypersequent calculus $\hcalcsv$ for the modal logic $\logicsv$, which is due to Poggiolesi~\cite{Pog08} though adapted to the language we are using 
for $\logicsv$.\footnote{See~\cite{BedInd15,Kuroka13,Res07} for alternative hypersequent systems for the modal logic $\logicsv$.} The hypersequent calculus $\hcalcsv$ contains the rules shown in \fig~\ref{fig:hyperseq-sfive} together with analogs for the rules $\disl$, $\disr$, $\conl$, $\conr$, $\impl$, and $\impr$ from the Gentzen calculus $\mathsf{S(CL)}$. These latter rules perform the same operation as their Gentzen calculus counterparts and are applied to components of hyperseuquents; for example, the $\impl$ and $\impr$ rules are defined as follows:
\begin{center}
\begin{tabular}{c c}
\AxiomC{$G \mid \Gamma\sar  \forma,  \Delta$}
\AxiomC{$G \mid \Gamma, \formb \sar \Delta$}
\RightLabel{$\impl$}
\BinaryInfC{$G \mid \Gamma, \forma \rightarrow \formb \sar \Delta$}
\DisplayProof

&

\AxiomC{$G \mid \Gamma, \forma\sar \formb, \Delta$}
\RightLabel{$\impr$}
\UnaryInfC{$G \mid \Gamma \sar \forma \rightarrow \formb, \Delta$}
\DisplayProof
\end{tabular}
\end{center}

As with Gentzen calculi, hypersequent calculi may contain axioms, logical rules, and structural rules. For instance, the hypersequent calculus $\hcalcsv$ contains the axioms $\id$ and $\botl$ and all remaining rules are logical rules. Moreover, similar to Gentzen systems, one can often find a selection of structural rules that are admissible in a hypersequent system. Due to the additional structure present in hypersequent calculi, structural rules can be classified into a wider variety of types. That is to say, the hypersequent structure makes it possible to
define new external structural rules that allow for the exchange of information between different components of a hypersequent. This increases the expressive power of hypersequent calculi compared to ordinary Gentzen systems.

As an example, for the hypersequent calculus $\hcalcsv$, one can define both \emph{internal} and \emph{external} structural rules. Internal structural rules strictly affect components; for instance, the following internal weakening $(iw)$ and internal contraction $(ic)$ rules apply weakenings and contractions only within components of hypersequents:
\begin{center}
\begin{tabular}{c c}
\AxiomC{$G  \hh \Gamma \hsar \Delta$}
\RightLabel{$(iw)$}
\UnaryInfC{$G \hh \Gamma, \Sigma \hsar \Pi, \Delta$}
\DisplayProof

&

\AxiomC{$G \mid \Gamma, \Sigma, \Sigma \hsar \Pi, \Pi, \Delta$}
\RightLabel{$(ic)$}
\UnaryInfC{$G \mid \Gamma, \Sigma \hsar \Pi, \Delta$}
\DisplayProof
\end{tabular}
\end{center}
On the other hand, external structural rules are more general and affect the overall structure of a hypersequent; for instance,  the following external weakening $(ew)$ and external contraction $(ec)$ rules weaken in new components and contract components, respectively:
\begin{center}
\begin{tabular}{c c}
\AxiomC{$G$}
\RightLabel{$(ew)$}
\UnaryInfC{$G \hh \Gamma \hsar \Delta$}
\DisplayProof

&

\AxiomC{$G \mid \Gamma \hsar \Delta \mid \Gamma \hsar \Delta$}
\RightLabel{$(ec)$}
\UnaryInfC{$G \mid \Gamma \hsar \Delta$}
\DisplayProof
\end{tabular}
\end{center}
We remark that all of the above rules are admissible in $\hcalcsv$ as is a hypersequent version of the cut rule~\cite{Pog08}.

It is well-known that the hypersequent formalism allows for the formulation of cut-free sequent-style systems for logics failing to possess a cut-free Gentzen system. The formalism also supports the algorithmic transformation of large classes of Hilbert axioms and frame properties into cut-free hypersequent calculi for wide classes of logics, including substructural logics~~\cite{CiaGalTer08}, intermediate logics~\cite{CiabattoniST09}, and modal logics~\cite{Lah13,Lellmann16}. Therefore, the hypersequent formalism can be seen to satisfy our five desiderata to a large degree: with only a basic increase in syntactic complexity from that of Gentzen sequents, hypersequent systems with favorable properties (e.g., analyticity) can be algorithmically generated for wide classes of logics. This demonstrates the generality, uniformity, modularity, and constructibility of such systems. Nevertheless, there are logics for which the hypersequent formalism is ill-suited for providing analytic systems (e.g., the tense logic $\logickt$ and some modal logics characterized by geometric frame conditions~\cite{Sim94}), showing that the generality of the formalism is still limited in scope.

\subsection{2-Sequents and Linear Nested Sequents}\label{subsec:2-sequent-LNS}

 The 2-sequent formalism was introduced by Masini~\cite{Mas92,Mas93} as a generalization of Gentzen's sequent formalism whereby an \emph{infinite list} of multisets of formulae implies another infinite list. For instance, an example of a 2-sequent is shown below left and an another example is shown below right:
\begin{center}
\begin{tabular}{c @{\hskip 3em} c @{\hskip 3em} c}
\xymatrix@=.5em{
 A, B & & D\\
 C & \sar & E, F\\
 & & G
}

&

\xymatrix@=.5em{
 & & \\
  & \leadsto & \\
 & & 
}

&

\xymatrix@=.5em{
 A, B & & D\\
 C & \sar & E, F, \Box G
}
\end{tabular}
\end{center}
In the 2-sequent above left, the antecedent consists of the list whose first element is the multiset $A,B$, second element is the singleton $C$, and where every other element is the empty multiset. By contrast, the consequent consists of a list beginning with the three multisets (1) $D$, (2) $E,F$, and (3) $G$, and where every other element is the empty multiset. The 2-sequent shown above right is derivable from the 2-sequent shown above left by `shifting' the $G$ formula up one level and introducing a $\Box$ modality, thus demonstrating how modal formulae may be derived in the formalism. Systems built with such sequents have been provided for various logics---e.g., modal logics~\cite{Mas92}, intuitionistic logic~\cite{Mas93} and tense logics~\cite{BarMas04}---and tend to exhibit desirable proof-theoretic properties such as generalized forms of cut-elimination and the subformula property.

More recently, a refined but equivalent re-formulation of 2-sequents was provided by Lellmann~\cite{Lel15}. Rather than employing sequents with infinite lists for antecedents and consequents, the formalism employs \emph{linear nested sequents}, which are finite lists of Gentzen sequents. For example, the 2-sequents shown above left and right may be re-written as the linear nested sequents shown below left and right respectively with the `$\sslash$' constructor separating the components (i.e., each Gentzen sequent) in each list:
$$
A,B \sar D \sslash C \sar E, F \sslash \emptyset \sar G
\qquad
A,B \sar D \sslash C \sar E, F, \Box G
$$
Linear nested sequent systems have been provided for a diverse selection of logics; e.g., modal logics~\cite{Lel15}, Gödel-Löb provability logic~\cite{Lyo24}, propositional and first-order G\"odel-Dummett logic~\cite{KuzLel18,Lyo20b}, the tense logic $\logickt$~\cite{GorLel19} and the tense logics with linear time~\cite{Indrz16,Indrz19a}. Moreover, such calculi have been used to write constructive interpolation proofs~\cite{KuzLel18} and decision procedures~\cite{GorLel19}.
 
As discussed by Lellmann~\cite{Lel15}, there is a close connection between Gentzen sequent calculi, nested sequent calculi (discussed in \sect~\ref{subsec:nested} below), and linear nested sequent calculi. In particular, certain linear nested sequent systems have been found to encode \emph{branches} within sequent calculus proofs as well as branches within nested sequents. For example, the standard $(\Box)$ rule that occurs in the sequent calculus for the modal logic $\logick$ (shown below left) corresponds to $|\Gamma|$ many applications of the $(\Box_{l})$ rule followed by an application of the $(\Box_{r})$ rule in the linear nested sequent calculus for $\logick$ (cf.~\cite{Lel15}).
\begin{center}
\begin{tabular}{c @{\hskip 1em} c @{\hskip 1em} c}
\AxiomC{$\Gamma \sar A$}
\RightLabel{$(\Box)$}
\UnaryInfC{$\Sigma, \Box \Gamma \sar \Box A, \Delta$}
\DisplayProof

&

$\leadsto$

&

\AxiomC{$\Sigma, \Box \Gamma \sar \Box A, \Delta \sslash \Gamma \sar A$}
\RightLabel{$(\Box_{l}) \times |\Gamma|$}
\UnaryInfC{$\Sigma, \Box \Gamma \sar \Box A, \Delta \sslash \emptyset \sar A$}
\RightLabel{$(\Box_{r})$}
\UnaryInfC{$\Sigma, \Box \Gamma \sar \Box A, \Delta$}
\DisplayProof
\end{tabular}
\end{center}
Observe that the top linear nested sequent in the inferences shown above right stores the conclusion and premise of the $(\Box)$ rule, thus demonstrating how linear nested sequents can encode branches (i.e., sequences of inferences) in sequent calculus proofs.
 
Due to the fact that linear nested sequents employ a relatively simple data structure, the formalism typically allows for complexity-optimal proof-search algorithms, similar to (depth-first) algorithms written within sequent and nested sequent systems. As such, the linear nested sequent formalism strikes a balance between complexity-optimality on the one hand, and expressivity on the other, since the formalism allows for many logics to be captured in a cut-free manner while exhibiting desirable invertibility and admissibility properties. Note that if we let the `$\sslash$' constructor be commutative, then it can be seen as the hypersequent `$\hh$' constructor, showing that every hypersequent calculus is technically a linear nested sequent calculus, i.e., the latter formalism generalizes the former~\cite{Lel15}. It can be seen that the linear nested sequent formalism satisfies the same desiderata as the hypersequent formalism, though improves upon generality as the formalism is known to capture logics lacking a cut-free hypersequent calculus, e.g., the tense logic $\logickt$~\cite{GorLel19}.

\subsection{Bunched Sequents}\label{sec:bunched-sequents}

As recalled in Section~\ref{subsec-gentzen-sequent}, a standard Gentzen sequent is an object of the form \SEQ{\SG}{\SD} where the contexts $\SG$ and $\SD$ usually are sets or multisets, sometimes (but less often\footnote{This is the case for logics lacking {\em all} structural rules.}) lists.
Those data structures are one dimensional and built from a single context forming operator usually written as a comma or a semi-colon.
An interesting extension of Gentzen sequents are \emph{bunched sequents}, which arise when the contexts are built from more than one context forming operators.
For example, in~\cite{Chaudhuri11a} two context forming operators, ``;'' and `,` are used to split the contexts into several zones the formulae of which are handled differently by a focused sequent calculus.
Nevertheless, inside a zone, the formulae are arranged as a one dimensional structure (usually a multiset) and the inference rules can be applied to any formula in that shallow structure (thus making the inference rules shallow).
Let us remark that bunched structures can also be used to extend the hypersequent framework to bunched hypersequents (forests of sequents) as illustrated in a recent work~\cite{CiabattoniR17}.
However, in the remaining of the section, we shall focus on the most representative witness of a bunched sequent calculus, which is undoubtedly the one given for~Bunched Implications logic ({\BI}) in~\cite{Pym02a}. Let us note that such kind of structured calculi were initially proposed in the field of relevant logics \cite{Dunn74,Min72}.

In \BI, formulae are arranged as \quo{bunches} which can be viewed as trees whose leaves are labeled with formulae and whose internal nodes are labeled with either \quo{$\BIasep$} or \quo{$\BImsep$}.
More formally, bunches are trees given by the following grammar:
 \[
  \SG ::= \VPA \mpipe 
          \BIanul \mpipe 
          \SG \BIasep \SG \mpipe 
          \BImnul \mpipe
          \SG \BImsep \SG
 \]
The notation $\SG(\SD)$ denotes a bunch~$\SG$ that 
contains the bunch~$\SD$ as a subtree.

Bunches are considered up to a structural equivalence $\BIequi$ given by commutative monoid equations for \quo{$\BIasep$} and \quo{$\BImsep$} with units $\BIanul$ and $\BImnul$ respectively, together with the substitution congruence for subbunches.
From a logical point of view, bunches relate to formulae as follows: let $\SG$ be a bunch, the corresponding formula is $\mathrm{\Phi}_{\SG}$ which is obtained from $\SG$ by replacing each $\BImnul$ with $\BImtop$, each $\BIanul$ with $\BIatop$, each \quo{$\BImsep$} with $\BImand$ and each \quo{$\BIasep$} with $\BIaand$.

The standard internal calculus for {\BI} is a single conclusion bunched sequent calculus called {\LBI}.
In {\LBI}, bunches arise (on the left-hand side only) from the two kinds of implications $\BIaimp$ and $\BImimp$, that respectively give rise to two distinct context forming operators \quo{$\BIasep$} and \quo{$\BImsep$} as follows:
\[
\begin{ebp}
	\hypo{\SEQ{\SG\BIasep\VPA}{\VPB}}
	\infer1[\PRNR{\BIaimp}]{\SEQ{\SG}{\VPA\BIaimp\VPB}}
\end{ebp}
\qquad\qquad
\begin{ebp}
	\hypo{\SEQ{\SG\BImsep\VPA}{\VPB}}
	\infer1[\PRNR{\BImimp}]{\SEQ{\SG}{\VPA\BImimp\VPB}}
\end{ebp}
\]
From a syntactic point of view, the main distinction between \quo{$\BIasep$} (associated with~$\BIaand$) and \quo{$\BImsep$} (associated with~$\BImand$) is that \quo{$\BIasep$} admits both weakening and contraction while \quo{$\BImsep$} does not.

\begin{figure}[t]
	\begin{center}
		\begin{ebp}
			\hypo{}
			\infer1[\PRN{id}]{\SEQ{\VPA}{\VPA}} 
		\end{ebp}
		\;\;
		\begin{ebp}
			\hypo{}
			\infer1[\PRNR{\BImtop}]{\SEQ{\BImnul}{\BImtop}}
		\end{ebp}
		\;\;
		\begin{ebp}
			\hypo{}
			\infer1[\PRNR{\BIatop}]{\SEQ{\BIanul}{\BIatop}}
		\end{ebp}
		\;\;
		\begin{ebp}
			\hypo{}
			\infer1[\PRNL{\BIabot}]{\SEQ{\SG(\BIabot)}{\VPA}}
		\end{ebp}
	\end{center}
	
	\begin{center}
		\begin{ebp}
			\hypo{\SEQ{\SG(\BImnul)}{\VPA}}
			\infer1[\PRNL{\BImtop}]{\SEQ{\SG(\BImtop)}{\VPA}}
		\end{ebp}
		\;\;
		\begin{ebp}
			\hypo{\SEQ{\SG(\BIanul)}{\VPA}}
			\infer1[\PRNL{\BIatop}]{\SEQ{\SG(\BIatop)}{\VPA}}
		\end{ebp}
		\;\;
		\begin{ebp}
			\hypo{\SEQ{\SG(\VPB)}{\VPA}}
			\hypo{\SEQ{\SG(\VPC)}{\VPA}}
			\infer2[\PRNL{\BIaor}]{\SEQ{\SG(\VPB\BIaor\VPC)}{\VPA}} 
		\end{ebp}
	\end{center}
	
	\begin{center}
		\begin{ebp}
			\hypo{\SEQ{\SG}{\VPA_{\mathrm{i}\, \in\, \ens{1,2}}}}
			\infer1[\PRNR{\BIaor^\mathrm{i}}]{\SEQ{\SG}{\VPA_1\BIaor\VPA_2}}
		\end{ebp}
        \;\;
		\begin{ebp}
			\hypo{\SEQ{\SD}{\VPB}}
			\hypo{\SEQ{\SG(\VPC)}{\VPA}}
			\infer2[\PRNL{\BImimp}]{\SEQ{\SG(\VPB\BImimp\VPC\BImsep\SD)}{\VPA}}
		\end{ebp}
		\;\;
		\begin{ebp}
			\hypo{\SEQ{\SG\BImsep\VPA}{\VPB}}
			\infer1[\PRNR{\BImimp}]{\SEQ{\SG}{\VPA\BImimp\VPB}}
		\end{ebp}
    \end{center}

    \begin{center}
		\begin{ebp}
			\hypo{\SEQ{\SG(\VPB\BImsep\VPC)}{\VPA}}
			\infer1[\PRNL{\BImand}]{\SEQ{\SG(\VPB\BImand\VPC)}{\VPA}}
		\end{ebp}
		\;\;
		\begin{ebp}
			\hypo{\SEQ{\SG}{\VPA}}
			\hypo{\SEQ{\SD}{\VPB}}
			\infer2[\PRNR{\BImand}]{\SEQ{\SG\BImsep\SD}{\VPA\BImand\VPB}}
		\end{ebp}
        \;\;
		\begin{ebp}
			\hypo{\SEQ{\SD}{\VPB}}
			\hypo{\SEQ{\SG(\VPC)}{\VPA}}
			\infer2[\PRNL{\BIaimp}]{\SEQ{\SG(\VPB\BIaimp\VPC\BIasep\SD)}{\VPA}}
		\end{ebp}
	\end{center}
	
	\begin{center}
		\begin{ebp}
			\hypo{\SEQ{\SG\BIasep\VPA}{\VPB}}
			\infer1[\PRNR{\BIaimp}]{\SEQ{\SG}{\VPA\BIaimp\VPB}}
		\end{ebp}
		\;\;
		\begin{ebp}
			\hypo{\SEQ{\SG(\VPB\BIasep\VPC)}{\VPA}}
			\infer1[\PRNL{\BIaand}]{\SEQ{\SG(\VPB\BIaand\VPC)}{\VPA}}
		\end{ebp}
		\;\;
		\begin{ebp}
			\hypo{\SEQ{\SG}{\VPA}}
			\hypo{\SEQ{\SD}{\VPB}}
			\infer2[\PRNR{\BIaand}]{\SEQ{\SG\BIasep\SD}{\VPA\BIaand\VPB}}
		\end{ebp}
	\end{center}
	
	
	\begin{center}
		\begin{ebp}
			\hypo{\SEQ{\SG(\SD_1)}{\VPA}}
			\infer1[\PRN{WK}]{\SEQ{\SG(\SD_1\BIasep\SD_2)}{\VPA}}
		\end{ebp}
		\;\;
		\begin{ebp}
			\hypo{\SEQ{\SG(\SD\BIasep\SD)}{\VPA}}
			\infer1[\PRN{CR}]{\SEQ{\SG(\SD)}{\VPA}}
		\end{ebp}
		\;\;
		\begin{ebp}
			\hypo{\SEQ{\SG}{\VPA}}
			\infer1[\PRN{\SG\BIequi\SD}]{\SEQ{\SD}{\VPA}}
		\end{ebp}
    \end{center}

    \begin{center}
		\begin{ebp}
			\hypo{\SEQ{\SD}{\VPB}}
			\hypo{\SEQ{\SG(\VPB)}{\VPA}}
			\infer2[\PRN{CUT}]{\SEQ{\SG(\SD)}{\VPA}}
		\end{ebp}
	\end{center}
	
	\caption{The Sequent Calculus {\LBI}.}
	\label{fig:lbi-rules}
\end{figure}

The {\LBI} sequent calculus, depicted in \Fig~\ref{fig:lbi-rules}, derives sequents of the form  $\SEQ{\SG}{\VPC}$, where $\SG$ is a bunch and $\VPC$ is a formula.
A formula $\VPC$ is a \emph{theorem} of $\LBI$ iff $\SEQ{\BImnul}{\VPC}$ is provable in {\LBI}.
Let us remark that the inference rules of {\LBI} are deep in that they can be applied to formulae anywhere in the tree structure of a bunch and not only at the root.
Let us also mention that the~\PRN{CUT} rule is admissible in {\LBI}~\cite{Pym02a} and that the contraction rule~\PRN{CR} may duplicate whole bunches and not just formulae.
Indeed, as shown in Example~\ref{fig:lbi-proof-example}, restricting contraction to single formulae would not allow to prove the end sequent.

\begin{example}
{\LBI}-proof of $((\VPP \BImimp (\VPQ \BIaimp \VPR)\BIaand \VPP \BImimp 
	\VPQ)\BImand \VPP)\BImimp\VPR$.
	
\begin{ebpd}
	\hypo{}
	\infer1[\PRN{id}]{\SEQ{\VPP}{\VPP}}
	\hypo{}
	\infer1[\PRN{id}]{\SEQ{\VPP}{\VPP}}
	\hypo{}
	\infer1[\PRN{id}]{\SEQ{\VPQ}{\VPQ}}
	\hypo{}
	\infer1[\PRN{id}]{\SEQ{\VPR \BIasep \VPQ}{\VPR}}
	\infer2[\PRNL{\BIaimp}]{%
		\SEQ{\VPQ \BIaimp \VPR \BIasep \VPQ}{\VPR}%
	}
	\infer2[\PRNL{\BImimp}]{%
		\SEQ{\VPQ \BIaimp \VPR \BIasep (\VPP \BImimp \VPQ \BImsep \VPP)}{\VPR}%
	}
	\infer2[\PRNL{\BImimp}]{%
		\SEQ{(\VPP \BImimp (\VPQ \BIaimp \VPR) \BImsep \VPP) \BIasep (\VPP \BImimp \VPQ \BImsep \VPP)}{\VPR}%
	}
	\infer1[\PRN{WK}]{%
		\SEQ{(\VPP \BImimp (\VPQ \BIaimp \VPR) \BImsep \VPP) \BIasep ((\VPP \BImimp (\VPQ \BIaimp \VPR) \BIasep \VPP \BImimp \VPQ) \BImsep \VPP)}{\VPR}%
	}
	\infer1[\PRN{WK}]{%
		\SEQ{((\VPP \BImimp (\VPQ \BIaimp \VPR) \BIasep \VPP \BImimp \VPQ) \BImsep \VPP) \BIasep ((\VPP \BImimp (\VPQ \BIaimp \VPR) \BIasep \VPP \BImimp \VPQ) \BImsep \VPP)}{\VPR}%
	}
	\infer1[\PRN{CR}]{%
		\SEQ{(\VPP \BImimp (\VPQ \BIaimp \VPR)\BIasep \VPP \BImimp \VPQ)\BImsep 
		\VPP}{\VPR}%
	}
	\infer1[\PRNR{\BIaand}]{%
		\SEQ{((\VPP \BImimp (\VPQ \BIaimp \VPR)\BIaand \VPP \BImimp 
		\VPQ)\BImsep\VPP)
		}{\VPR}%
	}
	\infer1[\PRN{\BIequi}]{%
		\SEQ{\BImnul \BImsep ((\VPP \BImimp (\VPQ \BIaimp \VPR)\BIaand \VPP \BImimp 
		\VPQ)\BImsep\VPP)
		}{\VPR}%
	}
	\infer1[\PRNL{\BImand}]{%
		\SEQ{\BImnul \BImsep ((\VPP \BImimp (\VPQ \BIaimp \VPR)\BIaand \VPP \BImimp 
		\VPQ)\BImand\VPP)
		}{\VPR}%
	}
	\infer1[\PRNR{\BImimp}]{%
		\SEQ{\BImnul 
		}{((\VPP \BImimp (\VPQ \BIaimp \VPR)\BIaand \VPP \BImimp 
		\VPQ)\BImand\VPP)\BImimp\VPR}%
	}
\end{ebpd}
\label{fig:lbi-proof-example}
\end{example}

\subsection{Nested Sequents}\label{subsec:nested}

Nested sequent calculi were originally defined by Kashima for tense logics~\cite{Kas94} and Bull for the fragment of PDL without the Kleene star~\cite{Bul92}.\footnote{We note that nested sequent calculi can be seen as `upside-down' versions of prefixed tableaux~\cite{Fit72,Fit14}. Furthermore, Leivant's 1981 paper~\cite{Lei81} introduces a calculus for PDL that is structurally equivalent to a nested sequent calculus. Both of these works predate the work of Kashima and Bull.} The characteristic feature of such calculi is the use of \emph{trees} of Gentzen sequents in proofs. This additional structure has led to the development of cut-free calculi for various logics not known to possess a cut-free Gentzen sequent calculus. This formalism is general in the sense that sizable classes of logics can be uniformly captured with such systems. For example, cut-free nested sequent calculi have been given for classical modal logics~\cite{Bru09,Pog09}, for intuitionistic modal logics~\cite{Str13,Lyo21b}, for classical tense logics~\cite{Kas94,GorPosTiu11}, and for first-order non-classical logics~\cite{Lyo23a,LyoOrl23}. Moreover, the rules of nested sequent calculi are usually invertible, which---as mentioned above---are useful in extracting counter-models from failed proof search (cf.~\cite{TiuIanGor12,LyoGom22}). We remark that nested sequents have also been referred to as \emph{tree-hypersequents}~\cite{Pog09,Pog11}; however, we will stick to the term \emph{nested sequent} as it is far more prevalent in the literature and is the original term given by Kashima~\cite{Kas94} and Bull~\cite{Bul92}.

Nested sequents generalize the syntax of one-sided Gentzen sequents via the incorporation of a nesting constructor. For instance, for the tense logic $\logickt$, nested sequents are generated via the following grammar in BNF:
$$
\ns ::= A \ | \ \emptyset \ | \ \ns, \ns \ | \ \wnest{\ns} \ | \ \bnest{\ns}
$$
 where $A \in \langmod$ and $\emptyset$ is the empty nested sequent. Examples of nested sequents generated in the above syntax include (1) $\ns_{1} = p, \dia q$, (2) $\ns_{2} = \blbox p, \wnest{p, \bnest{q}}$, and (3) $\ns_{3} = p \land r, \wnest{\emptyset, \wnest{\dia q, p}, \bnest{\neg q \lor q}}, \wnest{\neg q}$, which are graphically displayed below as trees with labeled edges in order from left to right. 
\begin{center}
\begin{tikzpicture}
     \node[] [] (w) [] {$\boxed{p \land r}$};
     \node[] [] (u) [below left=of w,xshift=2em,yshift=2em]{$\boxed{\emptyset}$};
     \node[] [] (v) [below right=of w,xshift=-2em,yshift=2em]{$\boxed{\neg q}$};
     \node[] [] (x) [below left=of u,xshift=2em,yshift=2em]{$\boxed{\dia q,p}$};
     \node[] [] (y) [below right=of u,xshift=-2em,yshift=2em]{$\boxed{\neg q \lor q}$};

    \draw[->] (w) -- (u) node[midway, above] {$\circ$};
    \draw[->] (w) -- (v) node[midway, above] {$\circ$};
    \draw[->] (u) -- (x) node[midway, above] {$\circ$};
    \draw[->] (u) -- (y) node[midway, above] {$\bullet$};
     
     \node[] [] (w1) [left=of w,xshift=-12em] {$\boxed{\blbox p}$};
     \node[] [] (u1) [below right=of w1,xshift=-2em,yshift=2em]{$\boxed{p}$};
     \node[] [] (v1) [below right=of u1,xshift=-2em,yshift=2em]{$\boxed{q}$};

     \draw[->] (w1) -- (u1) node[midway, above] {$\circ$};
     \draw[->] (u1) -- (v1) node[midway, above] {$\bullet$};
     
     \node[] [] (w2) [left=of w1,xshift=1em] {$\boxed{p, \dia q}$};

\end{tikzpicture}
\end{center}
 
 Nested sequent calculi typically exhibit a mode of inference referred to as \emph{deep-inference}, whereby inference rules may be applied to any node within the tree encoded by the sequent~\cite{Gug07}. This contrasts with \emph{shallow-inference}, where inference rules are only applicable to the root of the tree encoded by the sequent. We remark that shallow-inference is an essential feature of \emph{display calculi}, which will be discussed in \sect~\ref{subsec:display} below. Although nested calculi are typically formulated with deep-inference, shallow-inference versions have been introduced~\cite{GorPosTiu08,Kas94}.\footnote{We note that in shallow-inference nested calculi certain rules called \emph{display} or \emph{residuation} rules are required for completeness. This will be discussed in the next section.} Nevertheless, as has become standard in the literature, we understand the term \emph{nested sequent calculus} to mean \emph{deep-inference nested sequent calculus} as the shallow-inference variants are known to be subsumed by the display calculus formalism~\cite{CiaLyoRamTiu21} and will be considered as such.

\begin{figure}[t]\label{fig:labeled-calculus-K}

\begin{center}
\begin{tabular}{c c c}
\AxiomC{$\phantom{\Gamma}$}
\RightLabel{$\id$}
\UnaryInfC{$\ns\{\Gamma, p, \dual p\}$}
\DisplayProof

&

\AxiomC{$\ns\{\Gamma, \forma, \formb\}$}
\RightLabel{$\disru$}
\UnaryInfC{$\ns\{\Gamma, \forma \lor \formb\}$}
\DisplayProof

&

\AxiomC{$\ns\{\Gamma, \forma\}$}
\AxiomC{$\ns\{\Gamma, \formb\}$}
\RightLabel{$\conru$}
\BinaryInfC{$\ns\{\Gamma, \forma \land \formb\}$}
\DisplayProof
\end{tabular}
\end{center}
\begin{center}
\begin{tabular}{c c c}
\AxiomC{$\ns\{\Gamma, \wnest{\forma}\}$}
\RightLabel{$\boxru$}
\UnaryInfC{$\ns\{\Gamma, \Box \forma\}$}
\DisplayProof

&

\AxiomC{$\ns\{\dia \forma, \wnest{\Gamma, \forma}\}$}
\RightLabel{$\diarui$}
\UnaryInfC{$\ns\{\dia \forma, \wnest{\Gamma}\}$}
\DisplayProof

&

\AxiomC{$\ns\{\Gamma, \forma, \bnest{\Delta, \dia \forma}\}$}
\RightLabel{$\diaruii$}
\UnaryInfC{$\ns\{\Gamma, \bnest{\Delta, \dia \forma}\}$}
\DisplayProof
\end{tabular}
\end{center}

\begin{center}
\begin{tabular}{c c c}
\AxiomC{$\ns\{\Gamma, \bnest{\forma}\}$}
\RightLabel{$\blboxru$}
\UnaryInfC{$\ns\{\Gamma, \blbox \forma\}$}
\DisplayProof

&

\AxiomC{$\ns\{\bldia \forma, \bnest{\Gamma, \forma}\}$}
\RightLabel{$\bldiarui$}
\UnaryInfC{$\ns\{\bldia \forma, \bnest{\Gamma}\}$}
\DisplayProof

&

\AxiomC{$\ns\{\Gamma, \forma, \wnest{\Delta, \bldia \forma}\}$}
\RightLabel{$\bldiaruii$}
\UnaryInfC{$\ns\{\Gamma, \wnest{\Delta, \bldia \forma}\}$}
\DisplayProof
\end{tabular}
\end{center}

\caption{The nested sequent system $\ncalckt$ for the modal logic $\logickt$~\cite{Kas94}.}
\end{figure}

 An example of a nested sequent calculus for the tense logic $\logickt$, referred to as $\ncalckt$, is provided in \fig~\ref{fig:labeled-calculus-K} and is due to Kashima~\cite{Kas94}. The notation $\ns\{\Gamma\}$ is commonly employed in the formulation of nested inference rules and exhibits deep-inference. We read $\ns\{\Gamma\}$ as stating that the nested sequent $\Gamma$ occurs at some node in the tree encoded by the nested sequent $\ns$. For example, we can write the nested sequent $\ns_{2}$ above as $\ns_{2}\{p,\bnest{q}\}$, or the nested sequent $\ns_{3}$ above as $\ns_{3}\{\neg q\}$ in this notation, thus letting us refer to the displayed nodes and the data confined within. Similarly, we may refer to multiple nodes in a nested sequent $\ns$ simultaneously by means of the notation $\ns\{\Gamma_{1}\}\{\Gamma_{2}\} \cdots \{\Gamma_{n}\}$. For instance, we could write $\ns_{2}$ as $\ns_{2}\{p\}\{q\}$ or $\ns_{3}$ as $\ns_{3}\{p \land r\}\{\wnest{\dia q, p}\}$.
 
 The $\id$ rule in $\ncalckt$ states that any nested sequent containing both $p$ and $\neg p$ at a node is an axiom. The remaining rules tell us how complex logical formulae may be constructed within any given node of a derivable nested sequent. For example, $\disru$ states that $A, B$ can be replaced by $A \lor B$, and $\boxru$ states that $\wnest{A}$ can be replaced by $\Box A$. As an example of how derivations are constructed in nested sequent calculi, we show how the modal axiom $\mathsf{K}$ (in negation normal form) can be derived in $\ncalckt$ below.
\begin{center}
\AxiomC{}
\RightLabel{$\id$}
\UnaryInfC{$\dia (p \land \neg q), \dia \neg p, \wnest{p, \neg p,q}$}
\AxiomC{}
\RightLabel{$\id$}
\UnaryInfC{$\dia (p \land \neg q), \dia \neg p, \wnest{\neg q, \neg p,q}$}
\RightLabel{$\conru$}
\BinaryInfC{$\dia (p \land \neg q), \dia \neg p, \wnest{p \land \neg q, \neg p,q}$}
\RightLabel{$\diaru \times 2$}
\UnaryInfC{$\dia (p \land \neg q), \dia \neg p, \wnest{q}$}
\RightLabel{$\boxru$}
\UnaryInfC{$\dia (p \land \neg q), \dia \neg p, \Box q$}
\RightLabel{$\disru \times 2$}
\UnaryInfC{$\dia (p \land \neg q) \lor \dia \neg p \lor \Box q$}
\DisplayProof
\end{center}

\begin{figure}[t]

\begin{center}
\begin{tabular}{c c}
\AxiomC{$\phantom{\Gamma}$}
\RightLabel{$\id^{\dag}$}
\UnaryInfC{$\ns\{\Gamma_{1}, p \nsar \Delta_{1}\}_{w}\{\Gamma_{2} \nsar p, \Delta_{2}\}_{u}$}
\DisplayProof

&

\AxiomC{$\phantom{\Gamma}$}
\RightLabel{$\botl$}
\UnaryInfC{$\ns\{\Gamma, \bot \nsar \Delta\}_{w}$}
\DisplayProof
\end{tabular}
\end{center}

\begin{center}
\begin{tabular}{c c}
\AxiomC{$\ns\{\Gamma, \forma \nsar \Delta\}_{w}$}
\AxiomC{$\ns\{\Gamma, \formb \nsar \Delta\}_{w}$}
\RightLabel{$\disl$}
\BinaryInfC{$\ns\{\Gamma, \forma \lor \formb \nsar \Delta\}_{w}$}
\DisplayProof

&

\AxiomC{$\ns\{\Gamma \nsar \forma, \formb, \Delta\}_{w}$}
\RightLabel{$\disr$}
\UnaryInfC{$\ns\{\Gamma \nsar \forma \lor \formb, \Delta\}_{w}$}
\DisplayProof
\end{tabular}
\end{center}

\begin{center}
\begin{tabular}{c c}
\AxiomC{$\ns\{\Gamma, \forma, \formb \nsar \Delta\}_{w}$}
\RightLabel{$\conl$}
\UnaryInfC{$\ns\{\Gamma, \forma \land \formb \nsar \Delta\}_{w}$}
\DisplayProof

&

\AxiomC{$\ns\{\Gamma \nsar \forma, \Delta\}_{w}$}
\AxiomC{$\ns\{\Gamma \nsar \formb, \Delta\}_{w}$}
\RightLabel{$\conr$}
\BinaryInfC{$\ns\{\Gamma \nsar \forma \land \formb, \Delta\}_{w}$}
\DisplayProof
\end{tabular}
\end{center}

\begin{center}
\resizebox{\columnwidth}{!}{
\begin{tabular}{c}
\AxiomC{$\ns\{\Gamma_{1}, \forma \iimp \formb \nsar \Delta_{1}\}_{w}\{\Gamma_{2}, \formb \nsar \Delta_{2}\}_{u} \quad \ns\{\Gamma_{1}, \forma \iimp \formb \nsar \Delta_{1}\}_{w}\{\Gamma_{2} \nsar \forma, \Delta_{2}\}_{u}$}
\RightLabel{$\iimpl^{\dag}$}
\UnaryInfC{$\ns\{\Gamma_{1}, \forma \iimp \formb \nsar \Delta_{1}\}_{w}\{\Gamma_{2} \nsar \Delta_{2}\}_{u}$}
\DisplayProof
\end{tabular}
}
\end{center}

\begin{center}
\begin{tabular}{c}
\AxiomC{$\ns\{\Gamma \nsar \Delta, [\forma \nsar \formb]_{u}\}_{w}$}
\RightLabel{$\iimpr$} 
\UnaryInfC{$\ns\{\Gamma \nsar \Delta, \forma \iimp \formb\}_{w}$}
\DisplayProof
\end{tabular}
\end{center}

\noindent
\textbf{Side condition:} $\dag$ = $u$ must be reachable from $w$. 

\caption{The nested sequent system $\ncalcint$ for intuitionistic logic.\label{fig:nested-calculus-Int}}
\end{figure}

Nested sequent calculi admit a couple methods of construction, which have proven to be rather general. One method is due to Gor\'e et al.~\cite{GorPosTiu08,GorPosTiu11} and consists of extracting nested sequent calculi from display calculi. The second method, referred to as \emph{structural refinement}, is due to Lyon~\cite{Lyo21b,Lyo21thesis} and consists of extracting nested sequent calculi from labeled sequent calculi or semantic presentations of non-classical logics.\footnote{Labeled sequent calculi are discussed in \sect~\ref{subsec-labeled} below.} In fact, a general algorithm was recently defined for extracting (cut-free) nested sequent calculi from (Horn) labeled sequent calculi~\cite{LyoOst24}. Since methods of construction for display and labeled calculi are well-understood and general, these approaches have led to the formulation of broad classes of cut-free nested sequent calculi for a variety of logics, including bi-intuitionistic logics~\cite{GorPosTiu08,LyoShiTiu24}, intuitionistic modal logics~\cite{Lyo21b}, (deontic) agency logics~\cite{LyoBer19,Lyo21thesis}, and standpoint logic~\cite{LyoGom22} (used in knowledge integration). 
 
Both methods rely on the elimination of structural rules in a display or labeled calculus, replacing them with \emph{propagation rules}~\cite{CasCerGasHer97,Fit72,Sim94}, or the more general class of \emph{reachability rules}~\cite{Lyo21thesis,Lyo23a,LyoShiTiu24}. Propagation rules operate by (bottom-up) propagating data along paths within a sequent, whereas reachability rules have the added functionality that data can be searched for within a sequent, and potentially propagated elsewhere (see~\cite[Chapter 5]{Lyo21thesis} for a discussion of these types of rules). Since propagation and reachability rules play a crucial role in the formulation of nested sequent calculi, we will demonstrate their functionality by means of an example. More specifically, we will introduce the nested sequent calculus $\ncalcint$, shown in \fig~\ref{fig:nested-calculus-Int}, which employs the $\iimpl$ propagation rule and $\id$ reachability rule.\footnote{The calculus $\ncalcint$ is the propositional fragment of the nested calculi given for first-order intuitionistic logics in~\cite{Lyo21thesis}, and is a variation of the nested calculus given by Fitting~\cite{Fit14}.} 

Nested sequents in $\ncalcint$ are generated via the following grammars:
$$
\ns ::= \Gamma \nsar \Gamma \ | \ \ns, [\ns]_{w}
\qquad
\Gamma ::= A \ | \ \emptyset \ | \ \Gamma, \Gamma
$$
 where $A \in \langint$, $w$ is among a countable set of labels $w$, $u$, $v$ $\ldots$, and $\emptyset$ is the empty multiset. The notation used in the rules of $\ncalcint$ marks nestings with labels, e.g., in $\id$ and $\iimpl$ the labels $w$ and $u$ are used. It is assumed that each label is used once in a nested sequent and we note that such labels are merely a naming device used to simplify the formulation of certain inference rules. 
As stated in \fig~\ref{fig:nested-calculus-Int}, the $\id$ and $\iimpl$ rules have a side condition stating that each respective rule is applicable only if the node $u$ is reachable from the node $w$. This means that in the tree encoded by the nested sequent $\ns$, the rule is applicable only if there is a path (which could be of length $0$) from $w$ to $u$. For example, in the $\ncalcint$ proof below the $\iimpl$ rule recognizes the $p \iimp q$ in the $w$ nesting and propagates $q$ into $[p \nsar q]_{u}$ in the left premise and $p$ into $[p \nsar q]_{u}$ in the right premise when read bottom-up. Each premise of $\iimpl$ can be read as an instance of $\id$ since $u$ is reachable from $u$ with a path of length $0$.
\begin{center}
\AxiomC{}
\RightLabel{$\id$}
\UnaryInfC{$\nsar [p \iimp q \nsar [p, q \nsar q]_{u}]_{w}$}

\AxiomC{}
\RightLabel{$\id$}
\UnaryInfC{$\nsar [p \iimp q \nsar [p \nsar p, q]_{u}]_{w}$}

\RightLabel{$\iimpl$}
\BinaryInfC{$\nsar [p \iimp q \nsar [p \nsar q]_{u}]_{w}$}
\RightLabel{$\iimpr$}
\UnaryInfC{$\nsar [p \iimp q \nsar p \iimp q]_{w}$}
\RightLabel{$\iimpr$}
\UnaryInfC{$\nsar (p \iimp q) \iimp (p \iimp q)$}
\DisplayProof
\end{center}

In \sect~\ref{sec:organizing-jungle}, we show how 
$\ncalcint$ can be extracted from a labeled sequent calculus, thus exemplifying the structural refinement method~\cite{Lyo21thesis,Lyo21b}.

 We end this subsection with a brief discussion concerning the relationship between propagation/reachability rules and the property of \emph{modularity}, that is, the ease with which a calculus for one logic may be transformed into a calculus for another logic within a given class. An interesting feature of propagation and reachability rules concerns the means by which they introduce modularity into a proof calculus. It has been argued---most notably by Avron~\cite[Section~1]{Avr96} and Wansing~\cite[Section~3.3]{Wan94}---that modularity ought to be obtained via \emph{Do\v{s}en's Principle}, which is stated accordingly:

\begin{quote}
[T]he rules for the logical operations are never changed: all changes are made in the structural rules~\cite[p.~352]{Dos88}
\end{quote}

\noindent
 Although we agree that modularity is an important feature of a proof formalism, we argue that Do\v{s}en's principle is \emph{too strict}. This perspective is supported by the formulation of propagation/reachability rules within nested systems, which attain modularity by a different means. Since these types of rules generalize the functionality of logical rules by permitting data to be shifted or consumed along paths within a nested sequent, systems which include such rules possess a high degree of modularity, obtained by simply changing the paths considered, irrespective of structural rules. 

\subsection{Display Sequents}\label{subsec:display}

Introduced by Belnap~\cite{Bel82} (and originally called \emph{Display Logic}), the {\em Display Calculus} formalism generalizes Gentzen's sequent calculus by supplementing the structural connective ($,$) and the turnstile ($\seq$) with a host of new structural connectives---corresponding to pairs of dual connectives---and rules manipulating them. Incorporating structural connectives for pairs of dual connectives has proven fruitful for the construction of cut-free proof systems for large classes of logics, including modal and intuitionistic logics~\cite{Bel82}, tense logics~\cite{Kra96,Wan94}, bunched implication logics~\cite{Bro12}, resource sensitive logics~\cite{Gor98}, and bi-intuitionistic logic~\cite{Wol98}. Display calculi also admit algorithmic constructibility starting from Hilbert axioms~\cite{CiaRam16,ALBA}. To provide the reader with intuition concerning display systems and related concepts we accompany our general descriptions of such systems with concrete examples in the context of tense logics~\cite{Pri57} and which comes from the work in~\cite{Kas94,Kra96,Wan94}. 

First, to demonstrate the concept of a structural connective, let us define \emph{structures}, which serve as the entire antecedent or consequent of a display sequent and fuse together formulae by means of structural connectives. When defining display sequents for tense logics, we let a structure $X$ be a formula obtained via the following grammar in BNF:
$$
X ::= A \ | \ I \ | \ {\ast}X \ | \ {\bullet} X \ | \ (X \circ X)
$$
where $A$ is a formula in the language of tense logic, i.e., within the language $\langten$. 
Using $X$, $Y$, $Z$, $\ldots$ to represent structures, we define a \emph{display sequent} to be a formula of the form $X \dsar Y$. We provide the reader with an example of a display sequent as well as show the pair of graphs representing the structures present in the antecedent and consequent of the display sequent.

\begin{example} As can be seen in the example below the antecedent (shown bottom left) and consequent (shown bottom right) of a (tense) display sequent encodes a polytree; cf.~\cite{Lyo24arxivLD}.
\begin{flushleft}
$
\underbrace{\bullet (\forma_{1} \circ \ast \forma_{2}) \circ \bullet \forma_{3} \circ \forma_{4}}_{Antecedent} \dsar \underbrace{\forma_{5} \circ \bullet (\ast \forma_{6} \circ \ast \forma_{7}) \circ \bullet \forma_{8}}_{Consequent}
$
\end{flushleft}
\begin{flushright}
\begin{tikzpicture}
     \node[] [] (x) [] {$\boxed{A_{4}}$};
     \node[] [] (z) [left=of x,yshift=2em]{$\boxed{A_{1} \circ \ast A_{2}}$};
     \node[] [] (y) [left=of x,yshift=-2em]{$\boxed{A_{3}}$};

     \path[->,draw] (y) to [] (x);
     \path[->,draw] (z) to [] (x);
     
     
     \node[] [] (w) [right=of x] {$\boxed{A_{5}}$};
     \node[] [] (u) [right=of w,yshift=2em]{$\boxed{\ast A_{6} \circ \ast A_{7}}$};
     \node[] [] (v) [right=of w,yshift=-2em]{$\boxed{A_{8}}$};

     \path[->,draw] (w) to [] (u);
     \path[->,draw] (w) to [] (v);
\end{tikzpicture}
\end{flushright}
\end{example}

A characteristic feature of the display calculus is the display property,
which states that every occurrence of a substructure in a sequent can be
written (displayed) as the entire antecedent or succedent (but not both).
Rules enabling the display property are called {\em display rules} or {\em residuation rules}, and display sequents derivable from one another via such rules are called {\em display equivalent}. These rules are invertible and hence a sequent can be identified
with the class of its display equivalent sequents.
%
For example, the bullet $\bullet$ represents a $\bldia$ in the antecedent of a display sequent and a $\Box$ in the consequent, and since $\bldia A \rightarrow B$ and $A \rightarrow \Box B$ are equivalent in the setting of tense logics, the display sequents $\bullet A \dsar B$ and $A \dsar \bullet B$ are defined to be mutually derivable from one another. This gives rise to the following \emph{display rule} introduced by Wansing~\cite{Wan94}.
\begin{center}
\AxiomC{$\bullet X \dsar Y$}
\doubleLine
\RightLabel{$(\bullet)$}
\UnaryInfC{$X \dsar \bullet Y$}
\DisplayProof
\end{center}

As mentioned in the previous section, nested sequent calculi employing shallow-inference are also types of display calculi. As an example, if we take the nested calculus $\ncalckt$ and add the display/residuation rules $\rf$ and $\rp$ rules shown below left as well as replace the $\diarui$, $\diaruii$, $\bldiarui$, and $\bldiaruii$ rules with the $\diaru$ and $\bldiaru$ rules shown below right, then we obtain Kashima's shallow-inference (i.e., display) calculus $\dcalckt$ for the logic $\logickt$~\cite{Kas94}. The calculus $\dcalckt$ can be seen as a `one-sided' display calculus that equates nested sequents with structures.  
\begin{center}
\begin{tabular}{c c c c}
\AxiomC{$\Gamma, \wnest{\Delta}$}
\RightLabel{$\rf$}
\UnaryInfC{$\bnest{\Gamma}, \Delta$}
\DisplayProof

&

\AxiomC{$\Gamma, \bnest{\Delta}$}
\RightLabel{$\rp$}
\UnaryInfC{$\wnest{\Gamma}, \Delta$}
\DisplayProof

&

\AxiomC{$\Gamma, \dia A, \wnest{\Delta, A}$}
\RightLabel{$\diaru$}
\UnaryInfC{$\Gamma, \dia A, \wnest{\Delta}$}
\DisplayProof

&

\AxiomC{$\Gamma, \bldia A, \bnest{\Delta, A}$}
\RightLabel{$\bldiaru$}
\UnaryInfC{$\Gamma, \bldia A, \bnest{\Delta}$}
\DisplayProof
\end{tabular}
\end{center}
The rules $\rf$ and $\rp$ are similar to Wansing's display rule $(\bullet)$, however, they rely on the equivalence between $\blbox \neg A \lor B$ and $\neg A \lor \Box B$. 

Quite significantly, Belnap's seminal paper~\cite{Bel82} also proves a \emph{general cut-elimination theorem}, that is, if a display calculus satisfies a set of eight conditions, then cut-elimination follows as a corollary. 
The display formalism has been used to supply truly sizable classes of logics with cut-free proof systems, proving the formalism general. Furthermore, the formalism is highly uniform and modular as structural rules are used to capture distinct logics, and due to the algorithms permitting their construction~\cite{CiaRam16,ALBA}, they enjoy constructibility. These nice features come at the cost of syntactic parsimony however as display sequents utilize complex structures to facilitate reasoning.

\subsection{Labeled Sequents}\label{subsec-labeled}

Labeled sequents generalize Gentzen sequents by annotating formulae with labels and introducing semantic elements into the syntax of sequents. For example, the labeled sequents used by Simpson~\cite{Sim94} and Vigan{\`o}~\cite{Vig00} have the form shown below top in the following example and encode a binary graph of Gentzen sequents. Thus, labeled sequents properly generalize all sequents considered in previous sections.

\begin{example} A labeled sequent is shown below top and its corresponding graph is shown below bottom.
$$
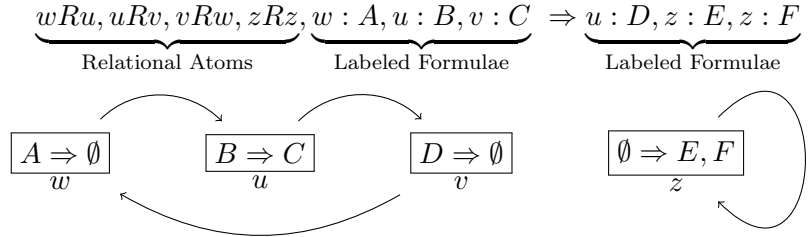
\underbrace{wRu, uRv, vRw, zRz}_{\text{Relational Atoms}}, \underbrace{w : A, u : B, v : C}_{\text{Labeled Formulae}} \ \lsar \underbrace{u : D, z : E, z : F}_{\text{Labeled Formulae}}
$$
\begin{center}
\vspace*{-3em}
\begin{tikzpicture}
     \node[] [] (w) [] {$\overset{\boxed{A \lsar \emptyset}}{w}$};
     \node[] [] (u) [right=of w]{$\overset{\boxed{B \lsar D}}{u}$};
     \node[] [] (v) [right=of u]{$\overset{\boxed{C \lsar \emptyset}}{v}$};
     \node[] [] (z) [right=of v]{$\overset{\boxed{\emptyset \lsar E, F}}{z}$};

     \path[->,draw,every loop/.style={looseness=5}] (z) to [out=45,in=315,loop] (z);
     \path[->,draw] (w) to [out=45,in=135] (u);
     \path[->,draw] (u) to [out=45,in=135] (v);
     \path[->,draw] (v) to [out=225,in=315,bend left=30] (w);
\end{tikzpicture}
\end{center}
\end{example}

\vspace*{-3em}

The idea of labeling formulae in sequents comes from Kanger~\cite{Kan57}, who made use of \emph{spotted formulae} to construct sequent systems for normal modal logics.\footnote{We note that labeling has also been used in tableaux for modal logics, e.g., the prefixed tableaux of Fitting~\cite{Fit72}.} The labeled sequent formalism is quite general and covers many logics; e.g., intuitionistic (modal) logics~\cite{DycNeg12,Sim94}, normal modal logics~\cite{Hei05,Min97,Vig00}, predicate modal logics~\cite{CasSma02,KusOka03}, relevance logics~\cite{Vig00}, and (deontic) agency logics~\cite{BerLyo19a,BerLyo21,LyoBer19}.

The labeled sequent formalism offers a high degree of uniformity and modularity. This is typically obtained by taking a \emph{base calculus} for a particular logic and showing that through the addition of structural rules the calculus can be extended into a calculus for another logic within a specified class. A favorable feature of the labeled sequent formalism is the existence of general theorems confirming properties such as cut-admissibility, invertibility of rules, and admissibility of standard structural rules~\cite{DycNeg12,Sim94,Vig00}. Another desirable characteristic concerns the ease with which labeled systems are constructed; e.g., for logics with a Kripkean semantics one straightforwardly obtains labeled calculi by transforming the semantic clauses and frame/model conditions into inference rules. Thus, the labeled formalism is highly general, uniform, modular, and constructible, but as with the display formalism, this comes at a cost of syntactic complexity.

To provide the reader with intuition about the generality, uniformity, and modularity of labeled sequent systems we consider a specific system for the tense logic $\logickt$ (cf.~\cite{CiaLyoRamTiu21,Bor08}). The labeled sequents used are defined to be expressions of the form $\rel \lsar \Gamma$, where $\rel$ is a set of \emph{relational atoms} $wRu$ and $\Gamma$ is a multiset of \emph{labeled formulae} $w : \forma$ with $\forma \in \langten$. The labeled sequent system $\lcalckt$ is presented in \fig~\ref{fig:labeled-calculus-K} and contains the axiom $\id$ as well as six logical rules; note that the $\boxru$ and $\blboxru$ rules is subject to a side condition, namely, the label $u$ must be \emph{fresh} and not occur in the conclusion of a rule application.


\begin{figure}[t]

\begin{center}
\begin{tabular}{c c}
\AxiomC{$\phantom{\Gamma}$}
\RightLabel{$\id$}
\UnaryInfC{$\rel \lsar w :p, w :\overline{p}, \Gamma$}
\DisplayProof

&

\AxiomC{$\rel \lsar w : \forma, w : \formb, \Gamma$}
\RightLabel{$\disru$}
\UnaryInfC{$\rel \lsar w : \forma \lor \formb, \Gamma$}
\DisplayProof
\end{tabular}
\end{center}

\begin{center}
\begin{tabular}{c}
\AxiomC{$\rel \lsar w : \forma, \Gamma$}
\AxiomC{$\rel \lsar w : \formb, \Gamma$}
\RightLabel{$\conru$}
\BinaryInfC{$\rel \lsar w : \forma \land \formb, \Gamma$}
\DisplayProof
\end{tabular}
\end{center}

\begin{center}
\begin{tabular}{c c}
\AxiomC{$\rel, wRu \lsar w : \Diamond \forma, u : \forma, \Gamma$}
\RightLabel{$\diaru$}
\UnaryInfC{$\rel, wRu \lsar w : \Diamond \forma, \Gamma$}
\DisplayProof

&

\AxiomC{$\rel, uRw \lsar w : \bldia \forma, u : \forma, \Gamma$}
\RightLabel{$\bldiaru$}
\UnaryInfC{$\rel, uRw \lsar w : \bldia \forma, \Gamma$}
\DisplayProof
\end{tabular}
\end{center}

\begin{center}
\begin{tabular}{c c}
\AxiomC{$\rel, wRu \lsar u : \forma, \Gamma$}
\RightLabel{$\boxru$}
\UnaryInfC{$\rel \lsar w : \Box \forma, \Gamma$}
\DisplayProof

&

\AxiomC{$\rel, uRw \lsar u : \forma, \Gamma$}
\RightLabel{$\blboxru$}
\UnaryInfC{$\rel \lsar w : \blbox \forma, \Gamma$}
\DisplayProof
\end{tabular}
\end{center}

\noindent
\textbf{Side condition:} $u$ must be fresh in $\boxru$ and $\blboxru$.

\caption{Labeled sequent system $\lcalckt$ for the tense logic $\logickt$~\cite{Bor08,CiaLyoRamTiu21}.\label{fig:labeled-calculus-K}}
\end{figure}

This calculus may be extended with structural rules to obtain labeled sequent systems for \emph{extensions} of $\logickt$ (e.g., tense $\logicsiv$ and $\logicsv$). Such rules encode frame properties corresponding to axioms. For example, to obtain a labeled sequent system for $\logickt$ with serial frames one can extend $\lcalckt$ with the $(ser)$ rule shown below, and to obtain a labeled sequent calculus for tense $\logicsiv$ one can extend $\lcalckt$ with the structural rules $(ref)$ and $(tra)$.

\begin{center}
\begin{tabular}{c @{\hskip 1em} c}
\AxiomC{$\rel, wRu \lsar \Gamma$}
\RightLabel{$(ser)$~\textit{with} $u$ \textit{fresh}}
\UnaryInfC{$\rel \lsar \Gamma$}
\DisplayProof

&

\AxiomC{$\rel, wRw \lsar \Gamma$}
\RightLabel{$(ref)$}
\UnaryInfC{$\rel \lsar \Gamma$}
\DisplayProof
\end{tabular}
\end{center}
\begin{center}
\AxiomC{$\rel, wRu, uRv, wRv \lsar \Gamma$}
\RightLabel{$(tra)$}
\UnaryInfC{$\rel, wRu, uRv \lsar \Gamma$}
\DisplayProof
\end{center}

In fact, most commonly studied frame properties for logics with Kripkean semantics (such as seriality, reflexivity, and transitivity) can be transformed into equivalent structural rules. Simpson showed that a large class of properties, referred to as \emph{geometric axioms}, could be transformed into equivalent structural rules, referred to as \emph{geometric rules}, in labeled sequent calculi~\cite{Sim94}. A geometric axiom is a formula of the form shown below, where each $\forma_{i}$ and $\formb_{j,k}$ is a first-order atom.
$$
\forall x_{1}, \ldots, x_{t} \big( A_{1} \land \cdots \land \forma_{n} \rightarrow \exists y_{1}, \ldots, y_{s} (\bigvee_{j = 1}^{m} \bigwedge_{k = 1}^{l_{j}} \formb_{1,k}) \big)
$$
Each geometric formula is equivalent to the geometric rule of the form shown below, where $\vec{\forma} := \forma_{1}, \ldots, \forma_{n}$, $\vec{\formb}_{j} := \formb_{j,1}, \ldots, \formb_{j,l_{j}}$, and each variable $y_{1}, \ldots, y_{s}$ is fresh, i.e., the rule can be applied only if none of the variables $y_{1}, \ldots, y_{s}$ occur in the conclusion.
\begin{center}
\AxiomC{$\rel, \vec{\forma}, \vec{\formb}_{1} \lsar \Delta$}
\AxiomC{$\cdots$}
\AxiomC{$\rel, \vec{\forma}, \vec{\formb}_{m} \lsar \Delta$}
\TrinaryInfC{$\rel, \vec{\forma} \lsar \Delta$}
\DisplayProof
\end{center}

\fig~\ref{fig:lab-struc-rules} gives a selection of structural rules that are typically admissible in labeled sequent systems with geometric rules. The weakening rule $\wk$ adds additional labeled formulae to the consequent of a labeled sequent, the contraction rule $\ctr$ removes additional copies of labeled formulae, the substitution rule $\lsub$ replaces a label $u$ by a label $w$ in a labeled sequent, and the $\cut$ rule encodes the transitivity of implication. The admissibility of these rules tends to hold generally for labeled sequent systems, along with all logical rules being invertible~\cite{DycNeg12,Hei05,LyoBer19,Sim94}. Beyond admissibility and invertibility properties, labeled systems allow for easy counter-model extraction due to the incorporation of semantic notions into the syntax of sequents, though termination of proof-search is not easily achieved as labeled sequents contain a large amount of structure.


\begin{figure}[t]

\begin{center}
\begin{tabular}{c c c}
\AxiomC{$\rel \lsar \Delta$}
\RightLabel{$\wkl$}
\UnaryInfC{$\rel, wRu \lsar \Delta$}
\DisplayProof

&

\AxiomC{$\rel \lsar \Delta$}
\RightLabel{$\wkr$}
\UnaryInfC{$\rel \lsar w: \forma, \Delta$}
\DisplayProof

&

\AxiomC{$\rel \lsar \Delta$}
\RightLabel{$\lsub$}
\UnaryInfC{$\rel[w/u] \lsar \Delta[w/u]$}
\DisplayProof
\end{tabular}
\end{center}

\begin{center}
\begin{tabular}{c c}
\AxiomC{$\rel \lsar w : \forma, w : \forma, \Delta$}
\RightLabel{$\ctr$}
\UnaryInfC{$\rel \lsar w : \forma, \Delta$}
\DisplayProof

&

\AxiomC{$\rel \lsar w : \forma, \Delta$}
\AxiomC{$\rel \lsar w : \overline{\forma}, \Delta$}
\RightLabel{$\cut$}
\BinaryInfC{$\rel \lsar \Delta$}
\DisplayProof
\end{tabular}
\end{center}

\caption{Labeled structural rules.\label{fig:lab-struc-rules}}
\end{figure}

Last, we note that although $\cut$ is usually admissible in labeled sequent systems, it is often the case that a \emph{strict} form of the subformula property fails to hold. 
This phenomenon arises due to the incorporation of geometric rules which may delete relational atoms from the premise when inferring the conclusion. Nevertheless, it is usually the case that labeled sequent systems possess a \emph{weak} version of the subformula property, i.e., it can be shown that every \emph{labeled formula} occurring in a derivation is a subformula of some labeled formula in the conclusion~\cite{Vig00}.

\section{Navigating the Proof-Theoretic Jungle}\label{sec:organizing-jungle}

As discussed in \sect~\ref{sec:formalisms} and shown in \fig~\ref{fig:hierarchy}, the data structure underlying sequents naturally imposes a hierarchy on sequent-style formalisms. At the base of this hierarchy sits Gentzen sequents, and each level of the hierarchy gets incrementally more general until labeled sequents are reached at the top. As we are interested in exploring this hierarchy, we present translations of proofs between systems in different proof-theoretic formalisms, thus letting us `shift' derivations up and down the hierarchy. The lesson we learn is that translating proofs down the hierarchy (usually) requires significantly more work than translating proofs up the hierarchy.


\subsection{Translations for $\logicsv$: Labeled and Hypersequent Calculi}\label{subsec:lab-hyper-s5}

We begin our demonstration of how to translate proofs between distinct formalisms by considering translations between hypersequent and labeled calculi for the modal logic $\logicsv$. In particular, we will explain how proofs are translated between the hypersequent calculus $\hcalcsv$ (see \fig~\ref{fig:hyperseq-sfive}) and the labeled sequent calculus $\lcalcsv$ shown in \fig~\ref{fig:s5-hpyersequent-rules}.

\begin{figure}[t]

\begin{center}
\begin{tabular}{c c}
\AxiomC{$\phantom{\Gamma}$}
\RightLabel{$\id$}
\UnaryInfC{$\Gamma, w : p \hsar w : p, \Delta$}
\DisplayProof

&

\AxiomC{$\phantom{\Gamma}$}
\RightLabel{$\botl$}
\UnaryInfC{$\Gamma, w : \bot \hsar \Delta$}
\DisplayProof
\end{tabular}
\end{center}

\begin{center}
\begin{tabular}{c c}
\AxiomC{$\Gamma \hsar w : \forma, w : \formb, \Delta$}
\RightLabel{$\disr$}
\UnaryInfC{$\Gamma \hsar w : \forma \lor \formb, \Delta$}
\DisplayProof

&

\AxiomC{$\Gamma, w : \forma \hsar \Delta$}
\AxiomC{$\Gamma, w : \formb \hsar \Delta$}
\RightLabel{$\disl$}
\BinaryInfC{$\Gamma, w : \forma \lor \formb \hsar \Delta$}
\DisplayProof
\end{tabular}
\end{center}

\begin{center}
\begin{tabular}{c c}
\AxiomC{$\Gamma, w : \forma, w : \formb \sar \Delta$}
\RightLabel{$\conl$}
\UnaryInfC{$\Gamma, w : \forma \land \formb \sar \Delta$}
\DisplayProof

&

\AxiomC{$\Gamma \hsar w : \forma, \Delta$}
\AxiomC{$\Gamma \hsar w : \formb, \Delta$}
\RightLabel{$\conr$}
\BinaryInfC{$\Gamma \hsar w : \forma \land \formb, \Delta$}
\DisplayProof
\end{tabular}
\end{center}

\begin{center}
\begin{tabular}{c c}
\AxiomC{$\Gamma\sar  w : \forma,  \Delta$}
\AxiomC{$\Gamma, w : \formb \sar \Delta$}
\RightLabel{$\impl$}
\BinaryInfC{$\Gamma, w : \forma \rightarrow \formb \sar \Delta$}
\DisplayProof

&

\AxiomC{$\Gamma, w : \forma \sar w : \formb, \Delta$}
\RightLabel{$\impr$}
\UnaryInfC{$\Gamma \sar w : \forma \rightarrow \formb, \Delta$}
\DisplayProof
\end{tabular}
\end{center}

\begin{center}
\begin{tabular}{c c}
\AxiomC{$\Gamma, w : \Box \forma, u : \forma \hsar \Delta$}
\RightLabel{$\boxl^{\dag_{1}}$}
\UnaryInfC{$\Gamma, w : \Box \forma \hsar \Delta$}
\DisplayProof

&

\AxiomC{$\Gamma \hsar u: \forma, \Delta$}
\RightLabel{$\boxr^{\dag_{2}}$}
\UnaryInfC{$\Gamma \hsar w : \Box \forma, \Delta$}
\DisplayProof
\end{tabular}
\end{center}

\textbf{Side conditions:} $\dag_{1}$ stipulates that $u \in \labset(\Gamma, \Delta)$ in $\boxl$ and $\dag_{2}$ stipulates that $u$ must be fresh in $\boxr$.

\caption{The labeled calculus $\lcalcsv$ for the modal logic $\logicsv$.\label{fig:s5-hpyersequent-rules}
}
\end{figure}

In this section, we define a labeled sequent to be an expression of the form $\Gamma \sar \Delta$ such that $\Gamma$ and $\Delta$ are finite multisets of labeled formulae $w : A$ with $w$ among a denumerable set $\labset := \{w, u, v, \ldots\}$ of labels and $A \in \langmod$. For a multiset $\Gamma$ of labeled formulae, we define $\labset(\Gamma)$ to be the set of all labels occurring in $\Gamma$, for a multiset $\set{A_{1}, \ldots, A_{n}}$ of formulae, we define $w : \set{A_{1}, \ldots, A_{n}} = \set{w : A_{1}, \ldots, w : A_{n}}$, and we let $\Gamma(w)$ be the multiset $\{A \mid w : A \in \Gamma\}$. 

A labeled sequent calculus $\lcalcsv$ for the modal logic $\logicsv$ is shown in \fig~\ref{fig:s5-hpyersequent-rules}. We remark that the labeled sequents used in $\lcalcsv$ have a simpler structure than those discussed in \sect~\ref{subsec-labeled}, namely, they do not use relational atoms. This is a special case and a byproduct of the fact that $\lcalcsv$ is a calculus for the modal logic $\logicsv$; in general, more complex modal logics require the use of relational atoms (cf.~\cite{Sim94,Vig00}). 

It is straightforward to define translations that map labeled sequents to hypersequents and vice-versa. To translate labeled sequents into hypersequents, we make use of the $h$ translation, defined as follows:
$$
h(\Gamma\lsar\Delta) := \Gamma(w_1) \lsar \Delta(w_1) \mid  \cdots \mid \Gamma(w_n) \lsar \Delta(w_n)
$$
 where $\labset(\Gamma, \Delta) := \set{w_{1}, \ldots, w_{n}}$. To translate hypersequents into labeled sequents, we make use of the $\ell$ translation, defined as follows:
$$
\ell(\Gamma_{1} \hsar \Delta_{1} \mid \cdots \mid \Gamma_{n} \hsar \Delta_{n}) := \bigcup_{1 \leq i \leq n} w_{i} : \Gamma_{i} \sar \bigcup_{1 \leq i \leq n} w_{i} : \Delta_{i}
$$
Using the above translations, we can confirm that all derivations (which, properly includes all proofs; see Section \ref{subsec-gentzen-sequent}) in $\lcalcsv$ and $\hcalcsv$ are isomorphic to each other.

\begin{proposition}
Every derivation in $\lcalcsv$ is isomorphic to a derivation in $\hcalcsv$ under the $h$ translation, and every derivation in $\hcalcsv$ is isomorphic to a derivation in $\lcalcsv$ under the $\ell$ translation.
\end{proposition}

\begin{proof} We prove the case for the $h$ translation by induction on the height of the given derivation $\deriv$, and remark that the case of translating proofs with the reverse translation $\ell$ is similar.

\textit{Base case.} If $\Gamma \sar \Delta$ is an axiom, i.e., an instance of $\id$ or $\botl$, then $h(\Gamma \sar \Delta)$ will be an axiom in $\hcalcsv$ as well. If $\Gamma \sar \Delta$ is a leaf in the derivation $\deriv$, but not an axiom, then $h(\Gamma \sar \Delta)$ trivially translates to a leaf in the hypersequent derivation.

\textit{Inductive step.} We show how to translate the $\boxl$ case. There are two cases to consider in the $\boxl$ case: in the first case, the label of the auxiliary formula $\forma$ is identical to the label of the principal formula $\Box \forma$. This is resolved as shown below where $G = h(\Gamma \setminus w : \Gamma(w) \sar \Delta \setminus w : \Delta(w))$.
\begin{center}
\AxiomC{$h(\Gamma, w : \Box \forma, w : \forma \hsar \Delta)$}
\RightLabel{=}
\UnaryInfC{$G \mid \Box \forma, \forma, \Gamma(w) \hsar \Delta(w)$}
\RightLabel{$\boxli$}
\UnaryInfC{$G \mid \Box \forma, \Gamma(w) \hsar \Delta(w)$}
\RightLabel{=}
\UnaryInfC{$h(\Gamma, w : \Box \forma \hsar \Delta)$}
\DisplayProof
\end{center}
In the second case, the label $u$ of the auxiliary formula is distinct from the label of the principal formula. This case is resolved as shown below where $G = h(\Gamma \setminus \set{w : \Gamma(w), u : \Gamma(u)} \sar \Delta \setminus \set{w : \Delta(w), u : \Delta(u)})$
\begin{center}
\AxiomC{$h(\Gamma, w : \Box \forma, u : \forma \hsar \Delta)$}
\RightLabel{=}
\UnaryInfC{$G \mid \Box \forma, \Gamma(w) \hsar \Delta(w) \mid \forma, \Gamma(u) \hsar \Delta(u)$}
\RightLabel{$\boxlii$}
\UnaryInfC{$G \mid \Box \forma, \Gamma(w) \hsar \Delta(w) \mid \Gamma(u) \hsar \Delta(u)$}
\RightLabel{=}
\UnaryInfC{$h(\Gamma, w : \Box \forma \hsar \Delta)$}
\DisplayProof
\end{center}
The remaining cases are easily resolved by applying IH and then the corresponding rule in $\hcalcsv$.
\end{proof}

\subsection{Translations for $\logickt$: Labeled and Display Calculi} \label{subsec:lab-dis-kt}

We show how to translate proofs from $\lcalckt$ into $\dcalckt$. The method of translation we present was first defined in~\cite{CiaLyoRamTiu21} and is strong enough to not only translate labeled proofs into display proofs for $\logickt$, but also for any extension of $\logickt$ with \emph{path axioms} of the form $\qdia_{1} \cdots \qdia_{n} p \rightarrow \qdia_{n+1} p$ with $\qdia_{i} \in \{\bldia,\dia\}$ for $1 \leq i \leq n+1$. A generalization of this technique is presented in~\cite{Lyo24arxivLD} and shows how to translate cut-free display proofs into cut-free labeled sequent proofs for the even wider class of \emph{primitive tense logics}~\cite{Kra96}. We note that the converse translation from $\dcalckt$ to $\lcalckt$ is simpler so we omit it, though the details can be found in~\cite{CiaLyoRamTiu21} for the interested reader.

The key to translating labeled proofs into display proofs is to recognize that `non-treelike' data (e.g., loops and cycles) cannot occur in proofs of theorems. We refer to these labeled sequents as \emph{labeled polytree sequents}~\cite{CiaLyoRamTiu21} and define them below. This insight is useful as labeled polytree sequents and display sequents are notational variants of one another, which facilitates our translation from $\lcalckt$ to $\dcalckt$. 

\begin{definition}[Labeled Polytree] Let $\Lambda := \rel, \Gamma \sar \Delta$  be a labeled sequent, and define the graph $G(\rel) = (V,E)$ such that $V$ is the set of labels ocurring in $\rel$ and $E = \{(w,u) \ | \ wRu \in \rel\}$. We define $\Lambda$ to be a \emph{labeled polytree sequent} \iffi $\rel$ forms a polytree, i.e., the graph $G(\rel)$ is connected and cycle-free, and all labels in $\Gamma, \Delta$ occur in $\rel$ (unless $\rel$ is empty, in which case every labeled formula in $\Gamma, \Delta$ must share the same label). We define a \emph{labeled polytree derivation} to be a derivation containing only labeled polytree sequents.
\end{definition}

\begin{lemma}\label{lem:lab-polytree-deriv}
Every derivation of a formula $\forma$ in $\lcalckt$ is a labeled polytree derivation.
\end{lemma}

\begin{proof} Suppose we are given a derivation of the labeled polytree sequent $\lsar w : \forma$ in $\lcalckt$. Observe that every rule of $\lcalckt$, if applied bottom-up to a labeled polytree sequent, yields a labeled polytree sequent since rules either preserve the set $\rel$ of relational atoms when applied bottom-up (e.g., $\disru$ and $\diaru$), or via $\boxru$ or $\blboxru$, add a new relational atom from a label occurring in the labeled sequent to a fresh label (which has the effect of adding a new forward or backward edge in the polytree encoded by the labeled sequent). Hence, the derivation of $\lsar w : \forma$ in $\lcalckt$ must be a labeled polytree derivation.
\end{proof}

We now define the $\ldkt$ function that maps labeled polytree sequents to display sequents, which can be stepwise applied to translate entire labeled polytree proofs into display proofs. As it will be useful here, and later on, we define the \emph{sequent composition} $\Lambda \seqcomp \Lambda'$ between two labeled sequents $\Lambda = \rel \sar \Gamma$ and $\Lambda' = \rel' \sar \Gamma'$ to be $\Lambda \seqcomp \Lambda' := \rel, \rel' \sar \Gamma, \Gamma'$.

\begin{definition}[Translation $\ldkt$]\label{def:lab-dis-trans} Let $\Lambda := \rel \sar \Gamma$ be a labeled polytree sequent containing the label $u$. We define $\Lambda' \subseteq \Lambda$ \iffi there exists a labeled polytree sequent $\Lambda''$ such that $\Lambda = \Lambda' \seqcomp \Lambda''$. Let us define $\Lambda_{u} := \rel' \sar \Gamma'$ to be the unique labeled polytree sequent rooted at $u$ such that $\Lambda_{u} \subseteq \Lambda$ and $\Gamma' \restriction u = \Gamma \restriction u$. We recursively define $\ldkt_{u}(\Lambda)$: 
\begin{itemize} 

\item[(1)] if $\rel = \empseq$, then $\ldkt_{v}(\Lambda) := (\lsar \Gamma \restriction v)$, and 

\item[(2)] if $vRx_{1}, \ldots vRx_{n}$ and $y_{1}Rv, \ldots y_{n}Rv$ are all relational atoms of the form $vRx$ and $yRx$, respectively, then
\end{itemize}
$$
\ldkt_{v}(\Lambda) := \Gamma \restriction v, \wnest{\ldkt_{x_{1}}(\Lambda_{x_{1}})}, \ldots, \wnest{\ldkt_{x_{n}}(\Lambda_{x_{n}})}, \bnest{\ldkt_{y_{1}}(\Lambda_{y_{1}})}, \ldots, \bnest{\ldkt_{y_{k}}(\Lambda_{y_{k}})}.
$$
\end{definition}

\begin{example} We let $\Lambda = \ wRv,vRu \lsar w : \dia q, w : r \lor q, v : p, v : q, u : \blbox p$ and show the output display sequent for $w$, $u$, and $v$.
\begin{eqnarray*}
 \ldkt_{w}(\Lambda) & =  \dia q, r \lor q, \wnest{p, q,, \wnest{\blbox p}}\\
\ldkt_{v}(\Lambda) & = \bnest{\dia q, r \lor q}, p, q, \wnest{\blbox p}\\
\ldkt_{u}(\Lambda) & = \bnest{\bnest{\dia q, r \lor q}, p, q}, \blbox p
\end{eqnarray*}
We find something interesting if we observe the display sequents $\ldkt_{w}(\Lambda)$, $\ldkt_{v}(\Lambda)$, and $\ldkt_{u}(\Lambda)$ above, namely, each display sequent is derivable from the other by means of the display rules $\rf$ and $\rp$. In fact, as stated in the following lemma, this relationship holds generally; its proof can be found in~\cite{CiaLyoRamTiu21}.
\end{example}

\begin{lemma}\label{lem:dis-eq-kt}
If $\Lambda = \rel \lsar \Gamma$ is a labeled polytree sequent with labels $w$ and $u$, then $\ldkt_{w}(\Lambda)$ and $\ldkt_{u}(\Lambda)$ are display equivalent, i.e., both are mutually derivable with the $\rf$ and $\rp$ rules.
\end{lemma}

Relying on Lemma~\ref{lem:lab-polytree-deriv} and~\ref{lem:dis-eq-kt}, we can define a proof translation from $\lcalckt$ to $\dcalckt$ as specified in the proof of the following theorem. 

\begin{theorem}\label{lem:lab-to-dis-kt}
Every proof of a formula $\forma$ in $\lcalckt$ can be step-wise translated into a proof of $\forma$ in $\dcalckt$.
\end{theorem}

\begin{proof} Suppose we are given a proof of a formula $\forma$ in $\lcalckt$, we know by Lemma~\ref{lem:refined-labeled-int-is-treelike} that the proof is a labeled polytree proof, and thus, $\ldkt$ is defined for every labeled sequent in the proof. We show that the proof can be translated into a proof in $\dcalckt$ by induction on the height of the proof. We only consider the $\boxru$ and $\bldiaru$ cases of the inductive step as the remaining cases are trivial or similar.

\begin{center}
\begin{tabular}{c c c}
\AxiomC{$\rel, wRu \lsar u : A, \Gamma$}
\RightLabel{$\boxru$}
\UnaryInfC{$\rel \lsar w : \Box A, \Gamma$}
\DisplayProof

&

$\leadsto$

&

\AxiomC{$\ldkt_{w}(\rel, wRu \lsar u : A, \Gamma)$}
\RightLabel{=}
\UnaryInfC{$\ldkt_{w}(\rel \lsar \Gamma), \wnest{A}$}
\RightLabel{$\boxru$}
\UnaryInfC{$\ldkt_{w}(\rel \lsar \Gamma), \Box A$}
\RightLabel{=}
\UnaryInfC{$\ldkt_{w}(\rel \lsar w : \Box A, \Gamma)$}
\DisplayProof
\end{tabular}
\end{center}

\begin{center}
\begin{tabular}{c c c}
\AxiomC{$\rel, wRu \lsar u : A, \Gamma$}
\RightLabel{$\bldiaru$}
\UnaryInfC{$\rel \lsar w : \bldia A, \Gamma$}
\DisplayProof

&

$\leadsto$

&

\AxiomC{$\ldkt_{u}(\rel, wRu \lsar w : A, u : \bldia A, \Gamma)$}
\RightLabel{=}
\UnaryInfC{$X, \bldia A, \bnest{Y, A}$}
\RightLabel{$\bldiaru$}
\UnaryInfC{$X, \bldia A, \bnest{Y}$}
\RightLabel{=}
\UnaryInfC{$\ldkt_{u}(\rel, wRu \lsar u : \bldia A, \Gamma)$}
\DisplayProof
\end{tabular}
\end{center}
The remaining cases of the translation can be found in~\cite{CiaLyoRamTiu21}.
\end{proof}


\subsection{Translations for $\logicint$: Labeled, Nested, and Sequent Calculi}\label{subsec:lab-nest-seq-int}

We now consider translating proofs between the sequent calculus $\scalcint$ and a labeled calculus $\lcalcint$ for intuitionistic logic shown in \fig~\ref{fig:labeled-calculus-Int}. The translation from the sequent calculus to the `richer' labeled sequent calculus is relatively straightforward and demonstrates the ease with which proofs may be translated up the proof-theoretic hierarchy (\fig~\ref{fig:hierarchy}). As traditional sequents are simpler than labeled sequents, the converse translation requires special techniques to remove extraneous structure from labeled proofs. To accomplish this task we utilize structural rule elimination (cf.~\cite{Lyo21thesis,Lyo21}) to first transform labeled proofs into nested proofs in $\ncalcint$ (see \fig~\ref{fig:nested-calculus-Int}), and then extract sequent proofs from these.

\subsubsection{From Sequents to Labeled Sequents} The labeled sequent calculus $\lcalcint$ (\fig~\ref{fig:labeled-calculus-Int}) makes use of labeled sequents of the form $\rel, \Gamma \lsar \Delta$, where $\rel$ is a (potentially empty) multiset of relational atoms of the form $w \leq u$ and $\Gamma$ and $\Delta$ are (potentially empty) multisets of labeled formulae of the form $w : A$ with $A \in \langint$. The theorem below gives a translation of proofs in $\scalcint$ into proofs in $\lcalcint$.

\begin{figure}[t]
\begin{center}
\begin{tabular}{c c}
\AxiomC{}
\RightLabel{$\id$}
\UnaryInfC{$\rel, w \leq u, \Gamma, w : p \lsar u :p, \Delta$}
\DisplayProof

&

\AxiomC{}
\RightLabel{$\botl$}
\UnaryInfC{$\rel, \Gamma, w : \bot \lsar \Delta$}
\DisplayProof
\end{tabular}
\end{center}

\begin{center}
\AxiomC{$\rel, w \leq u, \Gamma, u :\forma \lsar u : \formb, \Delta$}
\RightLabel{$\iimpr$}
\UnaryInfC{$\rel, \Gamma \lsar w : \forma \iimp \formb, \Delta$}
\DisplayProof
\end{center}

\begin{center}
\resizebox{\columnwidth}{!}{
\begin{tabular}{c c}
\AxiomC{$\rel, \Gamma, w : \forma \lsar \Delta$}
\AxiomC{$\rel, \Gamma, w : \formb \lsar \Delta$}
\RightLabel{$\disl$}
\BinaryInfC{$\rel, \Gamma, w : \forma \lor \formb \lsar w : \Delta$}
\DisplayProof

&

\AxiomC{$\rel, \Gamma \lsar w : \forma, w : \formb, \Delta$}
\RightLabel{$\disr$}
\UnaryInfC{$\rel, \Gamma \lsar w : \forma \lor \formb, \Delta$}
\DisplayProof
\end{tabular}
}
\end{center}

\begin{center}
\resizebox{\columnwidth}{!}{
\begin{tabular}{c c}
\AxiomC{$\rel, \Gamma, w : \forma, w : \formb \lsar \Delta$}
\RightLabel{$\conl$}
\UnaryInfC{$\rel, \Gamma, w : \forma \land \formb \lsar \Delta$}
\DisplayProof

&

\AxiomC{$\rel, \Gamma \lsar w : \forma, \Delta$}
\AxiomC{$\rel, \Gamma \lsar w : \formb, \Delta$}
\RightLabel{$\conr$}
\BinaryInfC{$\rel, \Gamma \lsar w : \forma \land \formb, \Delta$}
\DisplayProof
\end{tabular}
}
\end{center}

\begin{center}
\resizebox{\columnwidth}{!}{
\begin{tabular}{c}
\AxiomC{$\rel, w \leq u, \Gamma, w : \forma \iimp \formb, u : \formb \lsar \Delta$}
\AxiomC{$\rel, w \leq u, \Gamma, w : \forma \iimp \formb \lsar u : \forma, \Delta$}
\RightLabel{$\iimpl$}
\BinaryInfC{$\rel, w \leq u, \Gamma, w : \forma \iimp \formb \lsar \Delta$}
\DisplayProof
\end{tabular}
}
\end{center}

\begin{center}
\begin{tabular}{c c}
\AxiomC{$\rel, w \leq w, \Gamma \lsar \Delta$}
\RightLabel{$\refl$}
\UnaryInfC{$\rel, \Gamma \lsar \Delta$}
\DisplayProof

&

\AxiomC{$\rel, w \leq u, u \leq v, w \leq v, \Gamma \lsar \Delta$}
\RightLabel{$\trans$}
\UnaryInfC{$\rel, w \leq u, u \leq v, \Gamma \lsar \Delta$}
\DisplayProof
\end{tabular}
\end{center}

\textbf{Side conditions:} $u$ is fresh in $\iimpr$.

\caption{The labeled sequent calculus $\lcalcint$ for intuitionistic logic~\cite{DycNeg12}.}
\label{fig:labeled-calculus-Int}
\end{figure}

\begin{theorem}\label{thm:seq-to-lab-k}
Every proof of a sequent $\Gamma \sar \Delta$ in $\scalcint$ can be step-wise translated into a proof of $w : \Gamma \sar w: \Delta$ in $\lcalcint$.
\end{theorem} 

\begin{proof} By induction on the height of the given proof in $\scalcint$.

\textit{Base case.} The $\id$ rule is translated as shown below; translating the $\botl$ rule is similar.
\begin{center}
\begin{tabular}{c c c}
\AxiomC{}
\RightLabel{$\id$}
\UnaryInfC{$\Gamma, p \sar p, \Delta$}
\DisplayProof

&

$\leadsto$

&

\AxiomC{}
\RightLabel{$\id$}
\UnaryInfC{$w \leq w, w : \Gamma, w : p \lsar w : p, w : \Delta$}
\RightLabel{$\refl$}
\UnaryInfC{$w : \Gamma, w : p \lsar w : p, w : \Delta$}
\DisplayProof
\end{tabular}
\end{center}

\textit{Inductive step.} As the $\disl$, $\disr$, $\conl$, and $\conr$ cases are simple, we only show the more interesting cases of translating the $\iimpl$ and $\iimpr$ rules. 

$\iimpl$. For the $\iimpl$ case, we assume we are given a derivation in $\scalcint$ ending with an application of the $\iimpl$ rule, as shown below:
\begin{center}
\AxiomC{$\Gamma, \forma \iimp \formb, \formb \sar \Delta$}
\AxiomC{$\Gamma, \forma \iimp \formb \sar \forma, \Delta$}
\RightLabel{$\iimpl$}
\BinaryInfC{$\Gamma, \forma \iimp \formb \sar \Delta$}
\DisplayProof
\end{center}
 To translate the proof and inference into the desired proof in $\lcalcint$, we invoke IH, apply the admissible $\wkl$ rule, apply $\iimpl$, and finally, apply $\refl$ as shown below:
\begin{center}
\begin{tabular}{c @{\hskip 1em} c}
$\deriv =$

&

\AxiomC{}
\RightLabel{IH}
\UnaryInfC{$w : \Gamma, w : \forma \iimp \formb, w : \formb \sar w : \Delta$}
\RightLabel{$\wkl$}
\UnaryInfC{$w \leq w, w : \Gamma, w : \forma \iimp \formb, w : \formb \sar w : \Delta$}
\DisplayProof
\end{tabular}
\end{center}
\begin{center}
\AxiomC{$\deriv$}

\AxiomC{}
\RightLabel{IH}
\UnaryInfC{$w : \Gamma, w : \forma \iimp \formb \sar w : \forma, w : \Delta$}
\RightLabel{$\wkl$}
\UnaryInfC{$w \leq w, w : \Gamma, w : \forma \iimp \formb \sar w : \forma, w : \Delta$}

\RightLabel{$\iimpl$}
\BinaryInfC{$w \leq w, w : \Gamma, w : \forma \iimp \formb \sar w : \Delta$}
\RightLabel{$\refl$}
\UnaryInfC{$w : \Gamma, w : \forma \iimp \formb \sar w : \Delta$}
\DisplayProof
\end{center}

$\iimpr$. Translating the $\iimpr$ rule requires more effort. We must make use of the admissible weakening and label substitution rules $\wkl$ and $\lsub$ along with the following admissible lift rule:
\begin{center}
\AxiomC{$\rel, w \leq u, \Gamma, w : \forma, u : \forma \lsar \Delta$}
\RightLabel{$\lift$}
\UnaryInfC{$\rel, w \leq u, \Gamma, w : \forma \lsar \Delta$}
\DisplayProof
\end{center}
The translation is defined as shown below:
\begin{center}
\begin{tabular}{c @{\hskip 1em} c @{\hskip 1em} c}
\AxiomC{$\Gamma, \forma \sar \formb$}
\RightLabel{$\iimpr$}
\UnaryInfC{$\Gamma \sar \forma \iimp \formb, \Delta$}
\DisplayProof

&

$\leadsto$

&

\AxiomC{}
\RightLabel{IH}
\UnaryInfC{$w : \Gamma, w : \forma \sar w : \formb$}
\RightLabel{$\lsub$}
\UnaryInfC{$u : \Gamma, u : \forma \sar u : \formb$}
\RightLabel{$\wkl$}
\UnaryInfC{$w \leq u, w : \Gamma, u : \Gamma, u : \forma \sar u : \formb$}
\RightLabel{$\lift$}
\UnaryInfC{$w \leq u, w : \Gamma, u : \forma \sar u : \formb$}
\RightLabel{$\iimpr$}
\UnaryInfC{$w : \Gamma \sar w : \forma \iimp \formb$}
\DisplayProof
\end{tabular}
\end{center}
\end{proof}

\subsubsection{From Labeled Sequents to Sequents} We now consider the converse translation from $\lcalcint$ to $\scalcint$, which demonstrates the non-triviality of translating from the richer labeled sequent formalism to the sequent formalism. In this section, our main aim is to establish the following theorem:

\begin{theorem}\label{thm:lab-to-seq-int}
Every proof of a formula $\forma$ in $\lcalcint$ can be step-wise translated into a proof of $\forma$ in $\scalcint$.
\end{theorem} 

We prove the above theorem by establishing two lemmata: (1) we translate labeled proofs from $\lcalcint$ into nested proofs in $\ncalcint$, and (2) we translate nested proofs into sequent proofs in $\scalcint$. We then obtain the desired translation from $\lcalcint$ to $\scalcint$ by composing the two aforementioned ones. We first focus on proving the labeled to nested translation, and then argue the nested to sequent translation.

\begin{definition}[Labeled Tree] We define a \emph{labeled tree sequent} to be a labeled sequent $\Lambda := \rel, \Gamma \sar \Delta$ such that $\rel$ forms a tree and all labels in $\Gamma, \Delta$ occur in $\rel$ (unless $\rel$ is empty, in which case every labeled formula in $\Gamma, \Delta$ must share the same label). We define a \emph{labeled tree derivation} to be a proof containing only labeled tree sequents. We say that a labeled tree derivation has the \emph{fixed root property} \iffi every labeled sequent in the derivation has the same root.
\end{definition}

\begin{definition}[Translation $\lnint$]\label{def:lab-nested-trans} Let $\Lambda := \rel, \Gamma \sar \Delta$ be a labeled tree sequent with root $u$. We define $\Lambda' \subseteq \Lambda$ \iffi there exists a labeled tree sequent $\Lambda''$ such that $\Lambda = \Lambda' \seqcomp \Lambda''$. Let us define $\Lambda_{u} := \rel', \Gamma' \sar \Delta'$ to be the unique labeled tree sequent rooted at $u$ such that $\Lambda_{u} \subseteq \Lambda$, $\Gamma' \restriction u = \Gamma \restriction u$, and $\Delta' \restriction u = \Delta \restriction u$. We recursively define $\lnint(\Lambda) := \lnint_{u}(\Lambda)$: 
\[
  \lnint_{v}(\Lambda) :=
  \begin{cases}
  \Gamma \restriction v \sar \Delta \restriction v & \text{if $\rel = \empseq$}; \\
  \Gamma \restriction v \sar \Delta \restriction v, [\lnint_{z_{1}}(\Lambda_{z_{1}})], \ldots, [\lnint_{z_{n}}(\Lambda_{z_{n}})] & \text{otherwise}. 
  \end{cases}
\]
In the second case above, we assume that $v\leq z_{1}, \ldots v\leq z_{n}$ are all of the relational atoms occurring in the input sequent which have the form $v\leq x$.
\end{definition}
 

\begin{example} We let $\Lambda := w \leq v,v \leq u, v : p, u : p \sar w : p \iimp q, v : r, u : q$ and show the output nested sequent under the translation $\lnint$.
$$
\lnint(\Lambda) = \lnint_{w}(\Lambda) = \empseq \sar p \iimp q, [p \sar r, [ p \sar q ]]
$$
\end{example}
 
 As discussed in the section on the nested sequent formalism (\sect~\ref{subsec:nested}), propagation and reachability rules play a crucial role in the formulation of nested sequent calculi. As we aim to transform labeled proofs in $\lcalcint$ into nested sequent proofs in $\ncalcint$, we must define reachability rules in the context of labeled sequents. Toward this end, we define \emph{directed paths} in labeled sequents accordingly.
 
\begin{definition}[Directed Path~\cite{Lyo21}]\label{def:directed-path} Let $\Lambda = \rel, \Gamma \sar \Delta$ be a labeled sequent. We say that there exists a \emph{directed path} from $w$ to $u$ in $\rel$ (written $w \rpath u$) \iffi $w = u$, or there exist labels $v_{i}$ (with $i \in \{1, \ldots, n\}$) such that $w \leq v_{1}, \ldots, v_{n} \leq u \in \R$ (we stipulate that  $w \leq u \in \rel$ when $n = 0$).
\end{definition}

Directed paths are employed in the formulation of the labeled reachability rule $\idpr$ and the labeled propagation rule $\iimplpr$, shown in \fig~\ref{fig:propagation-rules-lab-int} and based on the work of~\cite{Lyo21,Lyo21thesis}. As the lemma below demonstrates, by adding these rules to $\lcalcint$, the structural rules $\refl$ and $\trans$ become eliminable. Since analogs of these structural rules do not exist in $\ncalcint$, showing their eliminability is a crucial step in translating proofs from the labeled setting to the (nested) sequent setting, as discussed later on.

\begin{figure}[t]\label{fig:reachability-rules-int}
\begin{center}
\begin{tabular}{c}
\AxiomC{}
\RightLabel{$\idpr$}
\UnaryInfC{$\rel, \Gamma, w : p \lsar u : p, \Delta$}
\DisplayProof
\end{tabular}
\end{center}
\begin{center}
\begin{tabular}{c}
\AxiomC{$\rel, \Gamma, w : \forma \iimp \formb \lsar u : \forma, \Delta$}
\AxiomC{$\rel, \Gamma, w : \forma \iimp \formb, u : \formb \lsar \Delta$}
\RightLabel{$\iimplpr$}
\BinaryInfC{$\rel, \Gamma, w : \forma \iimp \formb \lsar \Delta$}
\DisplayProof
\end{tabular}
\end{center}

\noindent
\textbf{Side conditions:} Both rules are applicable only if $w \rpath u$.
   
\caption{Reachability rules for $\lcalcint$.}
\label{fig:propagation-rules-lab-int}
\end{figure}

\begin{lemma}\label{lem:refine-labeled-int}
$\refl$ and $\trans$ are eliminable in $\lcalcint + \{\idpr,\iimplpr\}$.
\end{lemma}

\begin{proof} We argue the eliminability of both rules by induction on the height of the given derivation.

\textit{Base case.} We first argue the $\refl$ case. Note that $\refl$ is freely permutable above $\id$, except when the principal relational atom is auxiliary in $\refl$. This case is resolved by making use of the $\idpr$ propagation rule as shown below, where the side condition is satisfied since $w \rpath u$ holds (taking $w$ and $u$ to be equal).
\begin{center}
\begin{tabular}{c @{\hskip 1em} c @{\hskip 1em} c}
\AxiomC{}
\RightLabel{$\id$}
\UnaryInfC{$\rel, w \leq w, \Gamma, w : p \lsar w : p, \Delta$}
\RightLabel{$\refl$}
\UnaryInfC{$\rel, \Gamma, w : p \lsar w : p, \Delta$}
\DisplayProof

&

$\leadsto$

&

\AxiomC{}
\RightLabel{$\idpr$}
\UnaryInfC{$\rel, \Gamma, w : p \lsar w : p, \Delta$}
\DisplayProof
\end{tabular}
\end{center}

Similar to the $\refl$ case, the only non-trivial case of permuting $\trans$ above $\id$ is when the principal relational atom of $\id$ is auxiliary in $\trans$. Observe that the conclusion of the proof shown below left is an instance of $\idpr$ because $w \rpath v$ holds. Thus, the proof can be replaced by the instance of $\idpr$ as shown below right.
\begin{flushleft}
\begin{tabular}{c @{\hskip 1em} c}
\AxiomC{}
\RightLabel{$\id$}
\UnaryInfC{$\rel, w \leq u, u \leq v,  w \leq v, \Gamma, w : p \lsar v : p, \Delta$}
\RightLabel{$\trans$}
\UnaryInfC{$\rel, w \leq u, u \leq v, \Gamma, w : p \lsar v : p, \Delta$}
\DisplayProof

&

$\leadsto$
\end{tabular}
\end{flushleft}
\begin{flushright}
\AxiomC{}
\RightLabel{$\idpr$}
\UnaryInfC{$\rel, w \leq u, u \leq v, \Gamma, w : p \lsar v : p, \Delta$}
\DisplayProof
\end{flushright}

\textit{Inductive step.} With the exception of the $\iimpl$ rule, $\refl$ and $\trans$ freely permute above every rule of $\lcalcint$. Below, we show how to resolve the non-trivial cases where the relational atom principal in $\iimpl$ is auxiliary in $\refl$ or $\trans$. In the $\refl$ case below, observe that $\iimplpr$ can be applied after $\refl$ since $w \rpath w$ holds.
\begin{center}
\resizebox{\columnwidth}{!}{
\AxiomC{$\rel, w \leq w, \Gamma, w : \forma \iimp \formb \lsar w : \forma, \Delta$}
\AxiomC{$\rel, w \leq w, \Gamma, w : \forma \iimp \formb, w : \formb \lsar \Delta$}
\RightLabel{$\iimpl$}
\BinaryInfC{$\rel, w \leq w, \Gamma, w : \forma \iimp \formb \lsar \Delta$}
\RightLabel{$\refl$}
\UnaryInfC{$\rel, \Gamma, w : \forma \iimp \formb \lsar \Delta$}
\DisplayProof
}
\end{center}
 The above inference may be simulated with $\iimplpr$ as shown below:
\begin{center}
\begin{tabular}{c @{\hskip 1em} c}
$\deriv \ =$

&

\AxiomC{$\rel, w \leq w, \Gamma, w : \forma \iimp \formb \lsar w : \forma, \Delta$}
\RightLabel{$\refl$}
\UnaryInfC{$\rel, \Gamma, w : \forma \iimp \formb \lsar w : \forma, \Delta$}
\DisplayProof
\end{tabular}
\end{center}
\begin{center}
\AxiomC{$\deriv$}
\AxiomC{$\rel, w \leq w, \Gamma, w : \forma \iimp \formb, w : \formb \lsar \Delta$}
\RightLabel{$\refl$}
\UnaryInfC{$\rel, \Gamma, w : \forma \iimp \formb \lsar w : \forma, \Delta$}
\RightLabel{$\iimplpr$}
\BinaryInfC{$\rel, \Gamma, w : \forma \iimp \formb \lsar \Delta$}
\DisplayProof
\end{center}
 Let us consider the non-trivial $\trans$ case below: 
$$
\Lambda = \rel, w \leq u, u \leq v, w \leq v, \Gamma, w : \forma \iimp \formb \lsar v : \forma, \Delta
$$
\begin{center}
\AxiomC{$\Lambda$}
\AxiomC{$\rel, w \leq u, u \leq v, w \leq v, \Gamma, w : \forma \iimp \formb, v : \formb \lsar \Delta$}
\RightLabel{$\iimpl$}
\BinaryInfC{$\rel, w \leq u, u \leq v, w \leq v, \Gamma, w : \forma \iimp \formb \lsar \Delta$}
\RightLabel{$\trans$}
\UnaryInfC{$\rel, w \leq u, u \leq v, \Gamma, w : \forma \iimp \formb \lsar \Delta$}
\DisplayProof
\end{center}
 Observe that $\iimplpr$ can be applied after applying $\trans$ since $w \rpath v$ holds.
\begin{center}
\begin{tabular}{c c}
$\deriv \ =$

&

\AxiomC{$\rel, w \leq u, u \leq v, w \leq v, \Gamma, w : \forma \iimp \formb \lsar v : \forma, \Delta$}
\RightLabel{$\trans$}
\UnaryInfC{$\rel, w \leq u, u \leq v, \Gamma, w : \forma \iimp \formb \lsar w : \forma, \Delta$}
\DisplayProof
\end{tabular}
\end{center}
\begin{center}
\AxiomC{$\deriv$}

\AxiomC{$\rel, w \leq u, u \leq v, w \leq v, \Gamma, v : \forma \iimp \formb, w : \formb \lsar \Delta$}
\RightLabel{$\trans$}
\UnaryInfC{$\rel, w \leq u, u \leq v, \Gamma, w : \forma \iimp \formb \lsar w : \forma, \Delta$}

\RightLabel{$\iimplpr$}
\BinaryInfC{$\rel, w \leq u, u \leq v, \Gamma, w : \forma \iimp \formb \lsar \Delta$}
\DisplayProof
\end{center}
 Thus, the structural rules $\refl$ and $\trans$ are eliminable from every proof in $\lcalcint + \{\idpr, \iimplpr\}$.
\end{proof}

A consequence of the above elimination result is that $\lcalcintii := \lcalcint + \{\idpr, \iimplpr\} - \{\refl,\trans\}$\label{def:lcalcintii} serves as a complete calculus for intuitionistic logic. Moreover, the calculus has the interesting property that every proof of a theorem is a labeled tree derivation; the proof of the following lemma is similar to the proof of Lemma~\ref{lem:lab-polytree-deriv}.

\begin{lemma}\label{lem:refined-labeled-int-is-treelike}
Every derivation of a formula $\forma$ in $\lcalcintii$ is a labeled tree derivation.
\end{lemma}


 By means of the above lemmata, we may translate every proof of a theorem $\forma$ in $\lcalcint$ into a proof of the theorem in $\ncalcint$. The translation is explained in the lemma given below and completes part (1) of the translation from $\lcalcint$ to $\scalcint$.

\begin{lemma}\label{lem:lab-to-nest-int}
Every proof of a formula $\forma$ in $\lcalcint$ can be step-wise translated into a proof of $\forma$ in $\ncalcint$.
\end{lemma}

\begin{proof} By Lemma~\ref{lem:refine-labeled-int} above, we know that every proof of a formula $\forma$ in $\lcalcint$ can be transformed into a proof that is free of $\refl$ and $\trans$ inferences. By Lemma~\ref{lem:refined-labeled-int-is-treelike}, we know that this proof is a labeled tree derivation, and therefore, every labeled sequent in the proof can be translated via $\lnint$ into a nested sequent. One can show by induction on the height of the given proof that every rule directly translates to the corresponding rule in $\ncalcint$, though with $\idpr$ translating to $\id$ and $\iimplpr$ translating to $\iimpl$.
\end{proof}

There are two methods by which nested sequent proofs in $\ncalcint$ can be transformed into sequent proofs in $\scalcint$. The first method, discussed in~\cite{PimRamLel19}, shows how proofs within nested calculi of a suitable shape can be directly transformed into sequent calculus proofs. Alternatively, the second method~\cite{Lyo24} explains a \emph{linearization technique}, which first transforms nested sequent proofs into linear nested sequent proofs, which are then transformable into sequent calculus proofs. Both methods rely on restructuring nested sequent proofs by means of rule permutations and shedding the extraneous treelike structure inherent in nested sequents to obtain a proof in a sequent calculus. As the details of these procedures are tedious and involved, we omit them from the presentation and refer the interested reader to the papers~\cite{PimRamLel19} and~\cite{Lyo24}, noting that these methods imply the following lemma.

\begin{lemma}\label{lem:nest-to-seq-int}
Every proof of a formula $\forma$ in $\ncalcint$ can be step-wise translated into a proof of $\forma$ in $\scalcint$.
\end{lemma}

\newcommand {\fe} {\Vdash^{\exists}}
\newcommand{\Rcless}{\cless_r}
\newcommand{\Lcless}{\cless_l}
\newcommand{\Lc}{\subseteq_l}
\newcommand{\Nes}{\mathsf{nes}}
\newcommand{\calci}[1]{\mathcal{I}_{#1}^\mathsf{i}}
\newcommand{\calcl}[1]{LG({#1})}
\newcommand{\cblock}[2]{\left[ #1 \vartriangleleft #2 \right]}
\newcommand{\comi}{\mathsf{com}^\mathsf{i}}
\newcommand{\jumpi}{\mathsf{jump}}
\newcommand{\rulecpri}{\cless^\mathsf{i}_r}
\newcommand{\rulecpli}{\cless^\mathsf{i}_l}
\newcommand{\Mon}{\mathsf{mon}\exists}

\subsection{Translating Proofs for Conditional Logic}\label{subsec:cond-logic-trans}

\begin{figure}[t]

\begin{center}
	\begin{tabular}{c c}
        \AxiomC{$x \in a, x:A, \Gamma \Rightarrow \Delta$}
        \RightLabel{($\Vdash^{\exists}_l$)}
        \UnaryInfC{$a \Vdash^{\exists} A, \Gamma \Rightarrow \Delta$}
	\DisplayProof
	&
	           \AxiomC{$x \in a, \Gamma \Rightarrow \Delta, x:A, a \Vdash^{\exists}A$}
			\RightLabel{($\Vdash ^{\exists}_r$)}
			\UnaryInfC{$ x \in a, \Gamma \Rightarrow \Delta, a \Vdash^{\exists}A$}
			\DisplayProof
		\end{tabular}
\end{center}
		
		
\begin{center}
		\begin{tabular}{c}
		\AxiomC{$a \in S(x),a \fe B,  \Gamma \Rightarrow \Delta, a \fe A$}	
			\RightLabel{($\cless_r$)}
            \UnaryInfC{$\Gamma \Rightarrow \Delta, x: A\cless B $}			
			\DisplayProof
		\end{tabular}
\end{center}
\begin{center}
\resizebox{\columnwidth}{!}{
		\begin{tabular}{c}	
			\AxiomC{$a \in S(x), x: A\cless B, \Gamma \Rightarrow \Delta, a\fe B$}
			\AxiomC{$a \fe A, a \in S(x), x:A\cless B, \Gamma \Rightarrow \Delta$}
			\RightLabel{($\cless_l$)}
			\BinaryInfC{$a \in S(x), x: A \cless B, \Gamma \Rightarrow \Delta$}
			\DisplayProof
		\end{tabular}
}
\end{center}

\begin{center}
		\begin{tabular}{c}	
			\AxiomC{$ x \in a, a \subseteq b, x \in b, \Gamma \Rightarrow \Delta $}
			\RightLabel{($\subseteq_l$)}
			\UnaryInfC{$x\in a, a \subseteq b,  \Gamma \Rightarrow \Delta $}
			\DisplayProof
		\end{tabular}
\end{center}

\begin{center}
\resizebox{\columnwidth}{!}{
		\begin{tabular}{c}	
			
			\AxiomC{$a \subseteq b, a \in S(x), b \in S(x), \Gamma \Rightarrow \Delta$}
			\AxiomC{$b\subseteq a, a \in S(x), b \in S(x), \Gamma \Rightarrow \Delta$}
			\RightLabel{($\Nes$)}
			\BinaryInfC{$a \in S(x), b \in S(x), \Gamma \Rightarrow \Delta $}
			\DisplayProof
		\end{tabular}
}
\end{center}
		
\noindent
\textbf{Side conditions:} Label $x$ must be fresh in ($\fe_l$), and label $a$ must be fresh in ($\cless_r$). 
	
\caption{Some labeled calculus rules for the conditional logic $\VLe$.\label{fig:labeled Calculus V}}
\end{figure}

In this section, we discuss translations between two sequent-style calculi for the conditional logic $\VLe$ (introduced in Section \ref{condlog}). The first calculus is a labeled sequent calculus, dubbed $\lcalcv$, and consists of the rules shown in Figure \ref{fig:labeled Calculus V} along with the initial sequents and propositional rules for $\lor$, $\land$, and $\rightarrow$ from the labeled sequent calculus $\mathsf{G3K}$ for the modal logic $\logick$ (see~\cite{Neg05} for these latter rules). We remark that this labeled sequent calculus was introduced in \cite{GNO-aiml18}. 

The labeled calculus $\lcalcv$ uses two sorts of labels: $\mathsf{WLab} := \set{x, y, z, \dots}$ for worlds and $\mathsf{SLab} := \set{a,b,c,\dots}$ for spheres. We define a \emph{labeled formula} to be an expression of the form $x : A$ or $a\fe A$ with $x \in \mathsf{WLab}$ and $a \in \mathsf{SLab}$. Given a sphere model, labeled formulae of the form $x:A$ and $a\fe A$ are interpreted as $x\models A$ and $a\models^{\exists} A$ in the model, respectively. We define a \emph{relational atom} to be an expression of the form $x\in a$, $a\in S(x)$, or $a\subseteq b$ with $x \in \mathsf{WLab}$ and $a, b \in \mathsf{SLab}$. A \emph{labeled sequent} is an expression of the form $\Gamma \sar \Delta$ such that $\Gamma$ and $\Delta$ are finite multisets of relational atoms and labeled formulae.


The second sequent-style calculus we consider in this section is the \emph{structured calculus} $\calci{\VLe}$~\cite{GLO-Jelia16}, which is a kind of nested sequent calculus
\footnote{The $\mathsf{i} $ in $\calci{\VLe}$ stands for `invertible', as in~\cite{GLO-Jelia16}  a version of the same proof system with less invertible rules is also introduced. In $\calci{\VLe}$, the only non-invertible rule is $\jumpi$ (see \Cref{CalcInt}). }. 
Sequents in this calculus make use of a special structure called a \textit{block}, which is an expression of the form  $\cblock{\Sigma}{B}$ such that $\Sigma,B$ is a multiset of conditional formulae. In this setting, a \emph{sequent} is an expression $\Gamma \sar \Delta$, where $\Gamma$ is a multiset of conditional formulae, and $\Delta$ is a multiset of conditional formulae and blocks. 
 The  \emph{formula interpretation} $\iota(\Gamma \Rightarrow \Delta, \cblock{\Sigma_1}{B_1}, \ldots, \cblock{\Sigma_n}{B_n})$ of a sequent is taken to be equal to the following formula:
$$
\bigwedge \Gamma \to \bigvee \Delta' \lor \bigvee_{1\leq i\leq n}\bigvee_{A \in \Sigma_i}(A \cless B_i)
$$
Some interesting rules from the structured sequent calculus $ \calci{\VLe}$ are presented in Figure~\ref{CalcInt}; see~\cite{GLO-Jelia16} for the full list of rules. 

\begin{figure}[t]
	
	\begin{center}
		
		\begin{tabular}{c}		
			\AxiomC{$\Gamma \Rightarrow \Delta, \cblock{A}{B}$}
			\RightLabel{($\rulecpri$)}
			\UnaryInfC{$\Gamma \Rightarrow \Delta, A \cless B$}
			\DisplayProof		
		\end{tabular}

		\vspace{0.2cm}
        
\resizebox{\columnwidth}{!}{
\begin{tabular}{c}	
			\AxiomC{$\Gamma,  A \cless B \Rightarrow \Delta, \cblock{B,\Sigma}{C}$}
			\AxiomC{$\Gamma, A \cless B \Rightarrow \Delta, \cblock{\Sigma}{A},  \cblock{\Sigma}{C}$} 
			\RightLabel{($\rulecpli$)}		
			\BinaryInfC	{$\Gamma, A \cless B \Rightarrow \Delta,	\cblock{\Sigma}{C}$}
			\DisplayProof	
\end{tabular}
}
		
		\vspace{0.2cm}

\resizebox{\columnwidth}{!}{
\begin{tabular}{c}			
			\AxiomC{$\Gamma \Rightarrow \Delta, \cblock{\Sigma_1,\Sigma_2}{A}, \cblock{\Sigma_2}{B}$} 
			\AxiomC{$\Gamma \Rightarrow \Delta, \cblock{\Sigma_1}{A} , \cblock{\Sigma_1,\Sigma_2}{B}$}
			\RightLabel{$(\comi)$}		
			\BinaryInfC{$\Gamma \Rightarrow \Delta,\cblock{\Sigma_1}{A},\cblock{\Sigma_2}{B}$}					
			\DisplayProof							
\end{tabular}
}
		
		\vspace{0.2cm}
		
\begin{tabular}{c}		
	\AxiomC{$A \Rightarrow \Sigma$}
	\RightLabel{($\jumpi$)}
	\UnaryInfC{$\Gamma \Rightarrow \Delta, \cblock{\Sigma}{A}$}
	\DisplayProof	
\end{tabular}		
	\end{center}
\caption{Some structured calculus rules for the conditional logic $\VLe$.\label{CalcInt}}
\end{figure}

We remark that the translation between the labeled and structured calculi is not straightforward. The non-triviality of translating proofs between the two systems not only arises from the fact that both systems use a different language, but also from the fact that there is no direct correspondence between the relevant rules of the two calculi. In the following, we first discuss the translation from the structured calculus to the labeled one, and afterward, we discuss the reverse translation. See~\cite{GNO-aiml18} for a formal and complete description of both translations.

\subsubsection{From Structured Sequents to Labeled Sequents} We  illustrate the translation from the structured calculus to the labeled calculus.  
We adopt the following notational convention: given multisets of formulae $ \Gamma = \{ A_1,\dots, A_m  \} $ and $ \Sigma = \{D_1,\dots , D_k  \} $, we shall write $ \Gamma^x $ and $ a \fe \Sigma $ as abbreviations for $ x: A_1,\dots, x:A_m   $ and $ a\fe D_1,\dots ,a\fe  D_k $ respectively. 
To illustrate the translation, consider a sequent of the following shape, with  $\Gamma, \Delta$ multisets of formulae: 
$$S= \Gamma \Rightarrow \Delta, \cblock{\Sigma_1}{B_1}, \dots, \cblock{\Sigma_n}{B_n}$$
Then, fix as parameters a world label $x$ and a set of sphere  labels $\bar{a} = a_1,\ldots, a_n$. The translation $t(S)^{x,\bar{a}}$ of $S$ is the following labeled sequent: 
$$
\begin{array}{c c l}
t(S)^{x,\bar{a}} & := &a_{1} \in S(x), .. , a_{n} \in S(x),\  a_1 \fe B_1,.., a_n \fe B_n, \\
& &\Gamma^x \Rightarrow  \Delta^{x},  a_1 \fe  \Sigma_1,.., a_n\fe   \Sigma_n  
\end{array}
$$
 

The idea is that for each block $\cblock{\Sigma_i}{B_i}$ we introduce a new sphere label $ a_i $ such that $ a_i \in S(x) $, and formulae $ a_i \fe B_i$ in the antecedent and 
$ a_i\fe \Sigma_i $ in the consequent. These formulae correspond to the semantic condition for a block i.e., a disjunction of $ \cless $ formulae in sphere models. 

We can then define a formal translation of  any derivation $\mathcal{D}$ of a sequent $S$ in the structured calculus $ \calci{V}$  to a derivation  $\{\mathcal{D}\}^{x,\bar{a}}$ in the labeled calculus $\lcalcv$ of the translated sequent $t(S)^{x,\bar{a}}$. Some cases of the translation are reported in \fig~\ref{figtrans1}.

\begin{figure}[t!]
{\footnotesize

\noindent  
$
\left\{
\begin{matrix}
 \infer[(\rulecpri)]{\Gamma \Rightarrow \Delta, A\cless B   }{ \deduce{\Gamma \Rightarrow \Delta, \cblock{A}{B} }{\mathcal{D}_1} } 
 \end{matrix} \right\} ^{x,\bar{a}} \leadsto \quad   \begin{matrix}
 \infer[(\Rcless)] {t(\Gamma \Rightarrow \Delta, A\cless B )^{x,\bar{a}} }{ 
 \deduce{ t(\Gamma \Rightarrow \Delta, \cblock{A}{B})^{x,\bar{a}\, b} }{\{ \mathcal{D}_1\}^{x,\bar{a} \, b}  }
   } \end{matrix}$

\vspace{0.5cm}



\noindent 
$\left\{
  \begin{matrix}
   \infer[(\comi)]{\Gamma \Rightarrow \Delta, \cblock{\Sigma_1}{A}, \cblock{\Sigma_1}{B} }{ \deduce{\Gamma \Rightarrow \Delta, \cblock{\Sigma_1, \Sigma_2}{A}, \cblock{\Sigma_2}{B}}{\mathcal{D}_1} & \deduce{\Gamma \Rightarrow \Delta, \cblock{\Sigma}{A}, \cblock{\Sigma,\Pi}{B}}{\mathcal{D}_2} } 
   \end{matrix} \right\} ^{x,\bar{a} \, b\, c} $ $ \leadsto $
   
   

\vspace{0.3cm}

\noindent 
\begin{center}   
   $
     \begin{matrix}
    \infer[(\Nes)] {t(\Gamma \Rightarrow \Delta, \cblock{\Sigma_1}{A}, \cblock{\Sigma_1}{B})^{x,\bar{a}\, b\, c} }{ 
    \infer[(\Mon)]{ b \subseteq c, b \fe A, c\fe B,t(\Gamma)^{x,\bar{a}}  \Rightarrow t(\Delta)^{x,\bar{a}},  b \fe  \Sigma_1, c\fe  \Sigma_2 }{\infer[\wk]{b \subseteq c,  b \fe A, c\fe B, t(\Gamma)^{x,\bar{a} }  \Rightarrow t(\Delta)^{x,\bar{a}},  b \fe  \Sigma_1, b \fe  \Sigma_2, c\fe \Sigma_2}{
    \deduce{t(\Gamma \Rightarrow \Delta, \cblock{\Sigma_1, \Sigma_2}{A}, \cblock{\Sigma_2}{B})^{x, \bar{a} \, b \, c} }{ \{ \mathcal{D}_1 \}^{x, \bar{a} \, b \, c} } 
        } }
     & 
      \mathcal{E}
        }   \end{matrix} 
     $
     \end{center}
     
     \begin{minipage}{\linewidth}
In the above, $\mathcal{E}$ is a derivation of  sequent
$ c\subseteq b, b \fe A, c\fe B, t(\Gamma)^{x,\bar{a}}  \Rightarrow t(\Delta)^{x,\bar{a}}, b \fe \Sigma_1, c\fe  \Sigma_2   $ from the translation of the rightmost premiss of $(\comi)$. $\mathcal{E}$ is constructed similarly to the displayed derivation of the premiss of $(\Nes)$. 
\end{minipage}

\vspace{0.3cm}

\noindent 
$  \left\{
       \begin{matrix}
        \infer[(\jumpi)]{\Gamma \Rightarrow \Delta, \cblock{\Sigma}{A}}{ \deduce{x:\Sigma \Rightarrow x:A }{\mathcal{D}_1}  } 
        \end{matrix} \right\} ^{x,\bar{a}\, b} 
        $  $\leadsto $ \\
\vspace{-0.7cm}
\begin{flushright}
        $  \begin{matrix} 
           \infer[(\fe_l)] {t(\Gamma \Rightarrow \Delta, \cblock{\Sigma}{A})^{x,\bar{a}\, b } }{
           \infer[{(\fe_r)}\!\times \!n]{y \in b, b\in S(x), y :A, t(\Gamma)^{x,\bar{a}}  \Rightarrow t(\Delta)^{x,\bar{a}}, b \fe \Sigma }{
           \infer[\wk]{ y \in b, b\in S(x), y :A, t(\Gamma)^{x,\bar{a}}  \Rightarrow t(\Delta)^{x,\bar{a}}, y:\Sigma, b\fe \Sigma}{
          \deduce{t(x:\Sigma \Rightarrow x:A)^{x\,  [x/y]}  }{t\{\mathcal{D}_1\}^{x \, [x/y]}  }      
           }
           }
           }        
           \end{matrix}  $ 
\end{flushright}

}
    \caption{Some cases of the translation from  $ \calci{\VLe}$ to $\lcalcv$.\label{figtrans1}}
\end{figure}


The most interesting case is the translation of the rule ($\comi$). Since this rule encodes sphere nesting, it is worth noticing that its translation requires the ($\Nes$)-rule applied to 
$t(\Gamma \Rightarrow \Delta,  \cblock{\Sigma_1}{A},  \cblock{\Sigma_2}{B})^{x,\bar{a}, b, c}$,
which is derived by ($\Nes$) from the two sequents: 
\begin{quote}
${ b \subseteq c}, b \fe A, c \fe B, t(\Gamma)^{x,\bar{a}} \Rightarrow  t(\Delta)^{x,\bar{a}}, b\fe \Sigma_1, c\fe \Sigma_2$\\
${ c \subseteq b}, b \fe A, c \fe B, t(\Gamma)^{x,\bar{a}} \Rightarrow  t(\Delta)^{x,\bar{a}}, b\fe \Sigma_1, c\fe \Sigma_2$
\end{quote}
Thus, the ($\comi$) rule can be `mimicked' using the ($\Nes$) rule. 
Moreover, the translation uses the following rule ($\Mon$), admissible in $\lcalcv$: 
$$
\infer[{(\Mon)}]{b \subseteq a, \Gamma \Rightarrow \Delta, a \fe A}{ b \subseteq a, \Gamma \Rightarrow \Delta, a \fe A, b \fe A } 
$$
This rule propagates (in a backward semantic reading) a false $\fe$-statement from a larger to a smaller neighborhood. The translation is correct:

\begin{theorem}
Let $\mathcal{D}$ be a derivation  of a sequent $S$ in  $ \calci{\VLe}$, then   $\{\mathcal{D}\}^{x,\bar{a}}$ is a derivation  of $t(S)^{x,\bar{a}}$ in $\lcalcv$.
\end{theorem}

\begin{example}
As an example let us consider a derivation of $(A\cless B) \lor (B \cless A)$, one of the axioms of $\VLe$, in  $ \calci{\VLe}$: 
\begin{center}
		\AxiomC{$B\Rightarrow A,B$}
		\RightLabel{($\jumpi$)}
			\UnaryInfC{$\Rightarrow \cblock{A,B}{B},\cblock{B}{A}$}
		\AxiomC{$A\Rightarrow A,B$}
			\RightLabel{($\jumpi$)}
			\UnaryInfC{$\Rightarrow \cblock{A}{B},\cblock{A,B}{A}$}
		\RightLabel{$(\comi)$}
		\BinaryInfC{$\Rightarrow \cblock{A}{B},\cblock{B}{A} $}
		\RightLabel{($\rulecpri$)}
		\UnaryInfC{$\Rightarrow \cblock{A}{B}, B\cless A$}
		\RightLabel{($\rulecpri$)}
		\UnaryInfC{$\Rightarrow A\cless B, B\cless A$}
		\RightLabel{($\lor_r$)}
		\UnaryInfC{$\Rightarrow (A\cless B) \lor (B\cless A)$}
		\DisplayProof
\end{center}
The derivation above can be translated into a derivation in $\lcalcv$ as shown below. We only show the derivation of the left premise of ($\Nes$)  as the other is symmetric.
\begin{center}
		\resizebox{\columnwidth}{!}{
			\AxiomC{$ y:B \Rightarrow y:A, y:B$}
			\RightLabel{($Wk$)}
			\UnaryInfC{$a \in S(x), b \in S(x), y \in a,y \in b, y:B, b \fe A\Rightarrow a \fe A, a \fe B, b \fe B , y:A, y:B$}
			\RightLabel{($\Vdash^{\exists}_r    \times 2$)}
			\UnaryInfC{$a \in S(x), b \in S(x), y \in a,y \in b, y:B, b \fe A\Rightarrow a \fe A, a \fe B, b \fe B 
				$}
			\RightLabel{($\Vdash^{\exists}_l$)}
			\UnaryInfC{$a \in S(x), b \in S(x), a \fe B, b \fe A\Rightarrow a \fe A, a \fe B, b \fe B $}
			\RightLabel{($Wk$)}
			\UnaryInfC{$a \subseteq b, a \in S(x), b \in S(x), a\fe B, b \fe A\Rightarrow a \fe A, a\fe B, b \fe B $}
			\RightLabel{($\Mon$)}
			\UnaryInfC{$a \subseteq b, a \in S(x), b \in S(x), a\fe B, b \fe A\Rightarrow a \fe A, b \fe B $}
			\RightLabel{($\Nes$)}
			\UnaryInfC{$a \in S(x), b \in S(x), a \fe B, b \fe A \Rightarrow a \fe A, b \fe B $}
			\RightLabel{($\Rcless$)}
			\UnaryInfC{$a \in S(x), a \fe B \Rightarrow x:B\cless A, a \fe A$}
			\RightLabel{($\Rcless$)}
			\UnaryInfC{$\Rightarrow x: A\cless B, x: B\cless A$}
			\RightLabel{($\lor_r$)}
			\UnaryInfC{$\Rightarrow x: A\cless B \lor B\cless A$}
			\DisplayProof
		}
\end{center}
\end{example}

\subsubsection{From Labeled Sequents to Structured Sequents.} 
The translation from the labeled calculus $\lcalcv$ to the structured calculus  $ \calci{\VLe}$ is more difficult, as not  every sequent of the labeled calculus can be translated into a sequent of the structured calculus. Consequently, a derivation  in  $\lcalcv$ might contain steps that cannot be simulated in the  calculus $ \calci{\VLe}$. 
In this section we only describe the general strategy behind the translation; for a formal treatment we refer the reader to \cite{GNO-aiml18}.

More specifically, the translation only applies to labeled sequents of the form $t(\Gamma \Rightarrow \Delta)^x$ which are the image of the translation of a sequent $\Gamma \Rightarrow \Delta$ of the  structured calculus  $ \calci{\VLe}$. Then, since a proof of $t(\Gamma \Rightarrow \Delta)^x$  in $\lcalcv$ may involve sequents that are not translatable, the first step is to rearrange the proof in a specific \emph{normal form}, in which rules are applied in a certain order. Then, one shows that derivations in normal form can be `partitioned' into subderivations $\mathcal{S}$ such that, for each $\mathcal{S}$, the premisses of  $\mathcal{S}$ are translatable into premisses of a rule $\mathsf{r}$ of $ \calci{\VLe}$, and the conclusion of $\mathcal{S}$ can be translated into the conclusion of $\mathsf{r}$. Thus, the rules of the structured calculus $ \calci{\VLe}$ act as `macros' over the rules of the labeled calculus, `skipping' the untranslatable sequents.



	 
	
	
	
	

We illustrate the translation of labeled sequents into the sequents with blocks of $ \calci{\VLe}$ with an example.  Let $S$ be the following sequent, where $\Gamma,\Delta$ only contain formulae of the language:
\begin{eqnarray*}
	& a_1 \subseteq a_2 ,a_2 \subseteq a_3, a_1 \subseteq a_3,  a_1\in S(x), a_2 \in S(x), a_3 \in S(x), \\ 
	& a_1 \fe A_1, a_2 \fe A_2,  a_3 \fe A_3, x:\Gamma \Rightarrow    x:\Delta, a_1 \fe \Sigma_1,a_2 \fe \Sigma_2, a_3 \fe \Sigma_3
\end{eqnarray*}
The translation of $S$ is the following sequent with blocks: 
$$\Gamma \Rightarrow \Delta, \cblock{\Sigma_1, \Sigma_2, \Sigma_3 }{A_1}, \cblock{\Sigma_2, \Sigma_3}{A_2}, \cblock{\Sigma_3}{A_3}$$

\noindent Intuitively, the translation re-assembles the blocks from formulae labeled with the same sphere label. Furthermore, for each inclusion $ a_i \subseteq a_j$ we add to the corresponding block also formulae $ \Sigma_j $ such that $ a_j \fe \Sigma_j $ occurs in the consequent of the labeled sequent. 
Thus, each block in the internal calculus consists of $ \cless $-formulae relative to some sphere i.e., labeled with the same sphere label in $\lcalcv$.

Moreover, a labeled sequent is translatable only if it has a  {\em tree-like structure}. This tree-like structure is generated by the two spheres/worlds relations  $x\rightarrow a$ iff $a\in S(x)$, the relation $a\rightarrow y$ iff $y\in a$,  and their composition: $x \rightarrow y$ iff $x \rightarrow a\rightarrow y$ for some $a$. 
Intuitively, a labeled sequent $\Gamma \Rightarrow \Delta$ is tree-like if, for every label $x$ occurring in $\Gamma$, the set of labels $y$ reachable from $x$ by the transitive closure of the relation $x\rightarrow y$ forms a tree\footnote{To be precise, only a subset of sequents with a tree-like structure  can be translated in sequents of the language of $ \calci{\VLe}$, but we avoid giving full details here. }.


Concerning the translation of a normal form derivation $\mathcal{D}$, the idea is that one first translates, starting from the root, the rules whose sequents have a translation in $ \calci{\VLe}$, until labeled sequents that cannot be translated are reached. 
Next, we need to deal with untranslatable (but derivable!) sequents. 
For this, we show that 
a derivable untranslatable sequent $\Gamma \Rightarrow \Delta$  can be {\em replaced} by a derivable translatable sequent $\Gamma_1 \Rightarrow \Delta_1$ obtained by a  decomposition of  $\Gamma \Rightarrow \Delta$ determined by the tree-like structure associated to every label $x$ occurring in $\Gamma$: either $\Gamma_1 \Rightarrow \Delta_1$ is the subsequent containing only the labels in the tree rooted in $x$ and the formulae/relation involving these labels, or it is the subsequent obtained by {\em removing}  from $\Gamma \Rightarrow \Delta$ the labels and formulae of the tree of $x$, or it is obtained from the latter by iterating the process (on another label). If $\Gamma \Rightarrow \Delta$ is derivable, there exists a translatable subsequent $\Gamma_1 \Rightarrow \Delta_1$  of  $\Gamma \Rightarrow \Delta$ which is derivable too (with the same height).

Thus, in order to define  the translation of the whole derivation $\mathcal{D}$, when   an untranslatable sequent $\Gamma \Rightarrow \Delta$ is reached,  we consider then the translation of  the sub-derivation  $\mathcal{D}'$ of a subsequent $\Gamma_1 \Rightarrow \Delta_1$ obtained by decomposition of $\Gamma \Rightarrow \Delta$. 
Since the sequent $\Gamma_1 \Rightarrow \Delta_1$ is not determined in advance and it is not necessarily unique, the translation of $\mathcal{D}$ is not entirely {\em deterministic}.


\subsection{A More Difficult Case: Translating Bunched Logics}
\label{sec:translating-bunches}

Using the Kripke resource semantics of {\BI} it is not difficult to build a labeled sequent or labeled tableau proof system.
As usual, the first step is to devise a labeling algebra that reflects the properties of the semantics. 
The units $\neuS$, $\topS$ and $\absS$ are reflected into the labels units $\mneuL$, $\aneuL$ and $\abotL$.
The semantic properties of the binary operators $\mulS$, $\addS$ and the preodering relation $\leqS$ are reflected into the binary functors $\mulL$, $\addL$ and the binary relation $\leqL$.

\begin{definition}
	A countable set $\LabsL$ of symbols is a set of \emph{label letters} if it is disjoint from the set $\LabsU = \ens{ \mneuL, \aneuL, \abotL }$ of \emph{label units}.
	$\Labs^0_\LabsL = \LabsL \cup \LabsU$ is the set of \emph{atomic labels over $\LabsL$}.
	The set $\Labs_\LabsL$ of \emph{labels over $\LabsL$} is defined as $\bigcup_{n \in \NAT} \Labs^n_\LabsL$ where 
	\[
	\Labs^{n+1}_\LabsL := \Labs^n_\LabsL \cup 
	\ensc{ \RONL{\labl}{\labl'} }{\labl,\labl' \in \Labs^n_\LabsL \text{ and } 
		\ronL \in \ens{ \mulL, \addL }}.
	\]
	A \emph{label constraint} is an expression $\LEQL{\labl}{\labl'}$, where $\labl$ and $\labl'$ are labels. 
	A \emph{labeled formula} is an expression $\lf{\VPA}{\labl}$, where $\VPA$ is a formula and $\labl$ is a label. 
\end{definition}

The second step is to define labeled sequents (as in {\GBI}) of the form $\SEQ{\SG}{\SD}$, where $\SG$ is a multiset mixing both labeled formulae and label constraints and $\SD$ is a multiset of labeled formulae.

\begin{figure}[t]
	\begin{center}
		\begin{ebp}
			\hypo{}
			\infer1[\PRNR{\BIabot}]{\SEQ{\SG,\LEQL{\abotL}{\labl}}{\lf{\labl}{\VPA},\SD}}
		\end{ebp}
		\qquad
		\begin{ebp}
			\hypo{}
			\infer1[\PRN{id}]{\SEQ{\SG,\lf{\labl}{\VPA}}{\lf{\labl}{\VPA},\SD}}
		\end{ebp}
		\qquad
		\begin{ebp}
			\hypo{\phantom{\Gamma}}
			\infer1[\PRNR{\BImtop}]{\SEQ{\SG,\LEQL{\mneuL}{\labl}}{\lf{\labl}{\BImtop},\SD}}
		\end{ebp}
		
		\myskip
		
		\begin{ebp}
			\hypo{\SEQ{\SG,\LEQL{\abotL}{\labl}}{\SD}}
			\infer1[\PRNL{\BIabot}]{\SEQ{\SG,\lf{\labl}{\BIabot}}{\SD}}
		\end{ebp}
		\;\;
		\begin{ebp}
			\hypo{\SEQ{\SG,\LEQL{\mneuL}{\labl}}{\SD}}
			\infer1[\PRNL{\BImtop}]{\SEQ{\SG,\lf{\labl}{\BImtop}}{\SD}}
		\end{ebp}

        \myskip
        
		\begin{ebp}
			\hypo{\SEQ{\SG,\LEQL{\aneuL}{\labl}}{\SD}}
			\infer1[\PRNL{\BIatop}]{\SEQ{\SG,\lf{\labl}{\BIatop}}{\SD}}
		\end{ebp}
		\;\;
		\begin{ebp}
			\hypo{\phantom{\Gamma}}
			\infer1[\PRNR{\BIatop}]{\SEQ{\SG,\LEQL{\aneuL}{\labl}}{\lf{\labl}{\BIatop},\SD}}
		\end{ebp}
		
		\myskip
		
		\begin{ebp}
			\hypo{\SEQ{\LEQL{\ADDL{\labl}{\labl_1}}{\labl_2},\SG,\lf{\labl}{\VPA\BIaimp\VPB}}{\lf{\labl_1}{\VPA},\SD}}
			\hypo{\SEQ{\LEQL{\ADDL{\labl}{\labl_1}}{\labl_2},\SG,\lf{\labl}{\VPA\BIaimp\VPB},\lf{\labl_2}{\VPB}}{\SD}}
			\infer2[\PRNL{\BIaimp}]{%
				\SEQ{\LEQL{\ADDL{\labl}{\labl_1}}{\labl_2},\SG,\lf{\labl}{\VPA\BIaimp\VPB}}{\SD}%
			}
		\end{ebp}
		
		\myskip
		
		\begin{ebp}
			\hypo{\SEQ{\LEQL{\MULL{\labl}{\labl_1}}{\labl_2},\SG,\lf{\labl}{\VPA\BImimp\VPB}}{\lf{\labl_1}{\VPA},\SD}}
			\hypo{\SEQ{\LEQL{\MULL{\labl}{\labl_1}}{\labl_2},\SG,\lf{\labl}{\VPA\BImimp\VPB},\lf{\labl_2}{\VPB}}{\SD}}
			\infer2[\PRNL{\BImimp}]{%
				\SEQ{\LEQL{\MULL{\labl}{\labl_1}}{\labl_2},\SG,\lf{\labl}{\VPA\BImimp\VPB}}{\SD}%
			}
		\end{ebp}
		
		\myskip
		
		\begin{ebp}
			\hypo{%
				\SEQ{\LEQL{\ADDL{\labl}{\labl_1}}{\labl_2},\SG,\lf{\labl_1}{\VPA}}{\lf{\labl_2}{\VPB},\SD}%
			}
			\infer1[\PRNR{\BIaimp}]{\SEQ{\SG}{\lf{\labl}{\VPA\BIaimp\VPB},\SD}}
		\end{ebp}
		\qquad
		\begin{ebp}
			\hypo{%
				\SEQ{\LEQL{\MULL{\labl}{\labl_1}}{\labl_2},\SG,\lf{\labl_1}{\VPA}}{\lf{\labl_2}{\VPB},\SD}%
			}
			\infer1[\PRNR{\BImimp}]{\SEQ{\SG}{\lf{\labl}{\VPA\BImimp\VPB},\SD}}
		\end{ebp}
		
		\myskip
		
		\begin{ebp}
			\hypo{%
				\SEQ{\LEQL{\ADDL{\labl_1}{\labl_2}}{\labl},\SG,\lf{\labl_1}{\VPA},\lf{\labl_2}{\VPB}}{\SD}%
			}
			\infer1[\PRNL{\BIaand}]{\SEQ{\SG,\lf{\labl}{\VPA\BIaand\VPB}}{\SD}}
		\end{ebp}
		\qquad
		\begin{ebp}
			\hypo{%
				\SEQ{\LEQL{\MULL{\labl_1}{\labl_2}}{\labl},\SG,\lf{\labl_1}{\VPA},\lf{\labl_2}{\VPB}}{\SD}%
			}
			\infer1[\PRNL{\BImand}]{\SEQ{\SG,\lf{\labl}{\VPA\BImand\VPB}}{\SD}}
		\end{ebp}
		
		\myskip
		
		\begin{ebp}
			\hypo{\SEQ{\LEQL{\ADDL{\labl_1}{\labl_2}}{\labl},\SG}{\lf{\labl_1}{\VPA},\SD}}
			\hypo{\SEQ{\LEQL{\ADDL{\labl_1}{\labl_2}}{\labl},\SG}{\lf{\labl_2}{\VPB},\SD}}
			\infer2[\PRNR{\BIaand}]{%
				\SEQ{\LEQL{\ADDL{\labl_1}{\labl_2}}{\labl},\SG}{\lf{\labl}{\VPA \BIaand \VPB},\SD}%
			}
		\end{ebp}
		
		\myskip
		
		\begin{ebp}
			\hypo{\SEQ{\LEQL{\MULL{\labl_1}{\labl_2}}{\labl},\SG}{\lf{\labl}{\VPA 
			\BImand \VPB},\lf{\labl_1}{\VPA},\SD}}
			\hypo{\SEQ{\LEQL{\MULL{\labl_1}{\labl_2}}{\labl},\SG}{\lf{\labl}{\VPA 
			\BImand \VPB},\lf{\labl_2}{\VPB},\SD}}
			\infer2[\PRNR{\BImand}]{%
				\SEQ{\LEQL{\MULL{\labl_1}{\labl_2}}{\labl},\SG}{\lf{\labl}{\VPA \BImand \VPB},\SD}%
			}
		\end{ebp}
		
		\myskip
		
		\begin{ebp}
			\hypo{\SEQ{\SG,\lf{\labl}{\VPA}}{\SD}}
			\hypo{\SEQ{\SG,\lf{\labl}{\VPB}}{\SD}}
			\infer2[\PRNL{\BIaor}]{\SEQ{\SG,\lf{\labl}{\VPA\BIaor\VPB}}{\SD}}
		\end{ebp}
		\qquad
		\begin{ebp}
			\hypo{\SEQ{\SG}{\lf{\labl}{\VPA_{\mathrm{i}\,\in\,\ens{1,2}}},\SD}}
			\infer1[\PRNR{\BIaor^\mathrm{i}}]{%
				\SEQ{\SG}{\lf{\labl}{\VPA_1\BIaor\VPA_2},\SD}%
			}
		\end{ebp}
	\end{center}
	
	\textbf{Side conditions:}
	$\labl_1$ and $\labl_2$ must be fresh label letters in 
	\PRN{\BImand_L}, \PRN{\BIaand_L}, \PRN{\BImimp_R}, and \PRN{\BIaimp_R}.
	
	\caption{Logical Rules of {\GBI}.}
	\label{fig:gbi-logical-rules}
\end{figure}

\begin{figure}[t]
	\begin{center}
		
		\begin{ebp}
			\hypo{\SEQ{\LEQL{\labl}{\labl},\SG}{\SD}}
			\infer1[\PRN{R}]{\SEQ{\SG}{\SD}}
		\end{ebp}
		\;\;
		\begin{ebp}
			\hypo{%
				\SEQ{\LEQL{\labl_0}{\labl},\LEQL{\labl_0}{\labl_1},\LEQL{\labl_1}{\labl},\SG}{\SD}%
			}
			\infer1[\PRN{T}]{\SEQ{\LEQL{\labl_0}{\labl_1},\LEQL{\labl_1}{\labl},\SG}{\SD}}
		\end{ebp}
		\;\;

\medskip
        
		\begin{ebp}
			\hypo{\SEQ{\LEQL{\ADDL{\labl}{\labl}}{\labl},\SG}{\SD}}
			\infer1[\PRN{I_\addL}]{\SEQ{\SG}{\SD}}
		\end{ebp}
  
		\medskip
		
		\begin{ebp}
			\hypo{%
				\SEQ{\LEQL{\RONL{\labl}{\rneuL}}{\labl},\SG}{\SD}%
			}
			\infer1[\PRN{U^1_{\ronL}}]{%
				\SEQ{%
					\SG}{\SD}%
			}
		\end{ebp}
		\;\;
		\begin{ebp}
			\hypo{%
				\SEQ{\LEQL{\RONL{\rneuL}{\labl}}{\labl},\SG}{\SD}%
			}
			\infer1[\PRN{U^2_{\ronL}}]{%
				\SEQ{\SG}{\SD}%
			}
		\end{ebp}
		\;\;
		\begin{ebp}
			\hypo{\SEQ{\LEQL{\RONL{\labl_2}{\labl_1}}{\labl},\SG}{\SD}}
			\infer1[\PRN{E_{\ronL}}]{%
				\SEQ{\LEQL{\RONL{\labl_1}{\labl_2}}{\labl},\SG}{\SD}%
			}
		\end{ebp}
		
		\medskip
		
		\begin{ebp}
			\hypo{%
				\SEQ{%
					\LEQL{\RONL{\labl_3}{\labl_2}}{\labl_0},%
					\LEQL{\RONL{\labl_4}{\labl_0}}{\labl},\SG%
				}{\SD}%
			}
			\infer1[\PRN{A^1_\ronL}]{%
				\SEQ{%
					\LEQL{\RONL{\labl_4}{\labl_3}}{\labl_1},%
					\LEQL{\RONL{\labl_1}{\labl_2}}{\labl},\SG%
				}{\SD}%
			}
		\end{ebp}
		\;\;

\medskip

		\begin{ebp}
			\hypo{%
				\SEQ{%
					\LEQL{\RONL{\labl_1}{\labl_4}}{\labl_0},%
					\LEQL{\RONL{\labl_0}{\labl_3}}{\labl},\SG%
				}{\SD}%
			}
			\infer1[\PRN{A^2_\ronL}]{%
				\SEQ{%
					\LEQL{\RONL{\labl_4}{\labl_3}}{\labl_2},%
					\LEQL{\RONL{\labl_1}{\labl_2}}{\labl},\SG%
				}{\SD}%
			}
		\end{ebp}
		
		\medskip
		
		\begin{ebp}
			\hypo{%
				\SEQ{\LEQL{\labl_i}{\labl},\LEQL{\ADDL{\labl_1}{\labl_2}}{\labl},\SG}{\SD}%
			}
			\infer1[\PRN{P^i_\addL}]{%
				\SEQ{\LEQL{\ADDL{\labl_1}{\labl_2}}{\labl},\SG}{\SD}%
			}
		\end{ebp}
		\;\;
		\begin{ebp}
			\hypo{%
				\SEQ{\LEQL{\labl_i}{\labl},\LEQL{\MULL{\labl_1}{\labl_2}}{\labl},\SG}{\SD}%
			}
			\infer1[\PRN{P^i_\mulL}]{%
				\SEQ{\LEQL{\MULL{\labl_1}{\labl_2}}{\labl},\SG}{\SD}%
			}
		\end{ebp}
		
		\medskip
		
		\begin{ebp}
			\hypo{%
				\SEQ{%
					\LEQL{\RONL{\labl_0}{\labl_2}}{\labl},%
					\LEQL{\labl_0}{\labl_1},\LEQL{\RONL{\labl_1}{\labl_2}}{\labl},\SG%
				}{\SD}%
			}
			\infer1[\PRN{C^1_\ronL}]{%
				\SEQ{\LEQL{\labl_0}{\labl_1},\LEQL{\RONL{\labl_1}{\labl_2}}{\labl},\SG}{\SD}%
			}
		\end{ebp}
		\;\;
		\begin{ebp}
			\hypo{\SEQ{\LEQL{\labl}{\labl_1},\SG,\lf{\labl_1}{\VPA}}{\SD}}
			\infer1[\PRNL{K}]{\SEQ{\LEQL{\labl}{\labl_1},\SG,\lf{\labl}{\VPA}}{\SD}}
		\end{ebp}
		
		\medskip
		
		\begin{ebp}
			\hypo{%
				\SEQ{%
					\LEQL{\RONL{\labl_1}{\labl_0}}{\labl},%
					\LEQL{\labl_0}{\labl_2},\LEQL{\RONL{\labl_1}{\labl_2}}{\labl},\SG}%
				{\SD}%
			}
			\infer1[\PRN{C^2_\ronL}]{%
				\SEQ{\LEQL{\labl_0}{\labl_2},\LEQL{\RONL{\labl_1}{\labl_2}}{\labl},\SG}{\SD}%
			}
		\end{ebp}
		\;\;
		\begin{ebp}
			\hypo{\SEQ{\LEQL{\labl_1}{\labl},\SG}{\lf{\labl_1}{\VPA},\SD}}
			\infer1[\PRNR{K}]{\SEQ{\LEQL{\labl_1}{\labl},\SG}{\lf{\labl}{\VPA},\SD}}
		\end{ebp}
		
		\medskip
		
		\begin{ebp}
			\hypo{\SEQ{\SG_0}{\SD}}
			\infer1[\PRNL{W}]{\SEQ{\SG_0,\SG_1}{\SD}}
		\end{ebp}
		\;\;
		\begin{ebp}
			\hypo{\SEQ{\SG}{\SD_0}}
			\infer1[\PRNR{W}]{\SEQ{\SG}{\SD_0,\SD_1}}
		\end{ebp}
		\;\;
		\begin{ebp}
			\hypo{\SEQ{\SG_0,\SG_1,\SG_1}{\SD}}
			\infer1[\PRNL{C}]{\SEQ{\SG_0,\SG_1}{\SD}}
		\end{ebp}
		\;\;

		\medskip
        
		\begin{ebp}
			\hypo{\SEQ{\SG}{\SD_0,\SD_1,\SD_1}}
			\infer1[\PRNR{C}]{\SEQ{\SG}{\SD_0,\SD_1}}
		\end{ebp}
		
	\end{center}

	\textbf{Side conditions:}\\
	$i\,\in\,\ens{1,2}$ and $\ronL\,\in\,\ens{\mulL,\addL}$.
	\\
	$\labl_0$ is a fresh label letter in \PRN{A^i_\ronL}. 
	$\labl_{3-i}$ in \PRN{P^i_\mulL}
	must be in $\ens{ \mneuL, \abotL}$.
	\\
	$\labl$ in \PRN{R} and \PRN{I_\addL},
	$\labl_1,\labl_2$ in \PRN{P^{i}_\addL} and
	$\labl_i$ in \PRN{P^i_\mulL}
	must occur in $\SG$, $\SD$ or $\ens{ \mneuL, \aneuL, \abotL }$.
	\caption{Structural Rules of {\GBI}.}
	\label{fig:gbi-structural-rules}
\end{figure}

The third and final step is to devise logical rules capturing the meaning of the connectives and structural rules reflecting the properties of the underlying frame.
The logical rules of {\GBI} are given in \Fig~\ref{fig:gbi-logical-rules} and are direct translations of their semantic clauses.
The structural rules of {\GBI} are given in \Fig~\ref{fig:gbi-structural-rules} where $\ronL$ (resp. $\rneuL$) denotes either $\mulL$ or $\addL$ (resp. $\mneuL$ and $\aneuL$) in contexts where the multiplicative or additive nature of the functor (resp. unit) is not important (\eg, for properties that hold in both cases).

The structural rules \PRN{R} and \PRN{T} capture the reflexivity and transitivity of the accessibility relation.
Rules \PRN{U^i_\ronL} capture the identity of the functors $\mulL$ and $\addL$ \wrt\ $\mneuL$ and $\aneuL$.
The superscript $i \in \ens{1,2}$ in a rule name denotes which argument of an $\ronL$-functor is treated by the rule and can be dropped if we consider the $\ronL$-functor as implicitly commutative instead of having the explicit exchange rules~\PRN{E_\ronL} for commutativity.
The rules \PRN{A^i_\ronL} reflect the associativity of the $\ronL$-functors and \PRN{I_\addL} reflects the idempotency of $\addS$ into the $\addL$-functor.
The projection rules \PRN{P^i_\addL} reflect into the $\addL$-functor the fact that $\addS$ is increasing, \ie, $\LEQS{\wldm}{\ADDS{\wldm}{\wldn}}$.
The projection rules \PRN{P^i_\mulL} capture the fact that $\LEQS{\wldm}{\MULS{\wldm}{\wldn}}$ only holds if $\wldn$ is $\absS$ or $\neuS$.
The compatibility rules \PRN{C^i_\ronL} reflect that $\addS$ and $\mulS$ are both order preserving.

\begin{definition}
	\label{def:gbi-theoremhood}
	A formula $\VPA$ is a theorem of {\GBI} \iffi \SEQ{\LEQL{\mneuL}{\labl}}{\lf{\VPA}{\labl}} is provable in {\GBI} for
	some label letter $\labl$.
\end{definition}

\Fig~\ref{fig:gbi-example-one} gives an example of a proof in {\GBI}, where the notation \quo{$\gell$} subsumes all the elements we omit to keep the proof more concise. Let us also remark that in order to keep the proof shorter we do not explicitly represent the weakening steps before occurring before applying the axiom rule \PRN{id}.

\begin{figure}[p]
\centering
	\begin{ebp}
		\hypo{}
		\infer1[\PRN{id}]{
			\SEQZ{%
			}{%
				\lf{\VPP}{\labl_4}
			}{%
				\lf{\VPP}{\labl_4}
			}%
		}
		\hypo{}
		\infer1[\PRN{id}]{
			\SEQZ{%
			}{%
				\lf{\VPP}{\labl_4}
			}{%
				\lf{\VPP}{\labl_4}
			}%
		}
		\hypo{}
		\infer1[\PRN{id}]{
			\SEQZ{%
			}{%
				\lf{\VPQ}{\labl_1}
			}{%
				\lf{\VPQ}{\labl_1}%
			}%
		}
		\hypo{}
		\infer1[\PRN{id}]{
			\SEQZ{%
			}{%
				\lf{\VPR}{\labl_1}
			}{%
				\lf{\VPR}{\labl_1}%
			}%
		}
		\infer1[\PRNR{K}]{
			\SEQZ{%
				\LEQL{\labl_1}{\labl_2},\ldots,
			}{%
				\lf{\VPR}{\labl_1},\ldots 
			}{%
				\lf{\VPR}{\labl_2}%
			}%
		}
		\infer2[\PRNL{\BIaimp}]{
			\SEQZ{%
				\LEQL{\ADDL{\labl_1}{\labl_1}}{\labl_1},
				\\
				\LEQL{\labl_1}{\labl_2},
				\LEQL{\MULL{\mneuL}{\labl_1}}{\labl_2}
				\\
				\LEQL{\MULL{\labl_5}{\labl_4}}{\labl_1},
				\LEQL{\MULL{\labl_6}{\labl_4}}{\labl_1},
				\\
				\LEQL{\labl_5}{\labl_3},
				\LEQL{\labl_6}{\labl_3},
				\\
				\LEQL{\ADDL{\labl_5}{\labl_6}}{\labl_3},
				\LEQL{\MULL{\labl_3}{\labl_4}}{\labl_1},
				\\
				\LEQL{\MULL{\labl_0}{\labl_1}}{\labl_2},
				\LEQL{\mneuL}{\labl_0},
			}{%
				\lf{\VPQ}{\labl_1},
				\\
				\lf{\VPQ \BIaimp \VPR}{\labl_1},  
				\\
				\lf{\VPP \BImimp (\VPQ \BIaimp \VPR)}{\labl_5}, 
				\\
				\lf{\VPP \BImimp \VPQ}{\labl_6},
				\lf{\VPP}{\labl_4}%
			}{%
				\lf{\VPR}{\labl_2}%
			}%
		}
		\infer1[\PRN{I_{\addL}}]{
			\SEQZ{%
				\LEQL{\labl_1}{\labl_2},
				\LEQL{\MULL{\mneuL}{\labl_1}}{\labl_2}
				\\
				\LEQL{\MULL{\labl_5}{\labl_4}}{\labl_1},
				\LEQL{\MULL{\labl_6}{\labl_4}}{\labl_1},
				\\
				\LEQL{\labl_5}{\labl_3},
				\LEQL{\labl_6}{\labl_3},
				\\
				\LEQL{\ADDL{\labl_5}{\labl_6}}{\labl_3},
				\LEQL{\MULL{\labl_3}{\labl_4}}{\labl_1},
				\\
				\LEQL{\MULL{\labl_0}{\labl_1}}{\labl_2},
				\LEQL{\mneuL}{\labl_0},
			}{%
				\lf{\VPQ}{\labl_1},
				\\
				\lf{\VPQ \BIaimp \VPR}{\labl_1},  
				\\
				\lf{\VPP \BImimp (\VPQ \BIaimp \VPR)}{\labl_5}, 
				\\
				\lf{\VPP \BImimp \VPQ}{\labl_6},
				\lf{\VPP}{\labl_4}%
			}{%
				\lf{\VPR}{\labl_2}%
			}%
		}
		\infer1[\PRN{P_{\mulL}}]{
			\SEQZ{%
				\LEQL{\MULL{\mneuL}{\labl_1}}{\labl_2}
				\\
				\LEQL{\MULL{\labl_5}{\labl_4}}{\labl_1},
				\LEQL{\MULL{\labl_6}{\labl_4}}{\labl_1},
				\\
				\LEQL{\labl_5}{\labl_3},
				\LEQL{\labl_6}{\labl_3},
				\\
				\LEQL{\ADDL{\labl_5}{\labl_6}}{\labl_3},
				\LEQL{\MULL{\labl_3}{\labl_4}}{\labl_1},
				\\
				\LEQL{\MULL{\labl_0}{\labl_1}}{\labl_2},
				\LEQL{\mneuL}{\labl_0},
			}{%
				\lf{\VPQ}{\labl_1},
				\\
				\lf{\VPQ \BIaimp \VPR}{\labl_1},  
				\\
				\lf{\VPP \BImimp (\VPQ \BIaimp \VPR)}{\labl_5}, 
				\\
				\lf{\VPP \BImimp \VPQ}{\labl_6},
				\lf{\VPP}{\labl_4}%
			}{%
				\lf{\VPR}{\labl_2}%
			}%
		}
		\infer1[\PRN{C_{\mulL}}]{
			\SEQZ{%
				\LEQL{\MULL{\labl_5}{\labl_4}}{\labl_1},
				\LEQL{\MULL{\labl_6}{\labl_4}}{\labl_1},
				\\
				\LEQL{\labl_5}{\labl_3},
				\LEQL{\labl_6}{\labl_3},
				\\
				\LEQL{\ADDL{\labl_5}{\labl_6}}{\labl_3},
				\LEQL{\MULL{\labl_3}{\labl_4}}{\labl_1},
				\\
				\LEQL{\MULL{\labl_0}{\labl_1}}{\labl_2},
				\LEQL{\mneuL}{\labl_0},
			}{%
				\lf{\VPQ}{\labl_1},
				\\
				\lf{\VPQ \BIaimp \VPR}{\labl_1},
				\\
				\lf{\VPP \BImimp (\VPQ \BIaimp \VPR)}{\labl_5}, 
				\\
				\lf{\VPP \BImimp \VPQ}{\labl_6},
				\lf{\VPP}{\labl_4}%
			}{%
				\lf{\VPR}{\labl_2}%
			}%
		}
		\infer[separation=5ex]2[\PRNL{\BImimp}]{
            \hspace{15ex}%
			\SEQZ{%
				\LEQL{\MULL{\labl_5}{\labl_4}}{\labl_1},
				\LEQL{\MULL{\labl_6}{\labl_4}}{\labl_1},
				\\
				\LEQL{\labl_5}{\labl_3},
				\LEQL{\labl_6}{\labl_3},
				\\
				\LEQL{\ADDL{\labl_5}{\labl_6}}{\labl_3},
				\LEQL{\MULL{\labl_3}{\labl_4}}{\labl_1},
				\\
				\LEQL{\MULL{\labl_0}{\labl_1}}{\labl_2},
				\LEQL{\mneuL}{\labl_0},
			}{%
				\lf{\VPQ \BIaimp \VPR}{\labl_1},
				\\
				\lf{\VPP \BImimp (\VPQ \BIaimp \VPR)}{\labl_5}, 
				\\ 
				\lf{\VPP \BImimp \VPQ}{\labl_6},
				\lf{\VPP}{\labl_4}%
			}{%
				\lf{\VPR}{\labl_2}%
			}%
		}
		\infer[separation=-15ex]2[\PRNL{\BImimp}]{%
			\SEQZ{
				\LEQL{\MULL{\labl_5}{\labl_4}}{\labl_1},
				\LEQL{\MULL{\labl_6}{\labl_4}}{\labl_1},
				\\
				\LEQL{\labl_5}{\labl_3},
				\LEQL{\labl_6}{\labl_3},
				\\
				\LEQL{\ADDL{\labl_5}{\labl_6}}{\labl_3},
				\LEQL{\MULL{\labl_3}{\labl_4}}{\labl_1},
				\\
				\LEQL{\MULL{\labl_0}{\labl_1}}{\labl_2},
				\LEQL{\mneuL}{\labl_0},
			}{%
				\lf{\VPP \BImimp (\VPQ \BIaimp \VPR)}{\labl_5},
				\\
				\lf{\VPP \BImimp \VPQ}{\labl_6},
				\lf{\VPP}{\labl_4}
			}{
				\lf{\VPR}{\labl_2}
			}%
		}
		\infer1[\PRN{C_{\mulL}}]{%
			\SEQZ{
				\LEQL{\labl_5}{\labl_3},
				\LEQL{\labl_6}{\labl_3},
				\\
				\LEQL{\ADDL{\labl_5}{\labl_6}}{\labl_3},
				\LEQL{\MULL{\labl_3}{\labl_4}}{\labl_1},
				\\
				\LEQL{\MULL{\labl_0}{\labl_1}}{\labl_2},
				\LEQL{\mneuL}{\labl_0},
			}{%
				\lf{\VPP \BImimp (\VPQ \BIaimp \VPR)}{\labl_5},
				\\
				\lf{\VPP \BImimp \VPQ}{\labl_6},
				\lf{\VPP}{\labl_4}
			}{
				\lf{\VPR}{\labl_2}
			}%
		}
		\infer1[\PRN{P_{\addL}}]{%
			\SEQZ{
				\LEQL{\ADDL{\labl_5}{\labl_6}}{\labl_3},
				\LEQL{\MULL{\labl_3}{\labl_4}}{\labl_1},
				\\
				\LEQL{\MULL{\labl_0}{\labl_1}}{\labl_2},
				\LEQL{\mneuL}{\labl_0},
			}{%
				\lf{\VPP \BImimp (\VPQ \BIaimp \VPR)}{\labl_5},
				\\
				\lf{\VPP \BImimp \VPQ}{\labl_6}, 
				\lf{\VPP}{\labl_4}
			}{
				\lf{\VPR}{\labl_2}
			}%
		}
		\infer1[\PRNL{\BIaand}]{%
			\SEQZ{
				\LEQL{\MULL{\labl_3}{\labl_4}}{\labl_1},
				\\
				\LEQL{\MULL{\labl_0}{\labl_1}}{\labl_2},
				\LEQL{\mneuL}{\labl_0},
			}{%
				\lf{\VPP \BImimp (\VPQ \BIaimp \VPR)\BIaand \VPP \BImimp 
				\VPQ}{\labl_3},
				\\
				\lf{\VPP}{\labl_4}
			}{
				\lf{\VPR}{\labl_2}
			}%
		}
		\infer1[\PRNL{\BImand}]{%
			\SEQZ{
				\LEQL{\MULL{\labl_0}{\labl_1}}{\labl_2},
				\LEQL{\mneuL}{\labl_0},
			}{%
				\lf{(\VPP \BImimp (\VPQ \BIaimp \VPR)\BIaand \VPP \BImimp \VPQ)\BImand \VPP}{\labl_1}
			}{
				\lf{\VPR}{\labl_2}
			}%
		}
		\infer1[\PRNR{\BImimp}]{%
			\SEQZ{
				\LEQL{\mneuL}{\labl_0}
			}{%
			}{%
				\lf{((\VPP \BImimp (\VPQ \BIaimp \VPR) \BIaand \VPP \BImimp \VPQ) \BImand \VPP) \BImimp \VPR}{\labl_0}
			}%
		}
	\end{ebp}
	\caption{{\GBI}-proof of $((\VPP \BImimp (\VPQ \BIaimp \VPR) \BIaand \VPP \BImimp \VPQ) \BImand \VPP) \BImimp \VPR$.}
	\label{fig:gbi-example-one}
\end{figure}

\subsubsection{From Bunched to Labeled Proofs}

In order to highlight the relationships between the labels and the tree structure of bunches more easily let us use label letters of the form~$xs$ where $x$ is a non-greek letter and $s \in \ens{0,1}^*$ is a binary string that encodes the path of the node~$xs$ in a tree structure the root of which is~$x$.
Let us call~$x$ the root of a label letter~$xs$ and let us use greek letters to range over label letters with the convention that distinct greek letters denote label letters with distinct roots.

\begin{definition}
	\label{def:btol}
	Given a bunch\/~$\SG$ and a label letter $\LLD$, $\BTOL{\SG}{\LLD}$, the \emph{translation of\/~$\SG$} according to $\LLD$, is defined by induction on the structure of\/~$\SG$ as follows:
	\begin{itemize}[wide, labelwidth=!, labelindent=0pt]
		\item $\BTOL{\VPA}{\LLD} = \ens{ \lf{\VPA}{\LLD} }$,
		$\BTOL{\BIanul}{\LLD} = \ens{ \LEQL{\aneuL}{\LLD} }$,
		$\BTOL{\BImnul}{\LLD} = \ens{ \LEQL{\mneuL}{\LLD} }$,
		\item $\BTOL{(\SD_0 \BImsep \SD_1)}{\LLD} =
		\BTOL{\SD_0}{\LLD 0} \cup
		\BTOL{\SD_1}{\LLD 1} \cup
		\ens{ \LEQL{\MULL{\LLD 0}{\LLD 1}}{\LLD} }$,
		\item $\BTOL{(\SD_0 \BIasep \SD_1)}{\LLD} =
		\BTOL{\SD_0}{\LLD 0} \cup
		\BTOL{\SD_1}{\LLD 1} \cup
		\ens{ \LEQL{\ADDL{\LLD 0}{\LLD 1}}{\LLD} }$.
	\end{itemize}
	Given a sequent $\SEQ{\SG}{\VPA}$,
	$\BTOL{\SEQ{\SG}{\VPA}}{\LLD}$ is defined as
	$\SEQ{\BTOL{\SG}{\LLD}}{\lf{\VPA}{\LLD}}$.
\end{definition}
We write $\BT{\SG}{\LLD}$ as a shorthand for $\BTOL{\SG}{\LLD}$ so that
$
\BTOL{\SEQ{\SG}{\VPA}}{\LLD} = \SEQ{\BT{\SG}{\LLD}}{\lf{\VPA}{\LLD}}.
$
The following is an illustration of Definition~\ref{def:btol}:
\[
\begin{tikzpicture}[level distance=4em,text height=1em]
	\tikzstyle{level 1}=[sibling distance=6em]
	\tikzstyle{level 2}=[sibling distance=6em]
	\tikzstyle{edge from parent}=[->,draw]
	\node (sc) { \makebox[2em][c]{$\lf{\BIasep}{\LLD}$} } [grow'=up]
	child { node { \makebox[2em]{$\lf{\BImsep}{\LLD 0}$} }
		child { node (co) { \makebox[2em]{$\lf{\BImnul}{\LLD 00}$} } }
		child { node (p) { \makebox[2em]{$\lf{\VPP}{\LLD 01}$} } }
	}
	child { node (q) { \makebox[2em]{$\lf{\VPQ}{\LLD 1}$} } }
	;
	\draw[dashed] (-3.25em,2em) to (3.25em,2em);
	\node at (0em,3em) { $\scriptstyle \addL$ };
	\draw[dashed] (-6.5em,6.25em) to (0.5em,6.25em);
	\node at (-3em,7.25em) { $\scriptstyle \mulL$ };
	\node (lbis) at (14.5em,0em) {
		$
		\SEQ{\BT{((\BImnul \BImsep \VPP) \BIasep \VPQ)}{\LLD}}{\lf{\VPR}{\LLD}}%
		$ } [grow'=up]
	child { node {
			$
			\SEQ{%
				\LEQL{\ADDL{\LLD 0}{\LLD 1}}{\LLD},
				\BT{(\BImnul \BImsep \VPP)}{\LLD 0},
				\lf{\VPQ}{\LLD 1}
			}{%
				\lf{\VPR}{\LLD}%
			}
			$
		} child { node {
				$
				\SEQ{%
					\LEQL{\MULL{\LLD 00}{\LLD 01}}{\LLD 0},
					\LEQL{\ADDL{\LLD0}{\LLD1}}{\LLD},
					\lf{\VPP}{\LLD 01},
					\lf{\VPQ}{\LLD 1}
				}{%
					\lf{\VPR}{\LLD}%
				}
				$
		} }
	};
\end{tikzpicture}
\]
As a second example, the translation of the sequent $\SEQ{(\VPP \BImimp (\VPQ \BIaimp \VPR)\BIasep \VPP \BImimp \VPQ)\BImsep\VPP}{\VPR}$, which is the premiss of the $\PRNL{\BIaand}$ rule in the \LBI-proof presented in Example~\ref{fig:lbi-proof-example}, results in the following labeled sequent:
\[
\SEQZ{%
	\LEQL{\ADDL{\LLD 00}{\LLD 01}}{\LLD 0},
	\LEQL{\MULL{\LLD 0}{\LLD 1}}{\LLD},
}{%
	\lf{\VPP \BImimp (\VPQ \BIaimp \VPR)}{\LLD 01},
	\lf{\VPP \BImimp \VPQ}{\LLD 00},
	\lf{\VPP}{\LLD 1}
}{%
	\lf{\VPR}{\LLD}
}
\]

Using Definition~\ref{def:btol} it is not particularly difficult to translate {\LBI}-proofs into {\GBI}-proofs and we have the following result:
\begin{theorem}
	\label{thm:BI-TranslationThm-BTOL}
	If a sequent \SEQ{\SG}{\VPA} is provable in \LBI,
	then for any label letter~$\LLD$, the
	labeled sequent \SEQ{\BT{\SG}{\LLD}}{\lf{\VPA}{\LLD}} is provable
	in \GBI.
\end{theorem}
\begin{proof}
The proof is by induction on the structure of an \LBI-proof using an appropriate definition of label substitutions.
See \cite{Gal19a} for details.
\end{proof}

\subsubsection{From Labeled to Bunched Proofs}

Trying to translate labeled {\GBI}-proofs into bunched {\LBI}-proofs is much harder than the opposite way and is currently only known for a subclass of {\GBI}-proofs satisfying a conjunction of specific conditions called \emph{the tree property}.
Since describing the tree property in full technical details is out of the scope of this paper (see \cite{Gal18a} for details), we should focus here on giving an intuitive account of its content and discuss the related issues.
Let us also mention that the tree property arises from a careful inspection of the proof of Theorem~\ref{thm:BI-TranslationThm-BTOL} which shows that in all sequents of a translated {\LBI}-proof, the label constraints of {\GBI}-sequent describe a tree structure which allows the reconstruction of a bunch from the label of the formula on its right-hand side.

Let us write $\ltob(s,\labl)$ the function that translates a label sequent~$s$ to a bunch using the label $\labl$ (required to occur in $s$) as its reference point.
The result $s@l$ of $\ltob(s,\labl)$ is called ``the bunch translation of~$s$ at~$\labl$''.
For conciseness, we shall omit~$s$ when clear from the context.
Let us finally define $\ltob(s)$ as $\ltob(s,\labl)$ where~$\labl$ is the label of the formula on the right-hand side of~$s$.
In the light of Definition~\ref{def:btol}, let us make an intuitive attempt at algorithmically defining $\ltob(s,\labl)$.

\begin{definition}
    \label{def:BI-TranslationThm-LTOB}
$\ltob(s,\labl)$, the translation of a {\GBI} labeled sequent $s$ at label $\labl$, recursively constructs a bunch from the label-constraints in~$s$ as follows:
\begin{itemize}[wide, labelwidth=!, labelindent=0pt]
\item
if $\LEQL{\MULL{\labl_i}{\labl_j}}{\labl} \in s$, 
then $\ltob(\labl)= (\ltob(\labl_i) \BImsep \ltob(\labl_j))$
\item 
if $\LEQL{\ADDL{\labl_i}{\labl_j}}{\labl} \in s$, 
then $\ltob(\labl)= (\ltob(\labl_i) \BIasep \ltob(\labl_j))$
\item 
if $\LEQL{\labl'}{\labl} \in s$ and $\labl' \in \Labs^0$, 
then $\ltob(\labl) = \ltob(\labl')$
\item 
otherwise
$\ltob(\mneuL) = \BImnul, \ltob(\aneuL) = \BIanul$,
$\ltob(\labl) = \BIaand\ensc{\VPA}{\lf{\VPA}{\labl}}$ with $\BIaand \varnothing 
= \BIatop$
\end{itemize}
\end{definition}
The translation described in Definition~\ref{def:BI-TranslationThm-LTOB} is illustrated in Example~\ref{ex:example-ltob-with-tp}. Remark that label constraints of the form $\LEQL{\labl'}{\labl}$, where $\labl'$ is an atomic label, act as ``jumps'' that move the reference point from one label to another (such ``jumps'' are in fact reduction orderings described more precisely in~\cite{Gal19a}).

\begin{example}
\label{ex:example-ltob-with-tp}
Let $s$ be the labeled sequent which is the premiss of the $\PRNL{\BIaand}$ rule in the {\GBI}-proof presented in \Fig~\ref{fig:gbi-example-one}.
The starting point of $\ltob(s)$ is $\labl_2$, the label occurring of the right-hand side of~$s$. The translation proceeds as described in the following table, where the comment on the right-hand side at Step~$n$ is the justification of the result obtained on the left-hand side of Step~$n+1$. 
\[
	\begin{array}{l@{\quad}l@{\qquad}|@{\qquad}l}
	1. & \SEQ{\ltob(\labl_2)}{\VPR} & \LEQL{\MULL{\labl_0}{\labl_1}}{\labl_2},
	\\
	2. & \SEQ{(\ltob(\labl_0) \BImsep \ltob(\labl_1))}{\VPR} & 
	\LEQL{\mneuL}{\labl_0}
	\\
	3. & \SEQ{(\ltob(\mneuL) \BImsep \ltob(\labl_1))}{\VPR} & \ltob(\mneuL) = 
	\BImnul
	\\
	4. & \SEQ{(\BImnul \BImsep \ltob(\labl_1))}{\VPR} & 
	\LEQL{\MULL{\labl_3}{\labl_4}}{\labl_1}
	\\
	5. & \SEQ{(\BImnul \BImsep (\ltob(\labl_3) \BImsep \ltob(\labl_4))}{\VPR} & 
					\lf{\VPP \BImimp (\VPQ \BIaimp \VPR)\BIaand \VPP \BImimp 
					\VPQ}{\labl_3}
	\\
	6. & \SEQ{(\BImnul \BImsep (\VPP \BImimp (\VPQ \BIaimp \VPR)\BIaand \VPP 
	\BImimp \VPQ \BImsep \ltob(\labl_4))}{\VPR} & \lf{\VPP}{\labl_4}
	\\
	7. & \SEQ{(\BImnul \BImsep (\VPP \BImimp (\VPQ \BIaimp \VPR)\BIaand \VPP 
	\BImimp \VPQ \BImsep \VPP))}{\VPR} & \text{stop}

	\end{array}
\] 
\end{example}

Unfortunately, the translation in Definition~\ref{def:BI-TranslationThm-LTOB} only works when all of the sequents in a {\GBI}-proof satisfy the tree property.
As shown in~\cite{Gal19a}, all {\LBI}-translated {\GBI}-proofs satisfy the tree property, but at the cost of the flexibility of the labeled proof system in full generality. 
Indeed, translating contraction and weakening steps requires contrived labeled versions of the contraction and weakening rules that preserve the tree structure.
For instance, the tree-preserving contraction rule looks like this (the subtree root at $\BT{\ST}{\LLD s}$ is duplicated into two new subtrees $\BT{\ST}{\LLD s 0}$ and $\BT{\ST}{\LLD s 1}$ and linked as children of the old subtree):
\[
\begin{ebp}
	\hypo{%
		\SEQ{%
			\BT{%
				\SG(%
				\LEQL{\ADDL{\LLD s0}{\LLD s1}}{\LLD s},%
				\BT{\ST}{\LLD s0},%
				\BT{\ST}{\LLD s1}%
				)%
			}{%
				\LLD%
			}
		}{%
			\lf{\VPA}{\LLD}%
		}%
	}
	\infer1[\PRN{C_T}]{%
		\SEQ{%
			\BT{%
				\SG(%
				\BT{\ST}{\LLD s}%
				)%
			}{%
				\LLD%
			}
		}{%
			\lf{\VPA}{\LLD}%
		}%
	}
\end{ebp}
\]

\Fig~\ref{fig:gbi-example-one} and Example~\ref{fig:lbi-proof-example} respectively are {\GBI}- and {\LBI}-proofs of the same formula.
Comparing both proofs, we notice that they share the same logical proof plan, more precisely, they decompose the same logical connectives in the same order.
However, the {\GBI}-proof does not use any of the tree-preserving rules of~{\GBI} and thus does not correspond to an {\LBI}-translated {\GBI}-proof.
Translating the {\LBI}-proof would require the tree-preserving contraction rule discussed previously to perform the contraction step above the $\PRNL{\BIaand}$ rule.
Such tree-preserving rules are very restrictive, do not mimic the semantics and would not be naturally devised in a conventional labeled system.
Although conventional {\GBI} structural rules such as weakening, contraction and idempotency can easily break the tree property, they also allow more flexibility in the labeled proofs.
For example, let~$s$ be the sequent that is the premiss of the $\PRN{I_{\addL}}$ rule in the {\GBI}-proof depicted in \Fig~\ref{fig:gbi-example-one}.
Trying to compute $\ltob(s,\labl_2)$ would fail for the following reasons:
\begin{enumerate}[wide, labelwidth=!, labelindent=0pt]

\item[(1)] We have several distinct label-constraints with the same root (\ie, with the same label on the right-hand side).
For instance, we have $\LEQL{\labl_1}{\labl_2}$ and $\LEQL{\MULL{\labl_0}{\labl_1}}{\labl_2}$. 
Should we ``jump'' from $\labl_2$ to $\labl_1$ or should we recursively translate $\ltob(\labl_0)$ and $\ltob(\labl_1)$ ?
We could define a strategy for deterministically choosing between distinct label-constraints with the same root.
A reasonable one would be to choose the label-constraint that has been introduced the more recently (as being closer to the translated sequent might be more pertinent), but it emphasizes the fact that a suitable translation should take the global structure of the labeled proof into account and not just labeled sequents locally.  

\item[(2)] Anyway, whatever strategy we might come up with in the previously discussed point, the label-constraint $\LEQL{\ADDL{\labl_1}{\labl_1}}{\labl_1}$ clearly does not describe a tree structure, but a cycle forcing $\ltob(s,\labl_1)$ into an infinite loop.
We could place a bound on the number of loops allowed, but then which one? It is clear that the idempotency rule $\PRN{I_{\addL}}$ in {\GBI} is related with contraction in {\LBI}, but it is not currently clear to us how to predict the correct number of copies a bunch might need in a {\LBI}-proof using a general {\GBI}-proof that, on one hand, does not correspond to an {\LBI}-translated proof and, on the other hand, does not itself need any copy.  
\end{enumerate}

It is currently an open problem whether a general {\GBI}-proof can always be turned into a {\GBI}-proof satisfying the tree property.

\subsubsection{Lost in Translation: why it fails when it fails}

Bunched (and resource) logics exhibit a first notable difference with intuitionistic logic and modal logics like $\logick$ in that the corresponding semantics do not rely only on properties of an accessibility relation in a Kripke model, but also on world (resource) composition.
In particular, since {\BI} admits both an additive and a multiplicative  composition, the relational atoms $u R w$ are generalized into relations of the form $\LEQL{\RONL{\labl_1}{\labl_2}}{\labl}$ where $\ronL$ is one of the binary functors $\addL$ or $\mulL$.
Moreover, in intuitionistic logic or modal logics like $\logick$, $\logicsiv$, $\logicsv$, the semantic and the syntactic readings of a relational atom $u R w$ coincide when interpreted in terms of ordering relations ``successor'' and ``expanded after.'' More precisely, consider the rule for right implication in intuitionistic logic depicted in \Fig~\ref{fig:labeled-calculus-Int}. The semantic reading is that, when interpreted in a Kripke structure, $u$ should be the successor of $w$ \wrt\ the accessibility relation, which can be written as $w \leqslant_\text{succ} u$.
The syntactic reading of $u R w$ is that since $\VPA$ and $\VPB$ are labeled with~$w$ and $\VPA \BIaimp \VPB$ is labeled with~$u$, $\VPA$ and $\VPB$ are subformulae of $\VPA \BIaimp \VPB$ and should therefore necessarily appear (and be expanded) after $\VPA \BIaimp \VPB$ in a (shallow) proof system.
In other words, the subformula interpretation induces a rule application order in a syntactic proof system, which could be written as $f(w) \leqslant_\text{after} f(u)$ (the formulae labeled with~$u$ must be expanded after the ones labeled with~$w$).
Notice that $\leqslant_\text{succ}$ and $\leqslant_\text{after}$ are covariant ($w$ and $u$ occur on the same side in both orders).

However, a key problem in {\BI} (and in resource logics more generally) is that the syntactic and the semantic readings are contravariant and sometimes even fully lost.
Indeed, if we consider the rule for the left multiplicative conjunction~$\BImand$ given in \Fig~\ref{fig:gbi-logical-rules}, it is clearly seen that since $\VPA$ and $\VPB$ are subformulae of $\VPA \BImand \VPB$, we syntactically have $f(\labl) \leqslant_\text{after} f(\labl_i)$ (for $i \in \{1,2\})$, but we semantically have (reading $\leqL$ as $\leqslant_\text{succ}$)  $\MULL{\labl_1}{\labl_2} \leqslant_\text{succ} \labl$ (with $\labl_1$ and $\labl_2$ occurring on the opposite side compared with $\leqslant_\text{after}$).
Moreover, we do not even get any relation of the form $\labl \leqslant_\text{succ} \labl_i$ or $\labl_i \leqslant_\text{succ} \labl$ at all.

The immediate consequence of losing the general connection between the syntactic subformula ordering and the semantic successor ordering is that finding an extension of (the translation in) Definition~\ref{def:BI-TranslationThm-LTOB} that could work for unrestricted labeled proofs is not at all trivial and might even be impossible to achieve.


\subsection{Some Remarks on Translations}

The above translations substantiate our claim that translating up the proof-theoretic hierarchy tends to be `easier' than translating down. In particular, we found that 
structural rule elimination was needed to translate labeled proofs into nested proofs for intuitionistic logic (\sect~\ref{subsec:lab-nest-seq-int}). Moreover, translating labeled proofs to structured sequent proofs for conditional logics introduced non-determinism (\sect~\ref{subsec:cond-logic-trans}) and translating labeled proofs into bunched proofs (\sect~\ref{sec:translating-bunches}) was only possible given that the labeled proof was of a `treelike' shape. Converse translations were far simpler to obtain, e.g., translating sequent proofs into labeled proofs for intuitionistic logic (\sect~\ref{subsec:lab-nest-seq-int}) and translating display proofs into labeled proofs for the tense logic $\logickt$ (\sect~\ref{subsec:lab-dis-kt}). The sophistication required in translating proofs down the hierarchy supports the claim that formalisms higher up in the hierarchy are more expressive than those below them.

\section{The Internal and External Distinction}\label{sec:int-ext}

In the literature, proof formalisms and calculi have been classified into \emph{internal} or \emph{external}.\footnote{As discussed below, the distinction between internal and external systems is rather vague. Some interpretations of this distinction are essentially the same as the distinction between semantically polluted and syntactically pure proof systems; cf.~\cite{Rea15,PogRes12}.} Typically, a formalism or calculus is placed into one of these two classes based on the syntactic elements present within the sequents used and/or the interpretability of sequents as logical formulae. Various informal definitions have been given for `internal' and `external,' and are often expressed in one of two ways:
\begin{itemize}

\item[(1)] Internal calculi omit semantic elements from the syntax of their sequents, whereas external calculi explicitly include semantic elements.

\item[(2)] Internal calculi are those where every sequent is interpretable as a formula in the language of the logic, whereas external calculi are those without a formula translation.

\end{itemize}
For example, hypersequent and nested calculi are often considered internal since their sequents are (usually) interpretable as logical formulae~\cite{MarinMS21}. On the other hand, labeled calculi are often classified as external as they incorporate semantic information in their syntax and labeled sequents exist which resist interpretation as logical formulae~\cite{CiaLyoRam18}. We remark that sometimes extra machinery is inserted into a proof calculus for `bureaucratic' reasons (e.g., to correctly formulate proof-search algorithms); such machinery should be ignored when considering a calculus internal or external.
 
 A core motivation for separating formalisms/calculi into these two categories, is that internal and external formalisms/calculi are \emph{claimed} to possess distinct advantages over one another. It has been argued that internal calculi are better suited for establishing properties such as termination, interpolation, and optimal complexity, while external calculi are more easily constructed and permit simpler proofs of completeness, cut-admissibility, and counter-model generation (from terminating proof-search). However, we will argue that a large number of such claims are false.
 
In this section, we delve into the internal and external distinction, and discuss two main themes. First, we attempt to formally define the notions of internal and external, arguing that each candidate definition comes with certain drawbacks, or fails to satisfy our intuitions concerning internal and external systems in some way. Second, we aim to dispel myths about the claimed properties of internal and external systems, while identifying which attributes are genuinely useful for certain applications.

\subsection{Analyzing Definitions of Internal and External}

We begin by investigating definition (1) above, where external calculi are those which incorporate `semantic elements' into the syntax of their sequents while internal calculi are those which do not. 
 An immediate issue that arises with this definition is that it relies on an inherently vague notion: what do we take to be a `semantic element'? Admittedly, it seems clear that the labels and relational atoms used in labeled sequents should qualify as `semantic elements' as such syntactic objects encode features of relational models. Yet, via the translation from labeled to nested sequents (see \dfn~\ref{def:lab-nested-trans} in \sect~\ref{sec:organizing-jungle}), one can see that the tree structure encoded in a nested sequent also encodes features of relational models (with points in the tree corresponding to worlds and edges in the tree corresponding to the accessibility relation). Similarly, the components of linear nested sequents and hypersequents directly correspond to worlds in relational models with the linear nested structure `$\sslash$' and the hypersequent bar `$\vert$' encoding features of the accessibility relation (cf.~\cite{LelPim15,Lah13}). It seems that (linear) nested sequents and hypersequents should qualify as external systems then, contrary to the fact that such systems are almost always counted as internal. As another example, `semantic elements' are encoded in the language of the sequent calculi used for hybrid modal logics~\cite{Brau11}. Yet, many would qualify such proofs systems as internal since their sequents are straightforwardly interpretable as formulae in the language of the logic; in fact, it is the incorporation of `semantic elements' that allows this.

 The issue with the first proposed definition is that it is too vague to properly distinguish between internal and external systems as the concept of a `semantic element' is too vague. Thus, we find that definition (1) is unsuitable for distinguishing internality and externality.

Let us now investigate definition (2) above, where internal calculi are qualified as those with sequents interpretable as formulae in the language of the logic, and external calculi are those for which this property does not hold. A couple of questions come to the fore when we consider this definition. First, what does it mean for a sequent to be \emph{interpretable} as a formula? For instance, in the context of display calculi for modal and tense logics~\cite{Kra96}, display sequents are naturally translatable to tense formulae, yet, some of these tense formula can actually be reinterpreted as modal formulae. This shows that it is not always \emph{prima facie} clear that a sequent in fact translates to a formula in the language of the logic. A second question is: what \emph{properties} should such an interpretation possess?

We begin investigating these questions by considering a few examples of `internal' systems from the literature. Our aim is to extract general underlying patterns from the examples with the goal of supplying a formal definition of `internality' along the lines of definition (2) above. What we will find is that regardless of how we attempt to rigorously specify this definition, calculi (intuitively) recognized as `internal' and `external' exist which fail to satisfy the definition, thus witnessing its inadequacy.

Gentzen calculi, nested sequent calculi, and hypersequent calculi are normally characterized as internal systems. Typically, what is meant by an `interpretation of a sequent as a formula' is a \emph{translation} $\tau$ that maps every sequent to a (i) `structurally similar' and (ii) `logically equivalent' formula in the language of the logic. For instance, Gentzen sequents in $\mathsf{S(CP)}$, nested sequents in $\ncalcint$, and hypersequents for $\logicsv$ admit the following translations:
\begin{align*}
\tau(\Gamma \sar \Delta) & := \bigwedge \Gamma \rightarrow \bigvee \Delta\\
\tau(\Gamma \nsar \Delta, [\ns_{1}]_{w_{1}}, \ldots, [\ns_{n}]_{w_{n}}) & := \bigwedge \Gamma \iimp (\bigvee \Delta \lor \tau(\ns_{1}) \lor \cdots \lor \tau(\ns_{n}))\\
\tau(\Gamma_{1} \seq \Delta_{1} \hh \cdots \hh \Gamma_{n} \seq \Delta_{n}) & := \bigvee_{1 \leq i \leq n} \!\! \Box (\bigwedge \Gamma_{i} \rightarrow \bigvee \Delta_{i})
\end{align*}
We can see that the output of every translation produces a formula that is `structurally similar' to the input in the sense that it serves as a \emph{homomorphism} mapping every sequent into a formula of the same shape (by replacing all structural connectives in the sequent with logical connectives). Moreover, the input and output are `logically equivalent' by definition, i.e., a sequent is satisfied on a model of the underlying logic \iffi its output is. This indeed seems a promising candidate for formalizing the notion of `internality,' however, let us consider the labeled sequents from $\lcalcintii$ (defined on p.~\pageref{def:lcalcintii}). 
  
As witnessed by \dfn~\ref{def:lab-nested-trans}, every labeled sequent of a \emph{treelike} shape can be interpreted as a nested sequent, and thus, by the second translation above, can be interpreted as a `structurally similar' and `logically equivalent' formula in the language of the logic $\logicint$. Yet, $\lcalcintii$ is permitted to use labeled sequents of a \emph{non-treelike} shape (e.g., $w \leq u, u \leq w \lsar w : A$), despite that fact that such sequents play no role in deriving theorems of $\logicint$ as shown in Lemma~\ref{lem:refined-labeled-int-is-treelike}. Since it appears that labeled sequents of a non-treelike shape do not admit a `structurally similar' translation in the language of $\logicint$, we are forced to conclude by the above notion of internality that $\lcalcintii$ is external. Nevertheless, if we re-define $\lcalcintii$ slightly so that \emph{only} labeled tree sequents are permitted in proofs, then $\lcalcintii$ ceases to be external and becomes internal by what was said above. Hence, the above notion of `internality' implies that being internal or external is not a property of a formalism, but of the \emph{language} of a calculus, that is, the set of sequents that the calculus draws from to construct proofs. We should therefore speak of internal and external \emph{sequent languages} (i.e., the set of sequents used by a proof system) rather than internal or external calculi.

Based on the discussion above, we could identify calculi as internal or external if the sequent language of the calculus is internal or external, respectively. Nevertheless, two practical issues arise: first, confirming that a language is external is subject to the difficulty that one must confirm the non-existence of any translation mapping sequents to `structurally similar' and `logical equivalent' formulae. Although confirming the non-existence of such a translation is perhaps not impossible,\footnote{This has been confirmed, for instance, for the hypersequents in~\cite{MetcalfeOG05} for \L ukasiewicz logic---one of the main fuzzy logics.} it appears to be a relatively difficult feature. Second, if we define internal or external systems relative to an internal or external sequent language, then proof systems may `switch' from being internal or external simply based on expanding or contracting the sequent language associated with the calculus. 
 
However, it must be conceded that the above notion of `internal' is aligned with our intuition concerning what an internal system ought to be, and can be taken as a sufficient (but not necessary) criterion for applying the term `internal' to a proof system. What we find to be important however, is less about whether a proof system satisfies our intuitions concerning `internality,' and more about the existence of translations from sequents to `structurally similar' and `logically equivalent' formulae---a fact that will be discussed in more detail below.

\subsection{Purported Properties of Internal and External Systems}

Here we consider various properties attributed to internal and external calculi, and clarify how such claims are (in)correct. For ease of presentation, we first present each claim in italics, and after, provide our perspective of the claim. Although this section is intended to dispel myths about internal and external systems, we do put forth positive applications of `internal' calculi at the end of the section (which satisfy the sufficient criterion discussed at the end of the previous section). In particular, we explain how the existence of a translation from sequents to `structurally similar' and `logically equivalent' formulae can be practically leveraged in a few ways.\\
 
\noindent
\textit{Internal calculi are better suited than external calculi for decidability}. There are two standard methods in which decidability is obtained via proof-search in a sequent calculus, which we call (1) the brute-force method, and (2) the counter-model extraction method. In the former method, one establishes that every theorem has a proof of a certain form, and shows that only a finite number of such proofs exist. Decidability is then obtained by searching this finite space, and if a proof is found, the input is known to be valid; otherwise, the input is known to be invalid. In the latter method, one attempts to construct a proof of the input, and shows that if a proof-search fails, then a counter-model of the input can be extracted. The brute-force method is more easily applied to (analytic) Gentzen systems, which are typically characterized as internal systems. This is due to the simplicity of Gentzen sequents for which it is straightforward to establish an upper finite bound on the space of analytic derivations for a given formula. Nevertheless, external systems, e.g., those of Simpson~\cite{Sim94}, also admit decidability via the brute-force method.

When it comes to applying the counter-model extraction method, there appears to be a trade-off between using internal and external calculi. Note that this method consists of two components: (1) one must establish the termination of the proof-search procedure, and (2) one must extract a counter-model if proof-search fails. We point out that `internal' calculi seem better for securing termination while `external' calculi appear better suited for extracting a counter-model. First, since the sequents in internal systems tend to utilize simpler data structures, establishing the termination of proof-search tends to be more easily obtained than for external systems (which utilize more complex and difficult to control data structures). Second, extracting a counter-model from failed proof-search tends to be easier in external systems than internal systems as the former tend to encode model-theoretic information. 

Nevertheless, this observation merely points out that there are trade-offs in using one type of system as opposed to another, and does not outright prove that one type of system is more advantageous than another in establishing decidability. Indeed, there are many examples of decision/proof-search algorithms for wide classes of logics based on internal systems~\cite{LyoGom22,TiuIanGor12,Sla97} and external systems~\cite{HorSat04,Sim94,LyoBer24}, so we find that this claim is not warranted.\\

\noindent
\textit{Internal calculi are better suited than external calculi for interpolation}. The method of establishing interpolation via sequent-style systems is due to Maehara~\cite{Mae60}, and was originally introduced in the context of Gentzen systems. This method has been adapted to linear nested and hypersequent systems~\cite{KuzLel18}, nested systems~\cite{FitKuz15,LyoTiuGorClo20}, display systems~\cite{BroGor11}, and labeled systems~\cite{Kuz16}. If one compares such works on proof-theoretic interpolation, they will find that both internal and external systems alike are used in securing interpolation properties for a logics; e.g., truly sizable classes of logics have been shown to exhibit Craig and Lyndon interpolation with both internal (viz., nested) systems~\cite{Lyo21thesis} and external (viz., labeled) systems~\cite{Kuz16}. Therefore, the claim that internal calculi are better suited for establishing interpolation does not appear warranted.\\

\noindent
\textit{Internal calculi are harder to find/construct then external calculi}. We somewhat agree with the claim that internal calculi are more difficult to find/construct in contrast to external calculi. First, we note that it is rather straightforward to generate labeled calculi for diverse classes of logics~\cite{CiaMafSpe13,Sim94}. Nevertheless, techniques do exist for generating internal calculi as well. For example, numerous logics have been provided (internal) display calculi~\cite{Bel82,Wan02}, algorithms exist for producing sequent and hypersequent calculi from suitable Hilbert systems~\cite{CiaGalTer08}, and it is now understood how to transform certain semantic properties into nested sequent systems~\cite{Lyo21b,LyoOst24} or hypersequent systems~\cite{Lah13}. Even though such methods yield sizable classes of internal calculi, they are more involved than the method of generating labeled systems.\\

\noindent
\textit{Cut-admissibility is more difficult to establish for internal calculi}. The claim that cut-admissibility is more difficult to shown with internal calculi does not appear to be warranted. Both the labeled and display formalisms yield uniform and modular calculi for extensive classes of logics, yet, general cut-admissibility results exist for labeled calculi~\cite{Sim94} and a general cut-elimination theorem holds for display calculi~\cite{Bel82}.\\

In spite of the various properties attributed to `internal' and `external' systems, we have identified three ways in which `internal' systems (i.e., sequent-style systems with a `structurally similar' and `logically equivalent' formula translation) are useful. The first use concerns a relationship between \emph{formulaic completeness}, which is when every valid formula in a logic is provable in the proof system, and \emph{sequential completeness}, which is when every valid sequent is provable in the proof system. If the rules of an `internal' system are invertible and the system has formuliac completeness, then one can (typically) establish sequential completeness. It is straightforward to establish this property: if we assume a sequent is valid, then its formula translation is valid, meaning the formula translation is provable as the system has formulaic completeness. One can then apply the invertibility of the inference rules to the formula translation to prove the original sequent, which establishes sequential completeness. Although we do not claim that this property holds of any system that might be reasonably deemed `internal' we do note that this method of lifting formulaic completeness to sequential completeness works in a variety of cases; e.g., (linear) nested sequents~\cite{KuzLel18,Lyo21b}. 

A second favorable property of `internal' calculi concerns the lack of a `meta-semantics.' Since sequents are interpreted via their `structurally similar' and `logically equivalent' formula translations, there is no need to define a more general semantics as is done with labeled systems, for example. Third, it has been shown that `internal' (viz., nested) systems can be used to derive Hilbert systems, i.e., axiomatizations, for logics~\cite{IshKik07}. This is obviously beneficial for anyone interested in characterizing a logic purely in terms of its formulae with simple inference rules. 




\Acknowledgements{Tim S. Lyon was supported by the European Research Council, Consolidator Grant DeciGUT (771779).}


\bibliographystyle{BSLbibstyle}
\bibliography{bibliography_BSL}


\end{document}